\documentclass[a4paper,11pt]{article}
\pdfoutput=1
\usepackage[T1]{fontenc} 
\usepackage{jheppub,amsthm,bm,bbm,enumitem,amsmath,amssymb,amsfonts,slashed}

\usepackage{hyperref}
\allowdisplaybreaks

\newcommand{\der}{\partial}
\newcommand{\diag}{\mathrm{diag}}
\newcommand{\Tr}{\mathrm{Tr\,}}
\newcommand{\Str}{\mathrm{Str\,}}
\newcommand{\Sdet}{\mathrm{Sdet\,}}
\newcommand{\Pf}{\mathrm{Pf}}

\newcommand{\RE}{\mathrm{Re}}

\newcommand{\MAT}[2]{\begin{array}{#1}#2\end{array}}
\renewcommand{\bar}[1]{\overline{#1}}

\newcommand{\bep}{\begin{pmatrix}} 
\newcommand{\eep}{\end{pmatrix}}

\newcommand{\SU}{\text{SU}}

\newcommand{\U}{\text{U}}
\newcommand{\1}{\mathbbm{1}}

\DeclareMathOperator\RR{\mathbb{R}}
\DeclareMathOperator\ZZ{\mathbb{Z}}

\newcommand{\Herm}{\mathrm{Herm}}
\newcommand{\ms}{\hspace{-10pt}}

\def\ba#1\ea{\begin{align}#1\end{align}}

\title{Symmetry Crossover Protecting Chirality in Dirac Spectra}

\author[a]{Takuya Kanazawa}
\author[b]{and Mario Kieburg}
\affiliation[a]{Research and Development Group, Hitachi, Ltd., Kokubunji, Tokyo 185-8601, Japan}
\affiliation[b]{Department of Physics, Bielefeld University, PO-Box 100131, D-33501 Bielefeld, Germany}
\emailAdd{tkanazawa@nt.phys.s.u-tokyo.ac.jp}
\emailAdd{mkieburg@physik.uni-bielefeld.de}

\abstract{We consider a random matrix model in the hard edge limit (local spectral statistics at the origin in the limit of large matrix size) which interpolates between the Gaussian unitary ensemble (GUE) and the chiral Gaussian unitary ensemble (chGUE). We show that this model is equivalent to the low-energy limit of certain QCD-like theories in the epsilon-regime.
Moreover, we present a detailed derivation of the microscopic level density as well as the partially quenched and unquenched partition functions. Some of these results have been announced in a former letter by us. Our derivation relies on the supersymmetry method and is performed here step by step. Additionally, we  compute the chiral condensate and the pion condensate for the quenched as well as unquenched settings.  
We also investigate the limits to GUE and chGUE and confirm our conjecture that the non-uniformity of the GUE limit would carry over to the hard edge limit.
}

\begin{document}

\maketitle

\section{Introduction}\label{sec:intro}

Random matrix theory (RMT) is an extremely versatile tool in the statistical description of spectra in physical systems~\cite{handbook:2010}. This is especially true in QCD-like systems where RMT has been applied since the early 90's. Verbaarschot et al.~\cite{Shuryak:1992pi,Verbaarschot:1993pm,Verbaarschot:1994ip} have shown that non-linear sigma models emerge from RMT in the large-$N$ limit as a low-energy effective theory. Such sigma models exactly match those that arise in the $\varepsilon$-regime of QCD under appropriate conditions. For instance, the Compton wavelength of the Nambu-Goldstone modes have to exceed the system size. Then the path integral is dominated by their zero-momentum modes~\cite{Gasser:1987ah,Leutwyler:1992yt} and the contribution from the kinetic term factorizes. 

Usually the terms in the QCD chiral effective theory can be derived by spurion analysis and invoking local flavour symmetries as in~\cite{Gasser:1983yg,Leutwyler:1993iq}. In the same way one can create the corresponding random matrix models. To each quantity like quark mass, chemical potential or lattice spacing one can introduce a dimensionless counterpart in RMT.
The $\varepsilon$-regime of the partition function, then, uniquely fixes and identifies both sets, the physical variables in QCD and the dimensionless variables in RMT. This way one can derive an infinite number of spectral sum rules 
for the QCD Dirac operator along the lines of~\cite{Leutwyler:1992yt} in the QCD vacuum and~\cite{Luz:2006vu,Kanazawa:2009ks,Kanazawa:2011tt} at finite chemical potential, as long as the matrix model is in the same universality class as the considered physical QCD system.

We consider a chiral random matrix model that interpolates between the Gaussian unitary ensemble (GUE) and  the  chiral Gaussian unitary ensemble (chGUE) statistics. This model has been first proposed in~\cite{Bialas:2010hb} to describe the continuum limit of three dimensional staggered fermions. This is one of three possible applications which we discuss in detail in Sec.~\ref{sec:Motivation}. Beside this application, we also point out the possible use of our model to 3d continuum QCD with isospin chemical potential and to 4d gauge theories at high temperature. Furthermore, the considered model is also related to quantum chaos~\cite{Fyodorov:1997,Fyodorov:1997b} where the complex eigenvalues of the off-diagonal matrix block of the chiral matrix have been studied. In this topic the model is known as the elliptic complex Ginibre ensemble~\cite{Sommers,Fyodorov:1997,Akemann:2001bf,Akemann:2007rf,Kanazawa:2013}. In this paper we study its singular value statistics and, thus, a different aspect of this model.

Apart from analysing why our model might fit to QCD-like theories, we derive the low-energy effective partition function in Sec.~\ref{sec:chiL}, which is known as the hard edge scaling limit in RMT, that has been given by us in the letter~\cite{Kanazawa:2018}. By exploiting the fact that the considered random matrix model exhibits a Pfaffian point process~\cite{Kanazawa:2018kbo}, we concentrate on the partition functions of one and two flavours, either bosonic or fermionic, in Sec.~\ref{sec:hardedge}, because they are the building blocks of any spectral correlation.

Another quantity, which we have already reported without proof in the letter~\cite{Kanazawa:2018}, is the miscroscopic level density at the origin. Its derivation is outlined in Sec.~\ref{sec:susy.part}, and the lengthy details of the calculation are given in Appendix~\ref{sec:susy-part}. These computations are based on the supersymmetry method and we refer the reader to \cite{Zirnbauer,Guhr} for a pedagogical introduction. Note that this approach is different to the supersymmetric spurion analysis in~\cite{Bernard:1993,Golterman:2009}. In this section we also analyse the limits to GUE and chGUE and identify quantities which seem to be ideal to measure some low energy constants. Moreover, we study a quantity in Sec.~\ref{sec:observables}, that resembles the chiral condensate in 4d continuum QCD. Indeed it can be identified with the chiral condensate when considering the applications of 3d staggered fermions or 4d QCD with twisted boundary conditions. However for 3d continuum QCD with isospin chemical potential it is essentially the pion condensate, see Sec.~\ref{sec:Chem}.

Additionally to the sections pointed out above, we summarize our results in Sec.~\ref{sec:conclusion} and give details to several computations in Appendix~\ref{app:calc}.

\section{Motivations of the Model}\label{sec:Motivation}

We are interested in the spectral statistics of the chiral random matrix~\cite{Bialas:2010hb,Kanazawa:2018,Kanazawa:2018kbo} 
\begin{equation}\label{RMT-model}
\mathcal{D} = \left(\begin{array}{cc} 0 & iW \\ iW^\dagger & 0 \end{array}\right), \qquad W\equiv H_1+i\mu H_2\,, \quad H_1,H_2\in\Herm(N)\ \text{and}\ \mu\in\mathbb{R}
\end{equation}
drawn from the distribution
\begin{equation}\label{distribution}
P(\mathcal{D})= \frac{1}{2^N \pi^{N^2}}\exp\left[-\frac{1}{2}\Tr(H_1^2+H_2^2)\right].
\end{equation}
The set of Hermitian $N\times N$ matrices is denoted by $\Herm(N)$.

Our analysis starts with the partition function 
of $N_\mathrm{f}$ quarks,
\begin{equation}\label{eq:0Zdef}
		Z_N^{(N_{\rm f})} = \!\!\!\int\limits_{[\Herm(N)]^2}\ms  d  H_1d  H_2 \;P(\mathcal{D})\prod_{f=1}^{N_\mathrm{f}}\det\left(m_{f} \1_{2N}+\sum_{a=1}^3j_{a}\tau_a +\mathcal{D}\right)
\end{equation}
with masses $m_{f}$ and additional source variables $j_{a}$ (${f}=1,\ldots,N_{\rm f}$ and $a=1,2,3$). The source variables are helpful for calculating the observables in section~\ref{sec:observables}. The Pauli matrices $\tau_j$ are embedded in the $2N$-dimensional space as $\1_N\otimes\tau_j$ though we omit the tensor notation. The normalization ensures that $Z_N^{(0)}=1$.   The matrix size $N$ plays here the role of the space-time volume $V$.

When varying $\mu$ the level statistics of $\mathcal{D}$ interpolates between the Gaussian unitary ensemble (GUE)~\cite{Mehta_book,Verbaarschot:1994ip} at $\mu=0$ and the chiral Gaussian unitary ensemble (chGUE)~\cite{Mehta_book,Shuryak:1992pi,Verbaarschot:1993pm} at $\mu=1$, cf. Fig.~\ref{fg:arrow}. Due to the symmetries of this model we can restrict ourselves to the situation where $\mu\in[0,1]$, see~\cite{Kanazawa:2018kbo}.

When considering the spectral statistics of the complex matrix $W$ only, our model is also known as the complex elliptic ensemble~\cite{Sommers,Fyodorov:1997,Akemann:2001bf,Akemann:2007rf}. The spectrum of $W$ is generically complex and its support is given by an ellipse for large $N$, thus the name. The complex eigenvalues play an important role in the scattering at disordered or chaotic systems~\cite{Sommers,Fyodorov:1997,Fyodorov:1997b}.

Let us point out another model which interpolates between GUE and chGUE. It is of the form
\begin{equation}\label{Wilson}
\mathcal{D}_5 =\left(\begin{array}{cc} 0 & W \\ W^\dagger & 0 \end{array}\right)+\mu H,\ W\in\mathbb{C}^{N\times N}\ {\rm and}\ H\in{\rm Herm}(2N).
\end{equation}
This model describes the Hermitian Wilson-Dirac operator~\cite{Akemann:2011}, see also~\cite{Damgaard:2010cz} for an equivalent realization. There is a crucial difference between the models~\eqref{Wilson} and~\eqref{RMT-model}. While the eigenvalues of $\mathcal{D}$ come always in chiral pairs $\pm i \Lambda_n$ with $\Lambda_n\geq 0\;(n=1,\dots,N)$, it is only the case for $\mathcal{D}_5$ when $\mu=0$. As already explained in~\cite{Kanazawa:2018,Kanazawa:2018kbo} this difference is crucial for the behaviour of the eigenvalues close to the origin. When we have chiral pairs of eigenvalues we find a level repulsion from the origin regardless of how small $\mu$ is. This behaviour carries over to the microscopic limit about the origin when taking $N\to\infty$ as we will see below. In particular we are interested in the  scaling $\mu^2\propto 1/N$ because it is where the symmetry crossover sets in. For the scaling $\mu^2\propto1$ we will see that we always obtain the statistics of chGUE.

Before we come to the spectral statistics in the limit $N\to\infty$, let us point out some applications of this random matrix model in QCD. In Sec.~\ref{sec:stag} we discuss the relation to staggered fermions and their continuum limit. Also the high temperature limit of some quantum field theories can be modelled by $\mathcal{D}$, see Sec.~\ref{sec:hightemp}. The third application is presented in Sec.~\ref{sec:Chem} and deals with 3d QCD at finite isospin chemical potential.

\subsection{Staggered Fermions in 3d QCD}\label{sec:stag}
\subsubsection{Symmetries and the Continuum Limit}

Lattice Dirac operators of naive and staggered fermions~\cite{Susskind:1976jm} do not necessarily satisfy the same global symmetries as the continuum Dirac operator. Several lattice simulations~\cite{Damgaard:1998,Damgaard:2002,Bialas:2010hb} have shown this in three and four dimensions for different numbers of colors and fermions in the fundamental or adjoint representation. The good thing is that QCD in the Standard Model (4 dimensions, fermions in the fundamental representation and three colors) does not suffer from this behaviour. Only the rooting and the lack of a well-defined topological charge are problematic in this particular theory, which we will not address here. An attempt to understand this shift of symmetries in detail has been made in two-dimensional QCD-like theories~\cite{Kieburg:2014}. A detailed discussion of this phenomenon for gauge theories in general space-time dimension $d\geq2$ was done in~\cite{Kieburg:2017rrk} where a Bott-periodicity was revealed, not only in the space-time dimension but also in the number of directions with an even number (partition) of lattice sites. Let us briefly recount the situation.

For each direction, which has an even parity of lattice sites, a ``chirality'' operator $\Gamma_j$ ($\Gamma_j=\Gamma_j^\dagger=\Gamma_j^{-1}$ and $\Tr\Gamma_j=0$) can be defined. This operator assigns to an odd-lattice site (counted only in this direction) a ``$+1$'' and to an even lattice site a ``$-1$''. As a consequence, this operator anti-commutes with the naive Dirac operator $\mathcal{D}$ and, hence, generates an additional chiral symmetry. The collection of all these chiral operators $\Gamma_j$ build a Clifford algebra, i.e. $[\Gamma_k,\Gamma_l]_+=2\delta_{kl}$ with $[.,.]_+$ the anti-commutator. This has two consequences. First and foremost, the naive Dirac operator may become (highly) degenerate. Moreover, the Dirac operator changes its global symmetries along the Bott periodicity~\cite{Bott}.

What does this mean for staggered fermions? Staggered fermions are essentially naive fermions with an even parity in each direction. Due to the Bott periodic shift of global symmetries, it was shown in~\cite{Kieburg:2014} that staggered Dirac operators share always the global symmetries of the corresponding eight-dimensional continuum theory. This explains why QCD in the Standard Model does not suffer from this problem because the global symmetries of the four-dimensional and eight-dimensional continuum theory are the same. This is not true for 3d QCD  as well as in any other odd dimension. Staggered fermions of QCD with the gauge group $\SU(N_\mathrm{c}\geq3)$ in the fundamental representation yield always a chiral, complex, anti-Hermitian Dirac operator and, thus, shares the symmetries with the eight-dimensional and, thence, four-dimensional continuum theory. The question is whether the correct global symmetries are recovered when the continuum limit is taken.

\subsubsection{Matrix Model for the Symmetry Shift}

In~\cite{Bialas:2010hb} a mechanism of such a change from symmetries of even to odd dimensional $\SU(N_\mathrm{c}\geq3)$ gauge theory in the fundamental representation  was proposed. They considered the model~\eqref{RMT-model} and fitted Monte-Carlo simulations of this random matrix model~\eqref{RMT-model}  to lattice simulations for three-dimensional staggered fermions in the quenched theory for several coupling constants and lattice sizes. The comparison seems to look surprisingly good despite the fact that the degeneracy of the eigenvalues does not fit with the number of doubler fermions of unrooted staggered fermions. Without rooting the number of flavours of staggered fermions should be enhanced from $N_\mathrm{f}$ to $4N_\mathrm{f}$ in three dimensions. However the Dirac operator~\eqref{RMT-model} is for $\mu=0$ only doubly degenerate. We underline that the continuum limit is given by $\mu=0$ in this model.

A more appropriate model for the described situation would be
\begin{equation}\label{RMT-model-staggered}
\mathcal{D}_{\rm stag} = \left(\begin{array}{cc} 0 & i H\otimes\1_2+i\mu W \\ i H\otimes\1_2+i\mu W^\dagger & 0 \end{array}\right),\ H\in\Herm(N),\ W\in\mathbb{C}^{2N\times 2N}\ \text{and}\ \mu\in\mathbb{R},
\end{equation}
which obviously leads to more terms in the chiral Lagrangian compared to the one in the partition function~\eqref{eq:Zuniv} derived for the model~\eqref{RMT-model}.
We briefly show this, here. For this purpose, we want to consider the partition function
\begin{equation}
Z_{\rm stag}(M)=\int_{\Herm(N)} dH\int_{\mathbb{C}^{2N\times2N}}dW\;P_{\rm stag}(H,W)\prod_{f=1}^{N_{\rm f}}\det(\mathcal{D}_{\rm stag}+m_f\1_{4N})
\end{equation}
with $M=\diag(m_1,\ldots,m_{\rm f})$, splitting
\begin{equation}
W=\sum_{j=0}^3(W_{{\rm R},j}+iW_{{\rm I},j})\chi_j
\end{equation}
into a sum of Hermitian $N\times N$ matrices $W_{{\rm R},j}$ and $W_{{\rm I},j}$ with $\chi_0=\1_2$ and $\chi_{1,2,3}$ the three Pauli matrices. The distribution is chosen to be Gaussian
\begin{equation}
\begin{split}
P_{\rm stag}(H,W)\propto \exp\bigg[-\frac{1}{2}\Tr H^2-\sum_{j=0}^3\Tr \bigg(\frac{1}{2c_{{\rm R},j}^2}W_{{\rm R},j}^2+\frac{1}{2c_{{\rm I},j}^2} W_{{\rm I},j}^2\bigg)\bigg].
\end{split}
\end{equation}
The choice of different standard deviations $c_{{\rm R},j},c_{{\rm I},j}$ takes into account that the lattice is not necessarily  invariant when interchanging its axes. Thus, our model is more in the spirit of~\cite{Osborn:2011} where a random matrix model for 4d staggered fermions has been proposed.

Next, we introduce an $N\times 4N_{\rm f}$ rectangular matrix $ V$ whose matrix entries are independent complex Grassmann variables with a complex conjugation of the second kind, i.e. $V_{ab}V_{cd}=-V_{cd}V_{ab}$, $(V_{ab}V_{cd})^*=V_{ab}^*V_{cd}^*$ and $(V_{ab}^*)^*=-V_{ab}$, see~\cite{Berezin} for an introduction to superanalysis and superalgebra. Then, we can rewrite the product of determinants as a Gaussian integral over $ V$, i.e.
\begin{equation}
\begin{split}
&\prod_{f=1}^{N_{\rm f}}\det(\mathcal{D}_{\rm stag}+m_f\1_{4N})\\
\propto&\int d V \exp\bigg[\Tr  V^\dagger V M\tau_1-i\Tr H V V^\dagger-\mu\sum_{j=0}^3\Tr\big( iW_{{\rm R},j} V\chi_j V^\dagger+W_{{\rm I},j} V\tau_3\chi_j V^\dagger\big)\bigg]
\end{split}
\end{equation}
The Pauli matrices $\tau_j$ originate from the chiral structure in $D_{\rm stag}$ and are embedded in the $4N_{\rm f}$-dimensional space as a tensor $\tau_j\equiv\1_{N_{\rm f}}\otimes\tau_j\otimes\1_2$; similarly we embed $M\equiv M\otimes\1_2\otimes\1_2$ and $\chi_j\equiv\1_{N_{\rm f}}\otimes\1_{2}\otimes\chi_j$. The average over the random matrices yields
\begin{equation}
\begin{split}\label{SUSYBosTrick}
&Z_{\rm stag}(M)\\
\propto&\int d V \exp\bigg[\Tr  V^\dagger V M\tau_1+\frac{1}{2}\Tr( V^\dagger V)^2+\frac{\mu^2}{2}\sum_{j=0}^3\Tr\Big(c_{{\rm R},j}^2( V^\dagger V\chi_j)^2-c_{{\rm I},j}^2( V^\dagger V\tau_3\chi_j)^2\Big)\bigg]
\\
\propto&\int\limits_{\U(4N_{\rm f})}\hspace*{-3mm}\frac{d\mu(U')}{\det {U'}^N}\,\exp\bigg[\sqrt{N}\Tr U' M\tau_1+\frac{N}{2}\Tr {U'}^2+\frac{N\mu^2}{2}\sum_{j=0}^3\Tr\Big(c_{{\rm R},j}^2(U'\chi_j)^2-c_{{\rm I},j}^2(U'\tau_3\chi_j)^2\Big)\bigg],
\end{split}
\end{equation}
where in the last line we employed the bosonization formula~\cite{Sommers:2007,Zirnbauer:2008,Kieburg:2008} and replaced $ V^\dagger V\to\sqrt{N} U'$. The measure $d\mu(U')$ is the Haar measure of the unitary group $\U(4N_{\rm f})$.

In the end we take $N\to\infty$ while keeping $\widehat{M}=\sqrt{N}M$ and $\widehat{\mu}^2=N\mu^2$ fixed, which tells us that the integral concentrates on the manifold $U'=U\tau_3 U^\dagger$ with $U\in\U(4N_{\rm f})$. See the detailed discussion in Sec.~\ref{sec:hardedge} since the saddle point equation is exactly the same. Thence, we end up with
\begin{equation}
\begin{split}\label{eff.Lag.stag}
Z_{\rm stag}(M)\overset{N\gg1}{\propto}&\int_{\U(4N_{\rm f})}d\mu(U)\exp\biggl[\Tr U\tau_3U^\dagger \widehat{M}\tau_1-\frac{\widehat{\mu}^2c_{{\rm I},0}^2}{2}\Tr (U\tau_3U^\dagger\tau_3)^2\\
&+\frac{\widehat{\mu}^2}{2}\sum_{j=1}^3\Tr\left(c_{{\rm R},j}^2(U\tau_3U^\dagger\chi_j)^2-c_{{\rm I},j}^2(U\tau_3U^\dagger\tau_3\chi_j)^2\right)\biggl].
\end{split}
\end{equation}
Our model only describes the first two terms of the chiral Lagrangian, cf.~\eqref{eq:Zuniv}. The inclusion of all terms makes the model certainly more sophisticated but, as a drawback, it becomes also more analytically involved, in particular the finite $N$ discussion~\cite{Kanazawa:2018kbo} would have not been possible anymore.

 Nonetheless, the good qualitative agreement~\cite{Bialas:2010hb}  of the spectral statistics of the Dirac operators of the model~\eqref{RMT-model} and of three-dimensional staggered fermions shows that the random matrix model~\eqref{RMT-model} is worthwhile to study. In particular one might conjecture, that the missing terms proportional to $c_{{\rm R},j>0}^2$ and $c_{{\rm I},j>0}^2$ in the chiral Lagrangian play a sub-leading role, at least in the quenched theory. What happens for the unquenched theory has to be still perused.
	
\subsection{Gauge theories at high temperature}\label{sec:hightemp}
Euclidean quantum field theories generally undergo dimensional reduction at high temperature $T=1/\beta$ \cite{Ginsparg:1980ef,Appelquist:1981vg,Nadkarni:1982kb,Nadkarni:1988fh, Landsman:1989be,Kajantie:1995dw,Braaten:1995jr} where $\beta$ is the circumference of $S^1$. A rough perturbative picture of this phenomenon is as follows. Let us consider QCD-like theories with Dirac fermions $\psi$ on spacetime $\RR^3\times S^1$.  The temporal part of the Lagrangian of $\psi$ reads $\bar\psi\gamma_4[\der_4+iA_4(x)]\psi$. If we substitute $A_4(x)$ by a constant $\langle A_4\rangle$ via a mean field approximation, the eigenvalues of $\der_4+iA_4$ are given by $i[(2n\pi+\theta)/\beta+\langle A_4\rangle]$ with $n\in\ZZ$, where $\theta$ specifies the boundary condition, $\psi(x_4+\beta)=e^{i\theta}\psi(x_4)$. In a thermal phase with a trivial Polyakov loop $P=\1$, one can set $\theta=\pi$ and $\langle A_4\rangle=0$ so that the smallest (in magnitude) eigenvalue is $\pi/\beta$. Therefore in the limit $\pi/\beta\gg\Lambda_{\rm QCD}$ fermions decouple from the low-energy dynamics. In particular, the chiral condensate evaporates and chiral symmetry is restored, inhibiting applications of chiral random matrix theory to this hot phase. 

What happens in other phases is more interesting: when the mass gap in units of $T$, which is $\min_{n\in\ZZ}|2n\pi+\theta+\beta\langle A_4\rangle|$, is small, the chiral symmetry breaking tends to persist up to higher $T$. Let us consider a pure gauge theory, where the situation is simplest. Recall that in hot $\SU(N_c)$ pure gauge theory there are $N_c$ distinct vacua having $\Tr P\propto e^{i\phi}$ with $\phi N_c/2\pi\in\ZZ$. It has been known that $\langle\bar\psi\psi\rangle$ of valence quarks obeying $\theta=\pi$ exhibits a strong dependence on $\phi$ \cite{Chandrasekharan:1995gt,Stephanov:1996he,Chandrasekharan:1998yx}. In particular, when $N_c$ is even, $\langle\bar\psi\psi\rangle$ seems to remain nonzero up to \emph{arbitrarily high} $T$ in the vacuum with $\phi=\pi$. Indeed, substituting $\theta=\pi$ and $\langle A_4\rangle=\pi/\beta$ yields $\min_{n\in\ZZ}|2n\pi+\theta+\beta\langle A_4\rangle|=0$, implying that quarks acquire no perturbative mass gap at high $T$.%
\footnote{The same phenomenon occurs for \emph{any} $N_c$ if we instead impose $\theta=0$ (i.e., periodic boundary condition) and look at the sector with $\phi=0$. For more details on the interplay of chiral symmetry breaking and the fermionic boundary condition, we refer the reader to \cite{Gattringer:2002tg,DeGrand:2006qb,Bilgici:2008qy,Misumi:2014raa}.} 
Such ``high-temperature chiral symmetry breaking'' can also be explained on the basis of topological excitations of the gauge field, called instanton-monopoles or dyons, that carry fermion zero modes due to the index theorem \cite{Bornyakov:2008bg,Bornyakov:2008im,Shuryak:2017kct}. We conjecture that the near-zero region of the quenched Dirac spectrum in this phase would undergo a \emph{dimensional crossover} from chGUE in $d=4$ to non-chiral GUE in $d=3$. This transition can be studied via the model \eqref{eq:0Zdef} with $N_{\rm f}=0$. In this case the parameter $\mu$ signifies the effective size of the fourth dimension. This conjecture can be tested in Monte Carlo simulations. We emphasize that chiral symmetry of the Dirac operator is not explicitly broken throughout the symmetry crossover. 

Whether such a smooth dimensional crossover is possible or not in the presence of dynamical quarks is a highly nontrivial issue. We refer to \cite{Cherman:2016hcd,Kanazawa:2017mgw} for recent works on this subject, which were inspired by the idea of adiabatic continuity in \cite{Unsal:2007vu,Unsal:2007jx,Unsal:2008eg}.

\subsection{3d QCD at Finite Isospin Chemical Potential}\label{sec:Chem}

Let us consider again QCD in three dimensions with the gauge group $\SU(N_\mathrm{c}\geq3)$ in the fundamental representation. For two flavours the ground state will accommodate a pion condensate $\langle\bar{u}d \rangle\ne 0$ at $T=0$ when the isospin chemical potential $\mu_\mathrm{I}$ is large enough, that entails a spontaneous breaking of the $\U(1)$ isospin symmetry, see~\cite{Son:2000xc,Splittorff:2000mm,deForcrand:2007uz,Kanazawa:2011tt,BES} for a similar discussion in four dimensions.

	The fermionic part of the Euclidean Lagrangian with a source term $j_{\pi}$ for the pion condensate 
	(similar to the diquark source term in two-color QCD~\cite{Kogut:2000ek,Kogut:2001na,Dunne:2002vb,Kanazawa:2011tt}) is given by
	\begin{equation}\label{Lag.chem}
	 \mathcal{L} =  \bar{ \psi}\big[\sigma_\nu D_\nu
		+ \mu_{\rm iso} \sigma_3\tau_3+m_\mathrm{u}(\1_2+\tau_3)/2+m_\mathrm{d}(\1_2-\tau_3)/2+j_{\pi}\tau_1\big] \psi
	\end{equation}
	with the covariant derivative $D_\nu = \partial_\nu + iA_\nu$ ($\nu=1,2,3$),  the gauge vector field $A_\mu\in\mathrm{su}(N_\mathrm{c}\geq3)$ and the two quark fields $\bar{ V}=( \bar{u},\bar{d} )$. In this section, we use Einstein's summation convention. Let us recall that the $x_3$-direction is the imaginary time-direction in the 3 dim-theory. Furthermore the Pauli matrices $\sigma_\nu$ act on the spinor space while the Pauli matrices $\tau_l$ act on the flavour index. 
	
	Taking the derivative of Eq.~\eqref{det.chem} with respect to $j_\pi$ at $j_\pi=0$ and averaging over the gauge field configurations,
	one obtains a Banks-Casher-type relation that links the pion 
	condensate to the density of the smallest singular values of $\sigma_\nu D_\nu+m_q +\mu \sigma_3$ 
	\cite{Fukushima:2008su,Kanazawa:2011tt,BES}. Note that one has to set $m_{\rm u} = - m_{\rm d} = m$ in order to derive the
Banks-Casher-type relation.
	Statistical fluctuations of these singular values can be analyzed with 
	the model~\eqref{RMT-model} for one flavour $N_{\rm f}=1$. To see this we briefly show that the chiral Lagrangian of the physical system in the $\varepsilon$-regime and with degenerate quark masses 
	agrees with the one following from the random matrix model~\eqref{RMT-model}.
	
Let $M_{\rm q}=\diag(m_{\rm u},m_{\rm d})$. After the quark fields are integrated out, the partition function comprises the determinant
	\begin{equation}\label{det.chem}
	\begin{split}
	&\det\left(\sigma_\nu D_\nu+M_{\rm q}
		+ \mu_{\rm iso} \sigma_3\tau_3+j_{\pi}\tau_1\right)\\
		=&\pm \det\left[j_{\pi}\1+M_{\rm q}\tau_1+\biggl(\begin{array}{cc} 0 & \sigma_\nu D_\nu+\mu_{\rm iso} \sigma_3 \\ \sigma_\nu D_\nu-\mu_{\rm iso} \sigma_3  & 0 \end{array}\biggl)\right].
	\end{split}
   \end{equation}	
	Comparing with the random matrix model~\eqref{RMT-model} one can identify $\sigma_\nu D_\nu \leftrightarrow  i  H_1$, $\sigma_3\leftrightarrow-H_2$ and $j_{\pi}\leftrightarrow m_f$ for $N_\mathrm{f}=1$. Note that $j_{\pi}$ plays the role of the mass here. The random matrix model, which can be naturally associated with this determinant, can be chosen as
	\begin{equation}\label{RMT-model-isospin}
	\begin{split}
	\mathcal{D}_{\rm iso}=&\left(\begin{array}{cc} 0 & i H +\mu \sigma_3\\ iH-\mu \sigma_3  & 0 \end{array}\right),\quad P_{\rm iso}(H)\propto \exp(-\Tr H^2)
	\end{split}
	\end{equation}
	with $H$ a $2N\times 2N$ dimensional Hermitian matrix. Let us underline that the quantities $j$ and $\mu$ are not equal to their physical counterparts in~\eqref{Lag.chem} but need a rescaling with the low energy constants in the $\varepsilon$-regime. The same also holds for the quark masses $M_{\rm q}$, which will be represented by the dimensionless diagonal $2\times 2$ matrix $\widetilde{M}_{\rm q}$ in the random matrix setting.
	
	We now proceed as in Sec.~\ref{sec:stag} and derive the corresponding chiral Lagrangian in the $\varepsilon$-regime, which corresponds to the partition function
	\begin{equation}
	Z_{\rm iso}(j)=\int\limits_{{\rm Herm}(2N)}\hspace{-4mm}dH\;P_{\rm iso}(H)\det(j\1_{4N}+\widetilde{M}_{\rm q}\tau_1+\mathcal{D}_{\rm iso})\,.
	\end{equation}
	 Anew we introduce a Grassmann valued rectangular matrix $ V$, which is of dimension $2N\times 2$, to rewrite the determinant as a Gaussian integral 
	\begin{equation}
	\det(j\1_{4N}+\mathcal{D}_{\rm iso})\propto\int d V \exp\left[\Tr V^\dagger(j\1_2+\widetilde{M}_{\rm q}\tau_1) V+i\mu\Tr V^\dagger\sigma_3 V\tau_2-i\Tr H V\tau_1 V^\dagger\right].
	\end{equation}
	The integration over the random matrix $H$ can now be readily done and we obtain
	\begin{equation}\label{part.chem.finite}
	\begin{split}
	Z_{\rm iso}(j)\propto&\int d V \exp\left[\Tr V^\dagger(j\1_2+\widetilde{M}_{\rm q}\tau_1) V+i\mu\Tr V^\dagger\sigma_3 V\tau_2+\frac{1}{4}\Tr ( V^\dagger V\tau_1)^2\right]\\
	\propto&\int_{\Herm(2)} \!\!\!dQ\int d V\; e^{\Tr V^\dagger(j\1_2+\widetilde{M}_{\rm q}\tau_1) V+i\mu\Tr V^\dagger\sigma_3 V\tau_2-N\Tr Q^2+\sqrt{N}\Tr Q\tau_1 V^\dagger V}\\
	\propto&\int_{\Herm(2)} \!\!\!dQ\ e^{-N\Tr Q^2}\det\biggl(\frac{j\tau_1+\widetilde{M}_{\rm q}+\mu\tau_3}{\sqrt{N}}+Q\biggl)^N\det\biggl(\frac{j\tau_1+\widetilde{M}_{\rm q}-\mu\tau_3}{\sqrt{N}}+Q\biggl)^N.
	\end{split}
	\end{equation}
	A Hubbard--Stratonovich transformation~\cite{Hub,Strat} has been applied in the second step, where we introduced the integral over the $2\times2$ Hermitian matrix $Q$, and the Grassmann variables were integrated out in the last one.
	
	When keeping $\widehat{j}=\sqrt{N}j$, $\widehat{M}_{\rm q}=\sqrt{N}\widetilde{M}_{\rm q}$, and $\widetilde{\mu}=\mu$ fixed, we find the saddle point equation $Q^2=\1_2$ in  the limit $N\to\infty$. The solutions $Q=\pm\1_2$ are algebraically suppressed compared to $Q=U\tau_3U^\dagger$ with $U\in\U(2)$. Thence, the asymptotics of the partition function~\eqref{part.chem.finite} is
	\begin{equation}
	\begin{split}
	Z_{\rm iso}(j)\propto&\int_{\U(2)} d\mu(U) \exp\left[2\Tr U\tau_3U^\dagger(\widehat{j}\tau_1+\widehat{M}_{\rm q})-\widehat{\mu}^2\Tr(U\tau_3U^\dagger\tau_3)^2\right].
	\end{split}
	\end{equation}
	This is the same result when applying the spurion analysis for the physical QCD-model~\eqref{Lag.chem} where the chiral Lagrangian at the leading order is equal to
	\begin{equation}\label{eff.Lag}
	\mathcal{L}_\mathrm{eff} =\Sigma\Tr U\tau_3U^\dagger M_{\rm q}+W\,j_\pi\Tr U\tau_3U^\dagger\tau_1+\frac{F^2}{2}\,\mu_{\rm iso}^2\Tr(U\tau_3U^\dagger\tau_3)^2.
	\end{equation}
	$\Sigma$ is the condensate $\langle\bar{u}u-\bar{d}d\rangle$, $F$ is the pion decay constant, and $W$ is another low energy constant related to the pion condensate $\langle\bar{u}d \rangle$. Comparing~\eqref{part.chem.finite} and~\eqref{eff.Lag}, we notice that the dimensionless quantities from the random matrix model are given by the physical quantities as $2\widehat{M}_{\rm q}=V\Sigma M_{\rm q}$, $2\widehat{j}=VW\,j_\pi$, and $\widehat{\mu}^2=VF^2\mu_{\rm iso}^2/2$, that entails also the physical scaling $M_{\rm q},j_\pi,\mu_{\rm iso}^2\sim1/V$.
	
For degenerate quark masses $m_\mathrm{u}=m_\mathrm{d}$ the chiral Lagrangian reduces to the one, which results from the random matrix model~\eqref{RMT-model-staggered} considered by us, see Sec.~\ref{sec:hardedge}, though, the case $m_\mathrm{u}=-m_\mathrm{d}$ for the Banks-Casher-type relation~\cite{Fukushima:2008su,Kanazawa:2011tt,BES} can be readily obtained from a slight modification of our calculations in Appendix~\ref{sec:ferm-part}.

Indeed, at some point one sees the difference between the models~\eqref{RMT-model} and~\eqref{RMT-model-isospin}. When choosing the parameter $\mu\sim \sqrt{N}$ on the level of random matrix theory or $V\mu_{\rm iso}^2\propto V$  the spectrum of $\mathcal{D}_{\rm iso}$ develops a spectral gap about the origin, whereas $\mathcal{D}$ of~\eqref{RMT-model} does not. Thus, the relation between $\mathcal{D}$ and $\mathcal{D}_{\rm iso}$ is similar to the relation of the Osborn model~\cite{Osborn:2004} and the Stephanov model~\cite{Stephanov:1996ki} for the baryon chemical potential.

\section{\label{sec:chiL}Effective Lagrangians of the Pseudo-Scalar Mesons}

In Sec.~\ref{sec:hardedge} we derive the effective Lagrangian of the model~\eqref{RMT-model} in the hard edge limit, which is the RMT counterpart of the $\varepsilon$-regime in QCD. In this way we establish the intimate connection between our model and the applications delineated in Sec.~\ref{sec:Motivation}. As a cross check  we study the limit to the GUE (3d QCD) and the chGUE (4d QCD) result in Sec.~\ref{sec:limitLag}. Furthermore we derive the macroscopic level density of the model~\eqref{RMT-model} in Sec.~\ref{sec:macro}, to underline that one has to be careful in comparing the hard edge results of RMT with Monte Carlo simulations at finite matrix dimension $N$.

\subsection{Effective Lagrangian}\label{sec:hardedge}

The connection of RMT with QCD is given via the non-linear $\sigma$-model which is the chiral Lagrangian in the 4-dimensional continuum QCD. The reason why these two very different theories are related is due to same spontaneous breaking of global symmetries. Thus we consider the partition function~\eqref{eq:0Zdef} for $N_{\rm f}\geq 1$ in the large-$N$ limit. We first map this partition function
\begin{equation}\label{partition.b}
		Z_N^{(N_{\rm f})} = \frac{1}{2^N \pi^{N^2}}\!\!\!\int\limits_{[\Herm(N)]^2}\!\!\!\!\! d  H_1 d  H_2\; e^{-\Tr(H_1^2+H_2^2)/2}\prod_{f=1}^{N_\mathrm{f}}\det\Bigg(
			\MAT{cc}{ m_{f} \1_{N} & iH_1-\mu H_2 \\ iH_1+\mu H_2 & m_{f} \1_{N} }\Bigg) 
\end{equation}
 to a dual matrix space whose dimension only depends on $N_\mathrm{f}$. This method is called the supersymmetry method and introductions can be found in~\cite{Zirnbauer,Guhr}. In Secs.~\ref{sec:stag} and~\ref{sec:Chem}, we have shown two kinds of its procedure, namely the superbosonization and the Hubbard--Stratonovich approach, respectively. Here, we pursue the superbosonization approach~\cite{Sommers:2007,Zirnbauer:2008,Kieburg:2008}.
 
In the first step, we introduce a rectangular matrix $V$ of dimension $N\times 2N_\mathrm{f}$ whose matrix entries consist of independent complex Grassmann variables. We rewrite the product of determinants as  
\begin{equation}\label{det-ident}
\prod_{{f}=1}^{N_\mathrm{f}}\det\Bigg(\MAT{cc}{ m_{f} \1_{N} & iH_1-\mu H_2 \\ iH_1+\mu H_2 & m_{f} \1_{N} }\Bigg)=\frac{\displaystyle \int dV \exp(\Tr V^\dagger V M+ i  \Tr V^\dagger H_1 V\tau_1- i \mu\Tr V^\dagger H_2 V \tau_2)}{\displaystyle \int dV \exp(\Tr V^\dagger V)}
\end{equation}
where we define $M=\diag(m_1,\ldots,m_{N_\mathrm{f}})$. The denominator on the right hand side correctly normalizes the Gaussian integral over $V$. Let us emphasize that $M$ and the Pauli matrices $\tau_j$ act on two different components of a tensor space; we embed them as $M\equiv M\otimes\1_2$ and $\tau_j\equiv\1_{N_{\rm f}}\otimes\tau_j$ though this is an abuse of the notation.

The average over $H_1$ and $H_2$ in Eq.~\eqref{partition.b} can be readily performed and yields 
\begin{equation}\label{partition.c}
		Z_N^{(N_{\rm f})} =\frac{\displaystyle \int dV \exp\bigg[\Tr V^\dagger V M+\frac12\Tr  (V^\dagger V\tau_1)^2+\frac{\mu^2}{2}\Tr (V^\dagger V\tau_2)^2\bigg]}{\displaystyle \int dV \exp(\Tr V^\dagger V)}\;.
\end{equation}
We recall that we get additional signs in the exponential function due to the anti-commuting nature of the matrix entries of $V$.

Now we are ready to apply the superbosonization formula~\cite{Sommers:2007,Zirnbauer:2008,Kieburg:2008} and replace the nilpotent $2N_\mathrm{f}\times 2N_\mathrm{f}$ matrix $V^\dagger V$ by $\sqrt{N} U'$ with $U'\in\mathrm{U}(2N_\mathrm{f})$. The scaling factor $\sqrt{N}$ is needed to perform the saddle point analysis. As a price of this exchange, we not only replace the flat Berezin measure $dV$ by the Haar measure $d\mu(U')$ on $\mathrm{U}(2N_\mathrm{f})$, but also get a factor $\det^{-N} U'$, as it has been the case in Eq.~\eqref{SUSYBosTrick}. The additional term reflects the nature of the integral over the Grassmann valued matrix $V$, that picks out only the highest term of a Taylor expansion in $V^\dagger V$. The group integral with the term $\det^{-N} U'$ is a multidimensional contour integral selecting the correct terms of this Taylor expansion. This  way we arrive at
\begin{equation}\label{partition.d}
		Z_N^{(N_{\rm f})} =\frac{\displaystyle \int_{\mathrm{U}(2N_\mathrm{f})}\!\!\! d\mu(U')\,{\det}^{-N} U' \exp\bigg[\sqrt{N}\Tr U' M+\frac{N}{2}\Tr (U'\tau_1)^2+\frac{N\mu^2}{2}\Tr (U'\tau_2)^2\bigg]}{\displaystyle \int_{\mathrm{U}(2N_\mathrm{f})}\!\!\! d\mu(U')\,{\det}^{-N} U' \exp\big(\sqrt{N}\Tr U'\big)} \;,
\end{equation}
where the denominator is only a constant depending on $N$ and $N_\mathrm{f}$ but nothing else.

The double scaling we are looking into is given by $\widehat{M}=\sqrt{N}M$ and $\widehat{\mu}=\sqrt{N}\mu$ fixed when $N\to\infty$. This scaling is obviously different from the Stephanov-type model~\eqref{RMT-model-isospin} and originates from the fact that the parameter $\mu$ is the prefactor of a random matrix $H_2$ with fully occupied matrix entries, cf.~\eqref{RMT-model}, while for the model~\eqref{RMT-model-isospin} $\mu$ stands in front of a fixed diagonal matrix.

To get a finite result in this limit, we need to renormalize the partition function by a factor $C_{N,N_\mathrm{f}}$ that only depends on $N$ and $N_\mathrm{f}$, i.e. $\widehat{Z}_N^{(N_{\rm f})}(\widehat{M})=C_{N,N_\mathrm{f}}Z_N^{(N_{\rm f})}$. The explicit form of $C_{N,N_\mathrm{f}}$ is irrelevant for physics, because it depends on the random matrix model and is, hence, not universal. We assume that it is chosen such that the resulting partition function is given by an integral with the normalized Haar measure, cf.~\eqref{eq:Zuniv}. We also absorb the denominator in Eq.~\eqref{partition.d} in this constant.

Taking $N\to\infty$ we perform a saddle point approximation. We first have to solve the saddle point equation 
\begin{equation}
(U'\tau_1)^2=\1_{2N_\mathrm{f}}.
\end{equation}
Hence the eigenvalues of $U'\tau_1$ are $\pm 1$. Diagonalizing $U'\tau_1$ yields a Jacobian proportional to the modulus square of the Vandermonde determinant of the eigenvalues of $U'\tau_1$. This Vandermonde determinant implies that all solutions with $\Tr U'\tau_1\neq0$ are algebraically suppressed by factors of $1/\sqrt{N}$ to those solutions which have an equal number of eigenvalues $+1$ and $-1$. All of these contributing solutions are unitarily equivalent such that the general solution is given by $U'\tau_1=U\tau_3 U^\dagger$ with $U\in \mathrm{U}(2N_\mathrm{f})$. When plugging this result into Eq.~\eqref{partition.d}, we finally arrive at the effective partition function
\begin{equation}\label{eq:Zuniv}
\begin{split}
\widehat{Z}^{(N_{\rm f})}(\widehat{M})=&\int_{\mathrm{U}(2N_\mathrm{f})} \!\!\!\! d\mu(U)\exp\left[\Tr U\tau_3U^\dagger \widehat{M}\tau_1-\frac{\widehat{\mu}^2}{2}\Tr(U\tau_3U^\dagger\tau_3)^2\right],
\end{split}
\end{equation}
cf. Eqs.~\eqref{eff.Lag.stag} and~\eqref{eff.Lag}. This equation is the main result of this section.

Actually, the integration is effectively only over the coset $\mathrm{U}(2N_\mathrm{f})/[\mathrm{U}(N_\mathrm{f})\times\mathrm{U}(N_\mathrm{f})]$. It is the manifold for the Nambu-Goldstone bosons of the spontaneous symmetry breaking $\mathrm{U}(2N_\mathrm{f})\rightarrow\mathrm{U}(N_\mathrm{f})\times\mathrm{U}(N_\mathrm{f})$ agreeing with those of 3d continuum QCD~\cite{Verbaarschot:1994ip}.

\subsection{Limits of the Effective Lagrangians}\label{sec:limitLag}

There are two interesting limits of the effective partition function~\eqref{eq:Zuniv}. 
First and foremost, we can take $\mu\to 0$ followed by  
a $\pi/4$-rotation $U\to e^{-i\pi \tau_2/4}U$. Then, the effective partition function is given by  
\begin{equation}\label{eq:Zuniv.mu0}
\begin{split}
\widehat{Z}^{(N_{\rm f})}(\widehat{M})\overset{\widehat{\mu}=0}{=}&\int_{\mathrm{U}(2N_\mathrm{f})} \!\!\! d\mu(U)\exp\big(\Tr U\tau_3U^\dagger \widehat{M}\tau_3\big).
\end{split}
\end{equation}
This is exactly the finite volume partition function of QCD 
in three dimensions~\cite{Verbaarschot:1994ip,Damgaard:1997pw,Szabo:2005gi}. 

The second limit is given by $\widehat{\mu}\gg1$, which is slightly more involved. The ground state corresponds to the minimum of 
$\Tr(U\tau_3 U^\dagger \tau_3)^2$.  To determine it, we consider the Hermitian matrix $U\tau_3 U^\dagger \tau_3+\tau_3U\tau_3 U^\dagger $ that satisfies
\begin{equation}
0\leq \Tr(U\tau_3 U^\dagger \tau_3+\tau_3U\tau_3 U^\dagger)^2=2\Tr(U\tau_3 U^\dagger \tau_3)^2+4N_\mathrm{f}\ \Longleftrightarrow\ \Tr(U\tau_3 U^\dagger \tau_3)^2\geq-2N_\mathrm{f}.
\end{equation}
The equality holds if and only if $U\tau_3 U^\dagger \tau_3=-\tau_3U\tau_3 U^\dagger$. Hence $U\tau_3 U^\dagger$ has to have a chiral form. In addition, it is unitary and Hermitian, i.e.
\begin{equation}
U\tau_3 U^\dagger=\Bigg(\MAT{cc}{0&\widetilde{U} \\ \widetilde{U}^\dagger &0}\Bigg)\quad{\rm with}\ \widetilde{U}\in\mathrm{U}(N_\mathrm{f})
\end{equation}
We plug this into Eq.~\eqref{eq:Zuniv} and find
\begin{equation}\label{eq:Zuniv.muinf}
\begin{split}
\widehat{Z}^{(N_{\rm f})}(\widehat{M})\overset{\widehat{\mu}\gg1}{\propto}&\int_{\mathrm{U}(N_\mathrm{f})} \!\!\! d\mu(\widetilde{U})\exp\big[\Tr \widehat{M}(\widetilde{U} +\widetilde{U}^\dagger)\big].
\end{split}
\end{equation}
This is
the conventional $\varepsilon$-regime partition function of 
QCD in four dimensions~\cite{Gasser:1987ah,Leutwyler:1992yt}. 

\begin{figure}[t]
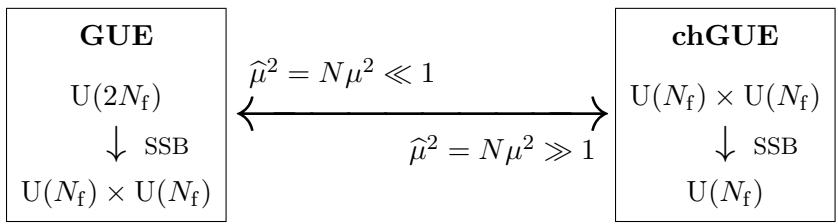

	\begin{eqnarray*}
		\put(-57,5){$\widehat{\mu}^2=N\mu^2\ll 1$}
		\put(3,-24){$\widehat{\mu}^2=N\mu^2\gg 1$}
		\put(-148,-10){\fbox{$\MAT{c}{
			\mathrm{\bf GUE}
			\vspace{8pt}\\
			\U(2N_{\rm f}) \vspace{3pt}\\ ~~~~~~\,
			\scalebox{1.4}{$\downarrow$}\;~\text{\footnotesize SSB} 
			\vspace{3pt}\\
			\U(N_{\rm f})\times \U(N_{\rm f})
		}$}}
		\put(80,-10){\fbox{$\MAT{c}{
			\mathrm{\bf chGUE}
			\vspace{8pt}\\ 
			\U(N_{\rm f})\times\U(N_{\rm f}) \vspace{3pt}\\ ~~~~~~\,
			\scalebox{1.4}{$\downarrow$}\;~\text{\footnotesize SSB} 
			\vspace{3pt}\\ \U(N_{\rm f})
		}$}}
		\put(-63,-13){\Huge $\overleftrightarrow{\hspace{142pt}}$}
	\end{eqnarray*}
	\caption{\label{fg:arrow}
	The large-$N$ crossover in the matrix model \eqref{eq:0Zdef} and its effective 
	theory \eqref{eq:Zuniv} from GUE to chGUE for varying $\widehat{\mu}^2=N\mu^2$. 
	The patterns of spontaneous symmetry breaking (SSB) in the massless limit are shown 
	in the two extremes.}
\end{figure}

As schematically shown in Fig.~\ref{fg:arrow}, 
the parameter $\widehat{\mu}=N\mu^2$ effectively controls the symmetry 
in the low-energy theory.  The coset manifold that represents the Nambu-Goldstone 
field also evolves with $\widehat{\mu}=N\mu^2$ accordingly; with increasing $\widehat{\mu}^2=N\mu^2$, 
some of the Nambu-Goldstone modes acquire a mass gap 
and gradually decouple from the low-energy physics. Note that it is indeed $[\mathrm{U}(N_\mathrm{f})\times \mathrm{U}(N_\mathrm{f})]/\mathrm{U}(N_\mathrm{f})\subset \mathrm{U}(2N_\mathrm{f})/[\mathrm{U}(N_\mathrm{f})\times \mathrm{U}(N_\mathrm{f})]$ which are the two Nambu-Goldstone manifolds corresponding to the two limits. 
We emphasize that the symmetry crossover in Fig.~\ref{fg:arrow} is different from the one realized by Wilson fermions~\cite{Akemann:2011,Damgaard:2010cz}, which break chiral symmetry explicitly. 

\subsection{Macroscopic Level Density}\label{sec:macro}

Let us turn to the macroscopic level density despite the lack of direct connection to continuum QCD. The reason why we want to calculate it comes from an observation while we have performed Monte Carlo simulations of the model~\eqref{RMT-model} (see Ref.~\cite{Kanazawa:2018} as well as Sec.~\ref{sec:susy.part}) that the magnitude of the level density at the origin varies with $\mu$ when $\mu$ becomes large. Hence, one has to be careful when simulating the random matrix model~\eqref{RMT-model} and comparing it with our results in the hard edge limit (microscopic spectral statistics about the origin).

To understand this behaviour let us take up the fermionic partition function~\eqref{partition.b} for $N_{\rm f}=1$ and take the limit $N\to\infty$ at fixed $M/\sqrt{N}=ix-\epsilon=ix^+$ and $\mu$. It is known that the macroscopic level density $R_1(x)$ is obtained as 
\begin{equation}
R_1(x)\propto \lim_{\epsilon\to0}{\rm Im}\,\lim_{N\to\infty}\partial_x{\rm ln}\,Z_N^{(N_{\rm f}=1)}(M=i\sqrt{N}x^+)\,.
\end{equation}
Note that the order of the limits and the derivative is important to select the correct solution. Moreover, we want to emphasize that this relation only works for the global spectral statistics. 
The calculation of the microscopic spectral density will be postponed until Sec.~\ref{sec:susy.part}.

The quotient $Z_N^{(N_{\rm f}=1)}(M=i\sqrt{N}x)/Z_N^{(N_{\rm f}=0)}$ is the skew-orthorgonal polynomial corresponding to the random matrix ensemble~\eqref{RMT-model}, see Ref.~\cite{Kanazawa:2018kbo} where the following formula has been derived
\begin{equation}
\frac{Z_N^{(N_{\rm f}=1)}(M=i\sqrt{N}x^+)}{Z_N^{(N_{\rm f}=0)}}=N!\oint\frac{dz}{2\pi iz^{N+1}}\frac{[1-(1+\mu^2)z]^{N+1}}{\sqrt{1-2(1+\mu^2)z+4\mu^2 z^2}}e^{N(x^+)^2 z}.
\end{equation}
Thus the level density is proportional to
\begin{equation}
R_1(x)\propto \lim_{\epsilon\to0}{\rm Im}\,\lim_{N\to\infty}x^+\frac{\displaystyle\oint\frac{dz}{2\pi iz^{N+1}}\,z\,\frac{[1-(1+\mu^2)z]^{N+1}}{\sqrt{1-2(1+\mu^2)z+4\mu^2 z^2}}e^{N(x^+)^2 z}}{\displaystyle\oint\frac{dz}{2\pi iz^{N+1}}\frac{[1-(1+\mu^2)z]^{N+1}}{\sqrt{1-2(1+\mu^2)z+4\mu^2 z^2}}e^{N(x^+)^2 z}}.
\end{equation}
For large $N$ one has to solve the saddle point equation
\begin{equation}
(x^+)^2-\frac{1}{z}-\frac{1+\mu^2}{1-(1+\mu^2)z}=0,
\end{equation}
which yields the two solutions $z_\pm=[x^+\pm\sqrt{(x^+)^2-4(1+\mu^2)}]/[2x^+(1+\mu^2)]$. Only $z_{-{\rm sign}(x)}$, with ${\rm sign}(x)=x/|x|$ the sign of $x$, survives the limit $N\to\infty$ due to the finite increment $\epsilon>0$ so that the macroscopic level density is
\begin{equation}
R_1(x)\propto \lim_{\epsilon\to0}{\rm Im}\bigg[
x^+\frac{x^+-{\rm sign}(x)\sqrt{(x^+)^2-4(1+\mu^2)}}{2x^+(1+\mu^2)}\bigg]=-\frac{\sqrt{4(1+\mu^2)-x^2}}{2(1+\mu^2)}
\end{equation}
or after proper normalization we find the Wigner semi-circle,
\begin{equation}
R_1(x)=2N\frac{\sqrt{4(1+\mu^2)-x^2}}{2\pi(1+\mu^2)},
\end{equation}
of radius $2\sqrt{1+\mu^2}$. It is this $\mu$-dependence of the radius which changes the height of the level density about the origin which is $1/[\pi\sqrt{1+\mu^2}]$.

The relation between the macroscopic level density $R_1(x)$ of the eigenvalues and the microscopic level density $\rho(\lambda)$ follows from the relation $x=\lambda/N$ between the global and local coordinates of the eigenvalues: 
\begin{equation}\label{densit.rel}
\lim_{\lambda\to\infty}\rho(\lambda)=\frac{1}{N}\lim_{x\to0}R_1(x)=\frac{2}{\pi\sqrt{1+\mu^2}}\,.
\end{equation}
We want to point out that quite often the relation between the two regimes is also chosen to be $x=\lambda/(2N)$, meaning divided by the whole matrix dimension of $\mathcal{D}$, which leads to the $1/\pi$ asymptotics for chGUE , see~\cite{Verbaarschot:1993pm,Verbaarschot}.

As a conclusion from the above discussion, in the hard edge limit, which is the one related to the $\varepsilon$-regime of QCD, the coupling parameter $\mu$ scales as $1/\sqrt{N}$ and, hence, the asymptotic height is $\lim_{\lambda\to\infty}\rho(\lambda)=2/\pi$. On the other hand, when $\mu$ is large enough we have to take into account a correction when comparing the microscopic limit results in Sec.~\ref{sec:susy.part} with finite $N$ Monte Carlo simulations. For example for $\widehat{\mu}=\sqrt{N}\mu=2$ with $N=50$
it is still of about 5\% and, thence, visible. Therefore one has to go to much larger matrix sizes when increasing the rescaled parameter $\widehat{\mu}$; noting that $N$ has to scale quadratically with $\widehat{\mu}$.

\section{Partition Function with Valence Quarks}\label{sec:hardsusy}

Now we turn to partition functions with valence quarks, i.e.
\begin{equation}\label{eq:Zval}
		Z_N^{(k_\mathrm{b},k_\mathrm{f})} = \frac{1}{2^N \pi^{N^2}}\!\!\!\int\limits_{[\Herm(N)]^2}  \ms d  H_1d  H_2\; e^{-\Tr(H_1^2+H_2^2)/2}\;\frac{\displaystyle \prod_{j=1}^{k_\mathrm{f}}\det\Bigg(
			\MAT{cc}{ \kappa_{\mathrm{f}, j} \1_{N} & iH_1-\mu H_2 \\ iH_1+\mu H_2 & \kappa_{\mathrm{f}, j}\1_{N} }\Bigg)}{\displaystyle \prod_{j=1}^{k_\mathrm{b}}\det\Bigg(
			\MAT{cc}{ \kappa_{\mathrm{b}, j} \1_{N} & iH_1-\mu H_2 \\ iH_1+\mu H_2 & \kappa_{\mathrm{b}, j}\1_{N} }\Bigg)}\,.
\end{equation}
The bosonic valence quarks have a non-vanishing real part, $\mathrm{Re}\,\kappa_{j,\mathrm{b}}\neq0$, to guarantee the integrability. Usually one sets $k_\mathrm{b}=k_\mathrm{f}-N_\mathrm{f}=k$ and chooses the first $N_\mathrm{f}$ masses $\kappa_{\mathrm{f}, j}$ equal to the masses of the dynamical quarks and the remaining $\kappa_{\mathrm{f}, j}$ and $\kappa_{\mathrm{b}, j}$ being the valence quark masses which might be complex as it is the case for calculating the $k$-point correlation function.

In~\cite{Kanazawa:2018kbo}, we have shown for the model~\eqref{RMT-model} at finite $N$ that the whole spectral statistics are completely described by the partition function of one and two bosonic or/and fermionic flavours. The reason is that the spectral statistics of the singular values of~\eqref{RMT-model} exhibit a Pfaffian point process, see~\cite{Pfaffian} for the definition. This carries over to the hard edge limit and, hence, to the physical QCD systems summarized in Sec.~\ref{sec:Motivation}. Therefore we especially concentrate on those quantities when giving explicit results.

In the following discussion, we pursue the same strategy as in Sec.~\ref{sec:hardedge}, but extend the ideas to superspace. First we arrange the masses in a diagonal $(k_\mathrm{b}|k_\mathrm{f})\times(k_\mathrm{b}|k_\mathrm{f})$-dimensional supermatrix, $\kappa=\diag(\kappa_{\mathrm{b}, 1} ,\ldots,\kappa_{\mathrm{b}, k_\mathrm{b}} ;\kappa_{\mathrm{f}, 1} ,$ $\ldots,\kappa_{\mathrm{f}, k_\mathrm{f}} )$. Here we use the notation that the boson-boson block is in the upper left block of a supermatrix and the fermion-fermion block in the lower right block. For the preparation of our calculation, we additionally need the diagonal supermatrix $L=\diag(L_1,\ldots, L_{k_\mathrm{b}};\1_{k_\mathrm{f}})$ with the signs $L_j=\mathrm{sign}( \mathrm{Re}\,\kappa_{\mathrm{b},j})=\pm 1$ to keep the integrability throughout the calculation.

In the first step we introduce a complex rectangular matrix $V$ which is this time supersymmetric with dimensions $N\times(2k_\mathrm{b}|2k_\mathrm{f})$. The supersymmetric counterpart of the identity~\eqref{det-ident} is
\begin{equation}\label{Sdet-ident}
\frac{\displaystyle \prod_{j=1}^{k_\mathrm{f}}\det\Bigg(
			\MAT{cc}{ \kappa_{\mathrm{f}, j} \1_{N} & iH_1-\mu H_2 \\ iH_1+\mu H_2 & \kappa_{\mathrm{f}, j}\1_{N} }\Bigg)}{\displaystyle \prod_{j=1}^{k_\mathrm{b}}\det\Bigg(
			\MAT{cc}{ \kappa_{\mathrm{b}, j} \1_{N} & iH_1-\mu H_2 \\ iH_1+\mu H_2 & \kappa_{\mathrm{b}, j}\1_{N} }\Bigg)}=\frac{\displaystyle \int dV\;e^{-\Str V^\dagger VL\kappa- i  \Str V^\dagger H_1 VL\tau_1+ i \mu\Str V^\dagger H_2 V L\tau_2}}{\displaystyle \int dV \exp[-\Str V^\dagger V]}\;.
\end{equation}
The supertrace of a supermatrix
\begin{equation}\label{susy:matrix}
\sigma=\left(\begin{array}{cc} \sigma_\mathrm{bb} & \sigma_\mathrm{bf} \\ \sigma_\mathrm{fb} & \sigma_\mathrm{ff} \end{array}\right)
\end{equation}
is $\Str\sigma=\Tr\sigma_\mathrm{bb}-\Tr\sigma_\mathrm{ff}$ and the related superdeterminant is given as $\Sdet\sigma=\det(\sigma_\mathrm{bb}-\sigma_\mathrm{bf}\sigma_\mathrm{ff}^{-1}\sigma_\mathrm{fb})/\det\sigma_\mathrm{ff}$.

Now we are able to integrate over the two Hermitian matrices $H_1$ and $H_2$ and  find
\begin{equation}\label{susy:partiton}
Z_N^{(k_\mathrm{b},k_\mathrm{f})} =\frac{1}{\pi^{2 Nk_\mathrm{b}}}\int dV \exp\bigg[
-\Str V^\dagger V L\kappa
-\frac{1}{2}\Str(V^\dagger VL\tau_1)^2
-\frac{\mu^2}{2}\Str (V^\dagger VL\tau_2)^2
\bigg]\,.
\end{equation}
Here we have chosen the normalization of an integral over a Grassmann variable $\eta$ as $\int d\eta=0$ and $\int \eta d\eta=1$ and the order in its measure is $d\eta d\eta^*$.

In the final step we replace the supermatrix $V^\dagger V L\tau_1$ by the supermatrix $\sqrt{N}U'\in\mathrm{Gl}_{\mathbb{C}}(2k_\mathrm{b}|2k_\mathrm{f})/\mathrm{U}(L_\mathrm{bb}\tau_1|2k_{\rm f})$ with the help of the superbosonization formula~\cite{Sommers:2007,Zirnbauer:2008,Kieburg:2008}.
What does this explicitly mean, especially the first part of the notation $\mathrm{U}(L_\mathrm{bb}\tau_1|2k_{\rm f})$? The matrix $U'$ has explicitly the form
\begin{equation}\label{susy:U}
U'=\left(\begin{array}{cc} U'_\mathrm{bb} & \eta^\dagger \\ \eta & U'_\mathrm{ff} \end{array}\right)
\end{equation}
with $U'_\mathrm{ff}\in\mathrm{U}(2k_\mathrm{f})$ being an ordinary unitary matrix and $U'_\mathrm{bb}L_\mathrm{bb}\tau_1$ being a positive definite Hermitian matrix.  Since the embedding of the non-compact, ordinary group $\mathrm{U}(k_{\rm b},k_{\rm b})$ in the supergroup $\mathrm{U}(L_\mathrm{bb}\tau_1|2k_{\rm f})$ is non-trivially given via the matrix $L_\mathrm{bb}\tau_1$ we have highlighted it in our notation. The off-diagonal block $\eta$ is a $2k_\mathrm{f}\times2k_\mathrm{b}$ rectangular matrix with independent complex Grassmann variables as matrix entries and $\eta^\dagger$ is its Hermitian adjoint. 

After exploiting the superbosonization formula, we arrive at
\begin{equation}\label{susy:partition.b}
		Z_N^{(k_\mathrm{b},k_\mathrm{f})} =C_N^{(k_\mathrm{b},k_\mathrm{f})} \int d\tilde{\mu}(U')\;\Sdet^{N}U' \exp\bigg[-\Str U'\widehat{\kappa}\tau_1-\frac{N}{2}\Str  {U'}^2+\frac{\widehat{\mu}^2}{2}\Str (U'\tau_3)^2\bigg]
\end{equation}
with $\widehat{\kappa}=\sqrt{N}\kappa$ and $\widehat{\mu}=\sqrt{N}\mu$. The (non-normalized) Haar measure $d\tilde{\mu}(U')$ can be expressed in terms of the flat measure $dU'$ (products of the differentials of independent matrix entries) as $d\tilde{\mu}(U')=\Sdet( U')^{2(k_\mathrm{f}-k_\mathrm{b})}dU'$. The normalization constant is
\begin{equation}\label{norm-SUSY}
\begin{split}
1/C_N^{(k_\mathrm{b},k_\mathrm{f})} =&\int d\tilde{\mu}(U')\;\Sdet^{N}U' \exp\big(-\sqrt{N}\Str U'L\tau_1\big)\\
&\hspace*{-2cm}=\frac{(2\pi)^{2k_\mathrm{f}}}{(-N)^{N(k_\mathrm{b}-k_\mathrm{f})}}\left(\prod_{j=0}^{2k_\mathrm{b}-1}\pi^j(N-j-1)!\right)\left(\prod_{l=0}^{2k_\mathrm{f}-1}\frac{\pi^l}{(N+l)!}\right)\left(\prod_{j=1}^{2k_\mathrm{b}}\prod_{l=1}^{2k_\mathrm{f}}(N+l-j)\right)
\end{split}
\end{equation}
which can be checked for $\kappa\to\infty$, where the partition function becomes $Z_N^{(k_\mathrm{b},k_\mathrm{f})}\to\Sdet^{-2N}\kappa$. We derived it in detail in Appendix~\ref{sec:norm-susy}.

Starting from Eq.~\eqref{susy:partition.b} we perform the hard edge limit for particular cases. In Secs.~\ref{sec:ferm} and~\ref{sec:bos} we consider the partition functions with either only fermionic or bosonic quarks, respectively. The partially quenched case is analyzed in Sec.~\ref{sec:susy.part} where we also derive the microscopic level density.

\subsection{The case $(k_\mathrm{b},k_\mathrm{f})=(0,N_\mathrm{f})$}\label{sec:ferm}

The case with only fermionic quarks has been already discussed in Sec.~\ref{sec:hardedge}. Here, we keep track of all constants and give very explicit results for the one and two flavour partition function. Especially the latter two functions are sufficient to construct all fermionic partition functions. For instance for an even number $N_{\rm f}\in2\mathbb{N}$ of flavours, it is~\cite{Kanazawa:2018kbo}
\begin{equation}\label{part-ferm-gen}
Z_N^{(0,N_\mathrm{f})}(M)\propto\frac{1}{\Delta_{N_{\rm f}}(M^2)}\Pf\left[(m_{a}^2-m_{b}^2)Z_{N+N_{\rm f}-2}^{(0,2)}(m_a,m_b)\right]_{a,b=1,\ldots,N_{\rm f}}.
\end{equation}
In the case of an odd $N_{\rm f}$ one introduces an additional mass $m_{N_{\rm f}+1}$ and applies Eq.~\eqref{part-ferm-gen} for $N_{\rm f}+1$ masses. Eventually, one lets  $m_{N_{\rm f}+1}\to\infty$ yielding in the last row and column the one-flavour partition function. In appendix~\ref{sec:ferm-part} we will derive Eq.~\eqref{part-ferm-gen} directly from the large-$N$ partition function~\eqref{eq:Zuniv}.

Before we arrive at the hard edge result~\eqref{eq:Zuniv}, we want to give a detailed list of contributions which result from the saddle point approximation. Let us take Eq.~\eqref{susy:partition.b} with no bosonic degrees of freedom, $k_\mathrm{b}=0$. In the hard edge limit, we expand the unitary matrix $U'$ as $U'=UzU^\dagger=U(\tau_3+\diag(i\delta z_1,-i\delta z_2)/\sqrt{N})U^\dagger$ with $\delta z_1$ and $\delta z_2$ two real diagonal $k_\mathrm{f}$-dimensional matrices para\-metrizing the massive modes and $U\in\mathrm{U}(2k_\mathrm{f})$ a unitary matrix distributed by the normalized Haar measure $d\mu(U)$. The different signs in front of $\delta z_1$ and $\delta z_2$ result from the opposite crossings of the contours of the eigenvalues of $U'$ through the saddle points $\pm1$. This change of variables yields the following approximation of the Vandermonde determinant: $\Delta_{2k_\mathrm{f}}^2(z)\approx 2^{2k_\mathrm{f}^2}N^{-k_\mathrm{f}(k_\mathrm{f}-1)}\Delta_{k_\mathrm{f}}^2(\delta z_1)\Delta_{k_\mathrm{f}}^2(\delta z_2)$. The differentials transform as $d z=N^{-k_\mathrm{f}}d\delta z_1 d\delta z_2$. Additionally, we get the constant $1/(2k_\mathrm{f})!\prod_{j=0}^{2k_\mathrm{f}-1}(\pi^j/j!)$ that is essentially the volume of the coset $\U(2k_\mathrm{f})/[\U^{2k_\mathrm{f}}(1)\times\mathbb{S}_{2k_\mathrm{f}}]$, where $\mathbb{S}_{2k_\mathrm{f}}$ is the symmetric group permuting the diagonal elements of $z$.
The latter has to be compensated since we ordered the signs in front of $\delta z$, which produces an additional factor of $(2k_\mathrm{f})!/(k_\mathrm{f}!)^2$. Moreover, we obtain the sign $(-1)^{Nk_\mathrm{f}}$ at the saddle points from the determinant $\Sdet^N U'={\det}^{-N} U$.

Collecting everything, we have for the fermionic partition function with an arbitrary $k_\mathrm{f}$
\begin{align}
\!\!
Z_N^{(0,k_\mathrm{f})}& \overset{N\gg 1}{\approx}  C_N^{(0,k_\mathrm{f})}\frac{(-1)^{Nk_\mathrm{f}}2^{2k_\mathrm{f}^2}}{N^{k_\mathrm{f}^2}(k_\mathrm{f}!)^2}\prod_{j=0}^{2k_\mathrm{f}-1}\frac{\pi^j}{j!}\int_{\mathbb{R}^{k_\mathrm{f}}}d\delta z_1\int_{\mathbb{R}^{k_\mathrm{f}}}d\delta z_2\ \Delta_{k_\mathrm{f}}^2(\delta z_1)\Delta_{k_\mathrm{f}}^2(\delta z_2)
\notag\\
&\quad \qquad \times\exp(Nk_\mathrm{f}-\Tr\delta z_1^2-\Tr\delta z_2^2)\widehat{Z}^{(k_\mathrm{f})}(\widehat{\kappa})\notag\\
& ~~=\frac{2^{k_\mathrm{f}^2}e^{Nk_\mathrm{f}}}{(2\pi)^{k_\mathrm{f}}N^{(N+k_\mathrm{f})k_\mathrm{f}}}\left(\prod_{j=0}^{k_\mathrm{f}-1}\frac{(j!)^2(N+2j)!(N+2j+1)!}{(2j)!(2j+1)!}\right)\widehat{Z}^{(k_\mathrm{f})}(\widehat{\kappa})\;.
\label{gen-flav-f-1}
\end{align}
with $\widehat{Z}^{(k_\mathrm{f})}$ given as in Eq.~\eqref{eq:Zuniv}. In the particular cases of one and two flavours this reads
\begin{equation}\label{one-flav-f-1}
\begin{split}
Z_N^{(0,1)}\overset{N\gg 1}{\approx}& 2N^{N+1}e^{-N}\int_{\mathrm{U}(2)} d\mu(U)\exp\left[\widehat{\kappa}\Tr U\tau_3U^\dagger \tau_1-\frac{\widehat{\mu}^2}{2}\Tr(U\tau_3U^\dagger\tau_3)^2\right]
\end{split}
\end{equation}
and
\begin{equation}\label{two-flav-f-1}
\begin{split}
Z_N^{(0,2)}\overset{N\gg 1}{\approx}&\frac{4N^{2N+4}}{3}e^{-2N}\int_{\mathrm{U}(4)} d\mu(U)\exp\left[\Tr U\tau_3U^\dagger \widehat{\kappa}\tau_1-\frac{\widehat{\mu}^2}{2}\Tr(U\tau_3U^\dagger\tau_3)^2\right],
\end{split}
\end{equation}
respectively, where we used Stirling's formula for the Gamma function in the limit of a large argument.

Comparing Eqs.~\eqref{part-ferm-gen} and~\eqref{gen-flav-f-1}, it is apparent that most of the group integral over $U\in\U(2k_{\rm f})$ can be performed, explicitly. In particular it yields a Pfaffian determinant. The remaining integrals in the one- and two-flavour partition function in Eqs.~\eqref{one-flav-f-1} and~\eqref{two-flav-f-1} are simplified further in Appendix~\ref{sec:ferm-part} so that we end up with
\begin{equation}\label{one-flav-f-2}
Z_N^{(0,1)}\overset{N\gg 1}{\approx} N^{N+1}e^{-N}\int_{-1}^1dx\; I_0\big(2\widehat{\kappa}\sqrt{1-x^2}\big)\exp[\widehat{\mu}^2(1-2x^2)]
\end{equation}
and
\begin{equation}\label{two-flav-f-2}
\begin{split}
Z_N^{(0,2)}\overset{N\gg 1}{\approx}&\frac{N^{2N+4}}{2}e^{-2N}\int_{-1}^1dx_1\int_{-1}^1dx_2\;\frac{x_1-x_2}{x_1+x_2}\exp[2\widehat{\mu}^2(1-x_1^2-x_2^2)]
\\
&\times\frac{I_0\big(2\widehat{\kappa}_{1}\sqrt{1-x_1^2}\big)I_0\big(2\widehat{\kappa}_{2}\sqrt{1-x_2^2}\big)-I_0\big(2\widehat{\kappa}_{1}\sqrt{1-x_2^2}\big)I_0\big(2\widehat{\kappa}_{2}\sqrt{1-x_1^2}\big)}{\widehat{\kappa}_2^2-\widehat{\kappa}_1^2}.
\end{split}
\end{equation}
These two results are reminiscent of the Bessel kernel~\cite{Verbaarschot},  which is deformed now. Indeed the limit $\widehat{\mu}\to\infty$ can be readily checked yielding the Bessel-kernel for vanishing topological charge, because the integrals~\eqref{one-flav-f-2} and~\eqref{two-flav-f-2} are evaluated at the saddle points $x=0$ and $x_1=x_2=0$, respectively.

Confirming the correct limit for $\widehat{\mu}\to0$ in Eqs.~\eqref{one-flav-f-2} and~\eqref{two-flav-f-2} is not that simple though one can set $\widehat\mu=0$ in the integrand. To see that the partition functions become indeed those of the sine-kernel~\cite{Mehta_book}, one has to reverse the Berezin-Karpelevich integral~\eqref{BK-int} and integrate over $U$ instead of $U\tau_3U^\dagger$ in Eqs.~\eqref{one-flav-f-1}  and~\eqref{two-flav-f-1} which is the Harish-Chandra-Itzykson-Zuber integral~\cite{Harish,IZ}. The result is the sine-kernel result which is for two or four flavours (masses always appear as chiral pairs $\pm\kappa_j$),
\begin{equation}
\widehat{Z}^{(k_\mathrm{f}=1)}(\widehat{\kappa})\Big|_{\widehat\mu=0}=\frac{\sinh(2\widehat{\kappa})}{2\widehat{\kappa}}
\end{equation}
and
\begin{equation}
\widehat{Z}^{(k_\mathrm{f}=2)}(\widehat{\kappa}_1,\widehat{\kappa}_2)\Big|_{\widehat\mu=0}=\frac{3}{4}\frac{\sinh(2\widehat{\kappa}_1)\sinh(2\widehat{\kappa}_2)(\widehat{\kappa}_1^2+\widehat{\kappa}_2^2)-2[\cosh(2\widehat{\kappa}_1)\cosh(2\widehat{\kappa}_2)-1]\widehat{\kappa}_1\widehat{\kappa}_2}{\widehat{\kappa}_1\widehat{\kappa}_2(\widehat{\kappa}_1^2-\widehat{\kappa}_2^2)^2},
\end{equation}
respectively.
 For $Z_N^{(0,1)}\propto\widehat{Z}^{(k_\mathrm{f}=1)}$, one can alternatively calculate this result more directly by employing the Taylor expansion of the Bessel function in Eq.~\eqref{one-flav-f-2} and then integrating each term separately.

\subsection{The case $(k_\mathrm{b},k_\mathrm{f})=(k_{\rm b},0)$}\label{sec:bos}

As for the fermionic partition functions, we want to determine the exact normalization constants as well as the partitions function of one and two bosonic flavours as those partition functions also satisfy~\cite{Kanazawa:2018kbo}
\begin{equation}\label{part-bos-gen}
Z_N^{(k_{\rm b},0)}(\kappa_{\rm b})\propto\frac{1}{\Delta_{k_{\rm b}}(\kappa_{\rm b}^2)}\Pf\left[(\kappa_{{\rm b},a}^2-\kappa_{{\rm b},b}^2)Z_{N-k_{\rm b}+2}^{(2,0)}(\kappa_{{\rm b},a},\kappa_{{\rm b},b})\right]_{a,b=1,\ldots,k_{\rm b}},
\end{equation}
for an even number $k_{\rm b}$ of bosonic flavours. The counterpart of this equation for odd $k_{\rm b}$ can be obtained in the same way as for fermionic flavours via introducing an additional auxiliary bosonic  quark with mass $\kappa_{{\rm b},k_{\rm b}+1}$ and sending this mass to infinity in the end.

To begin with, we again calculate the components of the saddle point approximation. The hard edge limit works the same way as in the fermionic case, in particular the matrix $U'$ can have only the eigenvalues $\pm1$ at the saddle point. However, we now have a restriction for the contours that is reflected by the fact that $U'L_\mathrm{bb}\tau_1$ is a positive definite Hermitian matrix. Hence, we can parametrize $U'=UzU^{-1}=U(L_\mathrm{bb}\tau_1+\delta z/\sqrt{N})U^{-1}$ with $U\in\mathrm{U}(L_\mathrm{bb}\tau_1)/[\mathrm{U}(k_\mathrm{b})\times\mathrm{U}(k_\mathrm{b})]$ and $\delta z$ being a $2k_\mathrm{b}\times2k_\mathrm{b}$ matrix which satisfies the following conditions. The matrix $\delta z L_\mathrm{bb}\tau_1$ is Hermitian and fulfills the commutation relation $[\delta z, L_\mathrm{bb}\tau_1]=0$. We want to emphasize that in the saddle point manifold we cannot arbitrarily permute the entries of $L_\mathrm{bb}\tau_1$ because of the nature of the non-compact group $\mathrm{U}(L_\mathrm{bb}\tau_1)$
\footnote{Let us recall that a matrix $U\in\mathrm{U}(L_\mathrm{bb}\tau_1)$ is pseudo-unitary as follows $UL_\mathrm{bb}\tau_1U^\dagger=L_\mathrm{bb}\tau_1$ and the notation shall only reflect the embedding of the non-compactness.} which is composed of disjoint parts. Therefore a combinatorial factor does not appear this time, the saddle point is unique. In spite of this, for all choices of $L_\mathrm{bb}$ the non-compact group $\mathrm{U}(L_\mathrm{bb}\tau_1)$ is unitarily equivalent to the non-compact unitary group $\mathrm{U}(k_\mathrm{b},k_\mathrm{b})$ because $\Tr L_\mathrm{bb}\tau_1=0$. More precisely, we can bring all matrices in  $\mathrm{U}(L_\mathrm{bb}\tau_1)$ to a standard form of matrices in $\mathrm{U}(k_\mathrm{b},k_\mathrm{b})$ by a single unitary transformation.

As $\mathrm{U}(L_\mathrm{bb}\tau_1)$ is a non-compact group, there is no normalizable Haar measure. Consequently, the normalization constant cannot be computed in the traditional way with the volumes of the groups, but we have to stick with the measure which is given by the pseudo-Riemannian length element,
\begin{equation}\label{riem-length}
\begin{split}
g(dU',dU')=\RE\,\Tr (dU')^2=&\frac{1}{N}\RE\,\Tr\delta z^2+\RE\,\Tr\left[U^{-1}dU,L_\mathrm{bb}\tau_1+\frac{\delta z}{\sqrt{N}}\right]^2\\
\approx&\frac{1}{N}\RE\,\Tr\delta z^2+\RE\,\Tr\left[U^{-1}dU,L_\mathrm{bb}\tau_1\right]^2.
\end{split}
\end{equation}
We denote the volume element resulting from the invariant length element by $d\hat\mu(U')$ which is given by the standard formula $d\hat\mu(U')=\sqrt{\det g}\,dU'$. Thence, we have for the flat measure
\begin{equation}
dU'=2^{-k_\mathrm{b}(2k_\mathrm{b}-1)}d\hat\mu(U')=2^{-k_\mathrm{b}(2k_\mathrm{b}-1)}N^{-k_\mathrm{b}^2} d\hat\mu(\delta z) d\hat\mu(U)=(2N)^{-k_\mathrm{b}^2} d\delta z d\hat\mu(U)
\end{equation}
The measure $d\hat\mu(U)$ is the pseudo-Riemannian volume element of the invariant length element $\RE\,\Tr\left[U^{-1}dU,L_\mathrm{bb}\tau_1\right]^2$.

We put again everything together and find for a general bosonic partition function
\begin{align}
Z_N^{(k_\mathrm{b},0)}& =C_N^{(k_\mathrm{b},0)}\int dU'\;{\det}^{N-2k_\mathrm{b}}U' \exp\bigg[-\Tr U'\widehat{\kappa}\tau_1-\frac{N}{2}\Tr{U'}^2+\frac{\widehat{\mu}^2}{2}\Tr (U'\tau_3)^2\bigg]
\notag \\
& \!\!\!\overset{N\gg 1}{\approx} (-1)^{Nk_\mathrm{b}}(2N)^{-k_\mathrm{b}^2} e^{-Nk_\mathrm{b}}C_N^{(k_\mathrm{b},0)}
\notag \\
& \qquad \times \int d\delta zd\hat{\mu}(U)\exp\left[-\Tr UL_\mathrm{bb}\tau_1U^{-1}\widehat{\kappa}\tau_1-\Tr \delta z^2+\frac{\widehat{\mu}^2}{2}\Tr (UL_\mathrm{bb}\tau_1U^{-1}\tau_3)^2\right]
\notag \\
& \!\!\!\overset{N\gg 1}{\approx} N^{Nk_\mathrm{b}}\frac{(2\pi)^{k_\mathrm{b}}}{(4\pi N)^{k_\mathrm{b}^2}} e^{-Nk_\mathrm{b}}\left(\prod_{j=0}^{2k_\mathrm{b}-1}\frac{1}{(N-j-1)!}\right)
\notag \\
&\qquad \times\int d\hat{\mu}(U)\exp\left[-\Tr UL_\mathrm{bb}\tau_1U^{-1}\widehat{\kappa}\tau_1+\frac{\widehat{\mu}^2}{2}\Tr (UL_\mathrm{bb}\tau_1U^{-1}\tau_3)^2\right].
\end{align}
This intermediate result is more suitable after performing a $\pi/4$-rotation, in particular $U\to \exp[i \pi L_\mathrm{bb}\tau_3/4]U\exp[-i \pi L_\mathrm{bb}\tau_3/4]$, resulting in
\begin{equation}\label{partition-b.1}
\begin{split}
Z_N^{(k_\mathrm{b},0)}\overset{N\gg 1}{\approx}&N^{Nk_\mathrm{b}}\frac{(2\pi)^{k_\mathrm{b}}}{(4\pi N)^{k_\mathrm{b}^2}} e^{-Nk_\mathrm{b}}\left(\prod_{j=0}^{2k_\mathrm{b}-1}\frac{1}{(N-j-1)!}\right)\\
&\times\int d\hat{\mu}(U)\exp\left[-\Tr U\tau_2U^{-1}L_\mathrm{bb}\widehat{\kappa}\tau_2+\frac{\widehat{\mu}^2}{2}\Tr (U\tau_2U^{-1}\tau_3)^2\right].
\end{split}
\end{equation}
Then the $2k_\mathrm{b}\times2k_\mathrm{b}$ matrix $U$ changes its symmetries to $U^{-1}=\tau_2U^\dagger\tau_2$ so that it becomes independent of $L_\mathrm{bb}$.
The cases of one and two flavours are given by
\begin{equation}\label{one-flav-b-1}
\begin{split}
Z_N^{(1,0)}\overset{N\gg 1}{\approx}& \frac{N^{1-N}e^N}{4\pi}\int d\hat{\mu}(U)\exp\left[-L_\mathrm{bb}\widehat{\kappa}\Tr U\tau_2U^{-1}\tau_2+\frac{\widehat{\mu}^2}{2}\Tr (U\tau_2U^{-1}\tau_3)^2\right]
\end{split}
\end{equation}
and
\begin{equation}\label{two-flav-b-1}
\begin{split}
Z_N^{(2,0)}\overset{N\gg 1}{\approx}&\frac{N^{4-2N}e^{2N}}{(4\pi)^4}\int d\hat{\mu}(U)\exp\left[-\Tr U\tau_2U^{-1}L_\mathrm{bb}\widehat{\kappa}\tau_2+\frac{\widehat{\mu}^2}{2}\Tr (U\tau_2U^{-1}\tau_3)^2\right].
\end{split}
\end{equation}

As before the integral~\eqref{partition-b.1} over the coset $\U(\1_{k_{\rm b}}\otimes\tau_2)/[\U(k_{\rm b})\times\U(k_{\rm b})]$ can be evaluated explicitly, even when it is of a non-compact type, because Eqs.~\eqref{part-bos-gen} and~\eqref{partition-b.1} must agree. Therefore, also this integral yields a Pfaffian determinant whose matrix entries are essentially given by Eqs.~\eqref{one-flav-b-1} and~\eqref{two-flav-b-1}. In Appendix~\ref{sec:boson-part} we simplify these two partition functions, which are explicitly
\begin{equation}\label{one-flavour-b-2}
\begin{split}
Z_N^{(1,0)}\overset{N\gg 1}{\approx}&\frac{N^{1-N}}{\pi} e^{N}e^{-\widehat{\mu}^2}\int_{-\infty}^\infty dx~e^{-2\widehat{\mu}^2x^2}K_0\big(2L\widehat{\kappa}\sqrt{1+x^2}\big)
\end{split}
\end{equation}
and
\begin{equation}\label{two-flavour-b-2}
\begin{split}
Z_N^{(2,0)}\overset{N\gg 1}{\approx}&\frac{N^{4-2N}}{2\pi^2} e^{2N}e^{-2\widehat{\mu}^2}\int_{-\infty}^\infty dx_1\int_{-\infty}^\infty dx_2~\frac{x_1-x_2}{x_1+x_2}\exp[-2\widehat{\mu}^2(x_1^2+x_2^2)]\\
&\hspace*{-1cm}\times\frac{K_0\big(2L_1\widehat{\kappa}_{1}\sqrt{1+x_1^2}\big)K_0\big(2L_2\widehat{\kappa}_{2}\sqrt{1+x_2^2}\big)-K_0\big(2L_1\widehat{\kappa}_{1}\sqrt{1+x_2^2}\big)K_0\big(2L_2\widehat{\kappa}_{2}\sqrt{1+x_1^2}\big)}{\widehat{\kappa}_{2}^2-\widehat{\kappa}_{1}^2}.
\end{split}
\end{equation}

As for the fermionic partition function, it is much simpler to show that the limit $\widehat{\mu}\to\infty$ yields the chGUE result compared to the limit $\widehat{\mu}\to0$ which is more subtle. For $\widehat{\mu}\to\infty$ we have only to rescale $x\to x/\widehat{\mu}$ in Eq.~\eqref{one-flavour-b-2} and $(x_1,x_2)\to(x_1,x_2)/\widehat{\mu}$ in Eq.~\eqref{two-flavour-b-2} so that the limit can be taken exactly inside the integral which gives the Bessel-kernel. The equations~\eqref{one-flav-b-1} and~\eqref{two-flav-b-1} are better suited for the limit $\widehat{\mu}\to0$ since one can easily omit the second term in the exponents without risking to lose convergence. The remaining integrals are Harish-Chandra--Itzykson--Zuber like integral over the non-compact cosets  $\U(\tau_2)/[\U(1)\times\U(1)]$ and $\U(\1_2\otimes\tau_2)/[\U(2)\times\U(2)]$, respectively, i.e.
\begin{equation}
Z_N^{(1,0)}\Big|_{\widehat{\mu}=0}\propto\frac{\exp(-2L\widehat{\kappa})}{L\widehat{\kappa}}
\end{equation}
and
\begin{equation}
Z_N^{(2,0)}\Big|_{\widehat{\mu}=0}\propto\frac{\exp(-2L_1\widehat{\kappa}_1-2L_2\widehat{\kappa}_2)}{L_1L_2\widehat{\kappa}_1\widehat{\kappa}_2(L_1\widehat{\kappa}_1+L_2\widehat{\kappa}_2)^2}.
\end{equation}
See~\cite{noncomp1,noncomp2} where these group integrals have been carried out; note that there another $\pi/4$-rotation $U\to \exp(i \pi \tau_1/4)U\exp(-i \pi \tau_1/4)$ is applied.

\subsection{The case $(k_\mathrm{b},k_\mathrm{f})=(1,1)$}\label{sec:susy.part}

The quenched partition function of one bosonic and one fermionic flavour is equal to
\begin{align}
Z_N^{(1,1)}=\;&\frac{1}{4\pi^4}\int dU' \Sdet^N U'
\exp\bigg[-\Str U'\widehat{\kappa}\tau_1-\frac{N}{2}\Str {U'}^2+\frac{\widehat{\mu}^2}{2}\Str(U'\tau_3)^2\bigg]
\notag\\
=\;&\frac{1}{4\pi^4} \int d{U'}_\mathrm{bb} d{U'}_\mathrm{ff}d\eta d\eta^* \det\big(\1_{2}-{U'}_\mathrm{bb}^{-1}\eta^\dagger {U'}_\mathrm{ff}^{-1}\eta\big)^N
\notag \\
&\times\exp\Big[-L({U'}_\mathrm{bb},\widehat{\kappa}_\mathrm{b})+L({U'}_\mathrm{ff},\widehat{\kappa}_\mathrm{f})-N\Tr\eta^\dagger\eta+\widehat{\mu}^2\Tr\eta^\dagger\tau_3\eta\tau_3\Big]
\label{partition-bf.1}
\end{align}
where we have defined the Lagrangian
\begin{equation}
L(V,x)=x\Tr V\tau_1-\frac{\widehat{\mu}^2}{2}\Tr(V\tau_3)^2+\frac{N}{2}\Tr V^2-N\Tr\mathrm{ln} V.
\end{equation}
This function should not be confused with the diagonal matrix $L$ consisting of signs defined right before Eq.~\eqref{Sdet-ident}.
The calculation of the large $N$-limit is quite lengthy and is deferred to Appendix~\ref{sec:susy-part}. Here we state the final result,
\begin{align}
\lim_{N\to\infty}Z_N^{(1,1)}&=\frac{1}{2\pi}\int_{-1}^{1} dx\int_{-\infty}^\infty dy~e^{-2\widehat{\mu}^2(y^2+x^2)}\biggl\{\bigl[8\widehat{\mu}^4\left(y^2-x^2+1\right)\left(y^2-x^2\right)-2\widehat{\mu}^2(x^2+y^2)
\notag \\
&\quad +2\widehat{\kappa}_\mathrm{b}^2(1+y^2)+2\widehat{\kappa}_\mathrm{f}^2(1-x^2)-1\bigl] I_0\big(2\widehat{\kappa}_\mathrm{f}\sqrt{1-x^2}\big)K_0\big(2L\widehat{\kappa}_\mathrm{b}\sqrt{1+y^2}\big)
\notag \\
&\quad +\left[8\widehat{\mu}^2(y^2-x^2+1)-1\right]\widehat{\kappa}_\mathrm{f}\sqrt{1-x^2}I_1\big(2\widehat{\kappa}_\mathrm{f}\sqrt{1-x^2}\big)K_0\big(2L\widehat{\kappa}_\mathrm{b}\sqrt{1+y^2}\big)
\notag \\
&\quad +\left[8\widehat{\mu}^2(y^2-x^2+1)+1\right]L\widehat{\kappa}_\mathrm{b}\sqrt{1+y^2}I_0\big(2\widehat{\kappa}_\mathrm{f}\sqrt{1-x^2}\big)K_1\big(2L\widehat{\kappa}_\mathrm{b}\sqrt{1+y^2}\big)
\notag \\
&\quad +4 L\widehat{\kappa}_\mathrm{b}\widehat{\kappa}_\mathrm{f}\sqrt{1-x^2}\sqrt{1+y^2}I_1\big(2\widehat{\kappa}_\mathrm{f}\sqrt{1-x^2}\big)K_1\big(2L\widehat{\kappa}_\mathrm{b}\sqrt{1+y^2}\big)\biggl\}\;.
\label{Susy-limit}
\end{align}
Despite this cumbersome result, the chGUE result can be readily regained in the limit $\widehat{\mu}\to\infty$. Then, only the second and third term contribute because the integration variables scale like $x,y\propto1/\widehat{\mu}$. Therefore the quenched partition function becomes
\begin{equation}
\lim_{\widehat{\mu}\to\infty}\lim_{N\to\infty}Z_N^{(1,1)}=2\widehat{\kappa}_\mathrm{f}I_1(2\widehat{\kappa}_\mathrm{f})K_0(2L\widehat{\kappa}_\mathrm{b})+2L\widehat{\kappa}_\mathrm{b}I_0(2\widehat{\kappa}_\mathrm{f})K_1(2L\widehat{\kappa}_\mathrm{b})
\end{equation}
which indeed agrees with the result of chGUE.

For $\widehat{\mu}=0$, the integrals can be performed explicitly as well,
\begin{equation}
\begin{split}
\lim_{\widehat{\mu}\to0}\lim_{N\to\infty}Z_N^{(1,1)}=\;&\frac{2L\widehat{\kappa}_\mathrm{b}\widehat{\kappa}_\mathrm{f}\cosh(2\widehat{\kappa}_\mathrm{f})+(\widehat{\kappa}_\mathrm{b}^2+\widehat{\kappa}_\mathrm{f}^2)\sinh(2\widehat{\kappa}_\mathrm{f})}{2L\widehat{\kappa}_\mathrm{b}\widehat{\kappa}_\mathrm{f}}e^{-2L\widehat{\kappa}_\mathrm{b}}.
\end{split}
\end{equation}
Its simple form in this limit is again reminiscent of the sine-kernel structure of the GUE.

After we gained the quenched partition function~\eqref{Susy-limit} we can calculate the microscopic level density by differentiating in $\widehat{\kappa}_\mathrm{f}$, thereafter setting $\widehat{\kappa}_\mathrm{b}=\widehat{\kappa}_\mathrm{f}=L\epsilon-i \lambda$, and eventually taking the real part in the limit $\epsilon\to0$, i.e.
\begin{equation}
\rho(\lambda)=\frac{1}{\pi}\lim_{\substack{N\to\infty \\ \epsilon\to0}}\mathrm{Re}\left.\partial_{\widehat{\kappa}_\mathrm{f}}Z_N^{(1,1)}\right|_{\widehat{\kappa}_\mathrm{b}=\widehat{\kappa}_\mathrm{f}=L\epsilon-i \lambda}.
\end{equation}
For this purpose we employ the identities~\cite{NIST:DLMF}
\begin{equation}
\begin{split}
\partial_{\widehat{\kappa}_\mathrm{f}} \widehat{\kappa}_\mathrm{f}^\nu I_\nu\big(2\widehat{\kappa}_\mathrm{f}\sqrt{1-x^2}\big)=\;&2\sqrt{1-x^2} \widehat{\kappa}_\mathrm{f}^\nu I_{\nu-1}\big(2\widehat{\kappa}_\mathrm{f}\sqrt{1-x^2}\big),\\
\lim_{\epsilon\to0}(\epsilon-i \lambda)^\nu I_\nu\big(2(\epsilon-i \lambda)\sqrt{1-x^2}\big)=\;&(-\lambda)^\nu J_\nu\big(2\lambda\sqrt{1+y^2}\big),\\
\lim_{\epsilon\to0}\mathrm{Im}(\epsilon-i \lambda)^\nu K_\nu\big(2(\epsilon-i \lambda)\sqrt{1+y^2}\big)=\;&\frac{\pi}{2}\big(\lambda\sqrt{1+y^2}\big)^\nu J_\nu\big(2\lambda\sqrt{1+y^2}\big)
\end{split}
\end{equation}
for $\lambda>0$. Therefore the microscopic level density of the quenched system is
\begin{align}
\rho(\lambda)=\;&\frac{2}{\pi}\int_{-1}^1 dx\int_{-\infty}^\infty dy\;\exp\left[-2\widehat{\mu}^2\left(x^2+y^2\right)\right]
\notag \\
&\times\biggl[\left(2\widehat{\mu}^4(y^2-x^2+1)(y^2-x^2)-\widehat{\mu}^2\frac{x^2+y^2}{2}-\frac{1}{2}\lambda^2(2+y^2-x^2)-\frac{1}{4}\right) 
\notag \\
&\times\sqrt{1-x^2} J_1\big(2|\lambda|\sqrt{1-x^2}\big)J_0\big(2\lambda\sqrt{1+y^2}\big)
\notag \\
&+\left(2\widehat{\mu}^2(y^2-x^2+1)+\frac{1}{4}\right)|\lambda|(1-x^2)J_0\big(2\lambda\sqrt{1-x^2}\big)J_0\big(2\lambda\sqrt{1+y^2}\big)
\notag \\
&+\left(2\widehat{\mu}^2(y^2-x^2+1)+\frac{1}{4}\right)|\lambda|\sqrt{1+y^2}\sqrt{1-x^2}J_1\big(2\lambda\sqrt{1-x^2}\big)J_1\big(2\lambda\sqrt{1+y^2}\big)
\notag \\
&+\lambda^2(1-x^2)\sqrt{1+y^2}J_0\big(2\lambda\sqrt{1-x^2}\big)J_1\big(2|\lambda|\sqrt{1+y^2}\big)\biggl]\,.
\label{level.density}
\end{align}
This is the main result of this section and has been published in the letter~\cite{Kanazawa:2018}, but without the detailed calculation in Appendix~\eqref{sec:susy-part}.\footnote{Let us emphasize that we normalized the level density to $\lim_{\lambda\to\infty}\rho(\lambda)=2/\pi$, cf. Eq.~\eqref{densit.rel}. Another common choice is the normalization to $1/\pi$, see~\cite{Verbaarschot}.}
The microscopic level density~\eqref{level.density} has also been compared with Monte Carlo simulations of the random matrix model~\eqref{RMT-model} in~\cite{Kanazawa:2018}.

\begin{figure}[t!]
	\centering 
	\includegraphics[width=\textwidth]{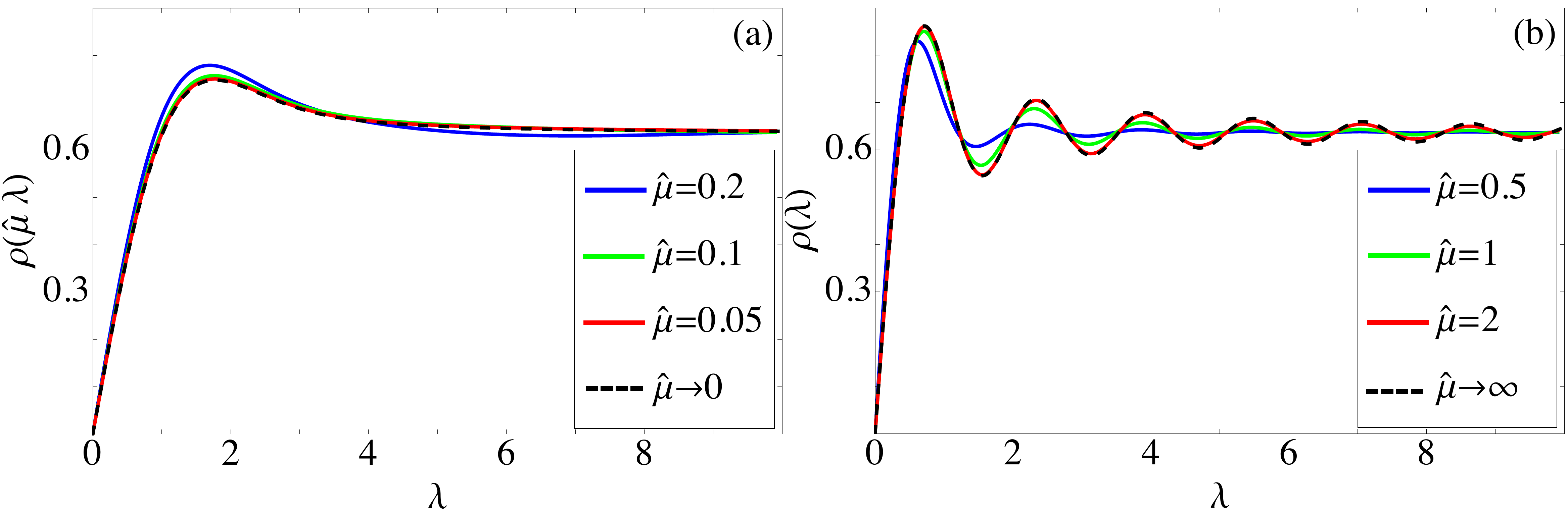}
	\caption{\label{fig.microlevel}The analytical result~\eqref{level.density} for the microscopic level density of the random matrix model~\eqref{RMT-model} for several coupling constants $\widehat{\mu}$. In the left plot we show how the asymptotic result~\eqref{level-dens.mu0} is approached while in the right plot the ``large'' $\widehat{\mu}$ behaviour is shown. Please note that in the left plot we have rescaled the singular values by $\widehat{\mu}$ so that what is shown lies on a very narrow strip about the origin and the remaining density can be well approximated by a constant $\rho(\lambda)\approx 2/\pi$.}
\end{figure}

It has been very surprising for us how fast it converges to the chGUE limit~\cite{Verbaarschot} ($\widehat{\mu}\to\infty$),
\begin{equation}
\begin{split}\label{level-dens.muinf}
\lim_{\widehat{\mu}\to\infty}\rho(\lambda)=\;&2\lambda\big[J_0^2(2\lambda)+J_1^2(2\lambda)\big],
\end{split}
\end{equation}
 since it is almost perfectly achieved for $\widehat{\mu}=2$, see~\cite{Kanazawa:2018} as well as the right plot in Fig.~\ref{fig.microlevel}. In contrast, the level density in the GUE limit ($\widehat{\mu}\to0$) has always a small but persistent peak close to the origin, see~\cite{Kanazawa:2018}, while it should be only a constant $\lim_{\widehat{\mu}\to0}\rho(\lambda)=2/\pi$. This has been already seen for the quenched staggered Dirac operator in three dimensions~\cite{Bialas:2010hb}. To understand this non-uniform behaviour we have to scale the spectrum on the scale $\widehat{\mu}$ where this peak lives. Thence we rescale $\lambda\to\widehat{\mu}\lambda$ in Eq.~\eqref{level.density} and take the limit $\widehat{\mu}\to0$. Before we can commute the limit with the integrals we have to rescale the integration variable $y\to y/\widehat{\mu}$, as well, i.e.
\begin{align}
\lim_{\widehat{\mu}\to0}\rho(\widehat{\mu}\lambda)& =\frac{8}{3\pi}\int_{-\infty}^\infty dy\exp(-2y^2)
\notag \\
&\quad \times\biggl[\left(4y^4+y^2-\lambda^2y^2-\frac{1}{4}\right)|\lambda|J_0(2\lambda y)+\left(4y^2+\frac{3}{2}\right)\lambda^2 yJ_1(2|\lambda| y)\biggl]
\nonumber\\
& = \sqrt{\frac{2}{\pi}}|\lambda| I_0\left(\frac{\lambda^2}{4}\right)\exp\left(-\frac{\lambda^2}{4}\right).
\label{level-dens.mu0}
\end{align}
We show the behaviour of this functions as well as how it is approached by the random matrix model~\eqref{RMT-model} in the left plot of Fig.~\ref{fig.microlevel}. The limit seems to be almost perfectly achieved for $\widehat{\mu}<0.1$. The persistent peak can be interpreted as the smallest singular value which is enforced to lie on the positive real line. The eigenvalue of the GUE closest to the origin has not this restriction, yet the smallest singular value of the model~\eqref{RMT-model} shall become exactly this eigenvalue. This obvious conflict is the origin of the non-uniform behaviour of the level density in the limit $\widehat{\mu}\to0$.

\section{Chiral Condensate/Pion Condensate}\label{sec:observables}

In this section we analyze the quantity
\begin{equation}\label{O.def}
\mathcal{O}(\widehat{\kappa})=\lim_{N\to\infty}\int_{[\Herm(N)]^2}\hspace{-5mm}dH_1dH_2\;P(\mathcal{D})\;\Tr\frac{1}{\sqrt{N}\mathcal{D}+\widehat{\kappa}\1_{2N}}=\int_{0}^\infty d\lambda\;\frac{2\widehat{\kappa}\rho(\lambda)}{\lambda^2+\widehat{\kappa}^2},
\end{equation}
where the prefactor $\sqrt{N}$ correctly rescales the eigenvalues for the hard edge scaling limit. This quantity is either proportional to the chiral condensate $\Sigma(\widehat{\kappa}=\widehat{m})=\langle\bar{u}u-\bar{d}d\rangle$ with $\widehat{m}=V\Sigma m$ the dimensionless quark mass (either of a sea or a valence quark) in the application to the staggered Dirac operator in three dimensions, see Sec.~\ref{sec:stag}, as well as in particular high temperature four dimensional quantum field theories, see Sec.~\ref{sec:hightemp}, or it is proportional to the pion condensate $\langle\bar{u}d\rangle$ with $\widehat{\kappa}=\widehat{j}=VWj_\pi$ with $\widehat{j}$ the corresponding dimensionless source variable in 3d QCD at finite isospin chemical potential as discussed in Sec.~\ref{sec:Chem}. Since the computation of $\mathcal{O}(\widehat{\kappa})$ differs between the quenched theory and a theory with dynamical quarks, we discuss it in separate Secs.~\ref{sec:quenched} and~\ref{sec:flavour}.

One last general comment on $\mathcal{O}(\widehat{\kappa})$: the asymptotics for large argument is determined by the asymptotics of $\rho(\lambda)$ which is $\lim_{\lambda\to\infty}\rho(\lambda)=2/\pi$. Hence the limit for $\mathcal{O}(\widehat{\kappa})$ is
\begin{equation}
\lim_{\widehat{\kappa}\to\pm\infty}\mathcal{O}(\widehat{\kappa})=2\,{\rm sign}(\widehat{\kappa}).
\end{equation}
This behaviour is also true with dynamical quarks because our  random matrix $\mathcal{D}$ only models the topologically trivial phase (topological charge  is $\nu=0$).

\subsection{Quenched theory}\label{sec:quenched}

\begin{figure}[t!]
	\centering 
	\includegraphics[width=\textwidth]{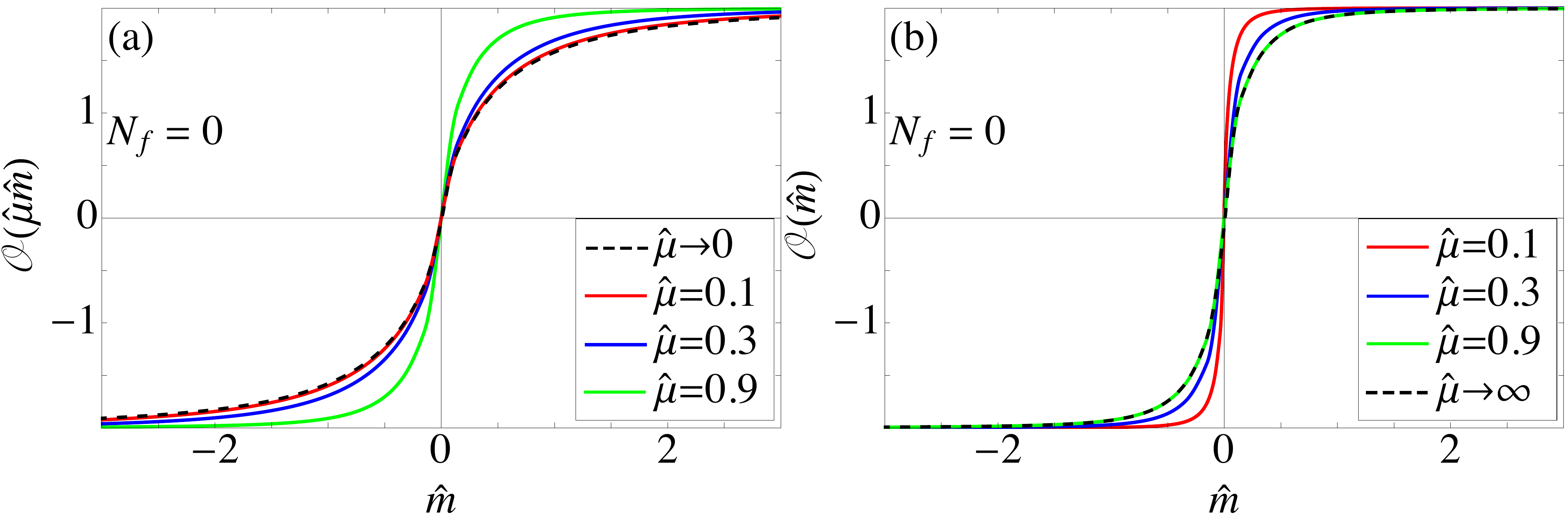}
	\caption{\label{fig.chiral.quenched} Illustrated is the convergence of the analytical result~\eqref{O.quenched} (solid coloured curves) to the  limits (dashed black curves) $\widehat{\mu}\to0$ (left plot) and $\widehat{\mu}\to\infty$ (right plot). The analytic form of the limits is given in Eqs.~\eqref{O.q.0} and~\eqref{O.q.inf}, respectively. We underline that in plot (a) we rescaled the mass by $\widehat{\mu}$ so that only the vicinity of the origin is shown. The asymptotics $2{\rm sign}(\widehat{m})$ for large mass $\widehat{m}$ lies at the top and bottom of the frame.}
\end{figure}

In the quenched case the observable~\eqref{O.def} can be computed via a derivative of the quenched partition function $Z_N^{(1,1)}$,
\begin{align}
\mathcal{O}^{(N_{\rm f}=0)}(\widehat{m})=&\lim_{N\to\infty}\partial_{\widehat{\kappa}_{\rm f}}\big|_{\widehat{\kappa}_{\rm b}=\widehat{\kappa}_{\rm f}=\widehat{m}}\int_{[\Herm(N)]^2}\hspace{-4mm}dH_1dH_2\;P(\mathcal{D})\;\frac{\det(\sqrt{N}
\mathcal{D} +\kappa_{\mathrm{f}}\1_{2N})}{\det(\sqrt{N}\mathcal{D} +\kappa_{\mathrm{b}}\1_{2N})}
\nonumber\\
=&\lim_{N\to\infty}\partial_{\widehat{\kappa}_{\rm f}}\big|_{\widehat{\kappa}_{\rm b}=\widehat{\kappa}_{\rm f}=\widehat{m}}Z_N^{(1,1)}\left(\frac{\widehat{\kappa}_{\rm b}}{\sqrt{N}},\frac{\widehat{\kappa}_{\rm f}}{\sqrt{N}}\right).
\end{align}
After plugging the result~\eqref{Susy-limit} into this equation, we find
\begin{align}
\mathcal{O}^{(N_{\rm f}=0)}(\widehat{m})=\;&\frac{{\rm sign}(\widehat{m})}{\pi}\int_{-1}^{1} dx\int_{-\infty}^\infty dy\;e^{-2\widehat{\mu}^2(y^2+x^2)}\sqrt{1-x^2}\biggl\{\bigl[8\widehat{\mu}^4\left(y^2-x^2+1\right)\left(y^2-x^2\right)
\notag \\
&-2\widehat{\mu}^2(x^2+y^2)+2\widehat{m}^2(2+y^2-x^2)-1\bigl] I_1\big(2|\widehat{m}|\sqrt{1-x^2}\big)K_0\big(2|\widehat{m}|\sqrt{1+y^2}\big)
\notag \\
&+\left[8\widehat{\mu}^2(y^2-x^2+1)+1\right]|\widehat{m}|\sqrt{1-x^2}I_0\big(2\widehat{m}\sqrt{1-x^2}\big)K_0\big(2|\widehat{m}|\sqrt{1+y^2}\big)
\notag \\
&+\left[8\widehat{\mu}^2(y^2-x^2+1)+1\right]|\widehat{m}|\sqrt{1+y^2}I_1\big(2|\widehat{m}|\sqrt{1-x^2}\big)K_1\big(2|\widehat{m}|\sqrt{1+y^2}\big)
\notag \\
&+4\widehat{m}^2\sqrt{1-x^2}\sqrt{1+y^2}I_0\big(2\widehat{m}\sqrt{1-x^2}\big)K_1\big(2|\widehat{m}|\sqrt{1+y^2}\big)\biggl\}.
\label{O.quenched}
\end{align}
Obviously, this result can be also derived via the definition~\eqref{O.def} when exploiting the result for the level density~\eqref{level.density}. We show the behaviour of $\mathcal{O}(\widehat{m})$ for several values of $\widehat{\mu}$ in Fig.~\ref{fig.chiral.quenched}. Especially the limits to the chGUE result
\begin{equation}\label{O.q.inf}
\begin{split}
&\lim_{\widehat{\mu}\to\infty}\mathcal{O}^{(N_{\rm f}=0)}(\widehat{m})=4\widehat{m}\big[I_0(2\widehat{m})K_0(2|\widehat{m}|)+I_1(2|\widehat{m}|)K_1(2|\widehat{m}|)\big]
\end{split}
\end{equation}
and the small $\widehat{\mu}$ result
\begin{equation}\label{O.q.0}
\begin{split}
&\lim_{\widehat{\mu}\to0}\mathcal{O}^{(N_{\rm f}=0)}(\widehat{\mu}\widehat{m})=\sqrt{\frac{2}{\pi}}\widehat{m}K_0\left(\frac{\widehat{m}^2}{4}\right)\exp\left(\frac{\widehat{m}^2}{4}\right) 
\end{split}
\end{equation}
are highlighted in the right and left plot of Fig.~\ref{fig.chiral.quenched}, respectively. Note that Eq.~\eqref{O.q.0} is not the GUE result, i.e., $\mathcal{O}_{\rm GUE}^{(N_{\rm f}=0)}(\widehat{m})=2{\rm sign}(\widehat{m})$ because the microscopic level density is a constant in this case. Indeed Eq.~\eqref{O.q.0} lies on the scale of $\widehat{\mu}$ so that it shows the behaviour in a very narrow region around the origin. Without this rescaling we would certainly find $\lim_{\widehat{\mu}\to0}\mathcal{O}^{(N_{\rm f}=0)}(\widehat{m})=2{\rm sign}(\widehat{m})$, yet this limit is again not uniform in contrast to the chGUE limit~\eqref{O.q.inf}.

Interestingly, both limits are reached extremely quickly. For $\widehat{\mu}\to0$ differences can be hardly seen at $\widehat{\mu}=0.1$, whilst for $\widehat{\mu}\to\infty$ it is already in good agreement with $\widehat{\mu}=0.9$. Thus the convergence is even faster than that for the microscopic level density (cf. Sec.~\ref{sec:susy.part}), which can be actually expected because integrals usually show an improved convergence. The impressive convergence for ``large'' $\widehat{\mu}$ can be understood by the Gaussian factor in the integral~\eqref{O.quenched} that suppresses very rapidly any deviation from the saddle point $x=y=0$. For ``small'' $\widehat{\mu}$ it helps that everything only depends on its square. Therefore the corrections of $\widehat{\mu}=0.1$ is only of about $1\%$.

\subsection{One and Two flavours}\label{sec:flavour}

When we have dynamical quarks we can set the auxiliary variable $\widehat{\kappa}$ equal to one of the quark masses, say $\widehat{m}_1$, so that we can express it in terms of a derivative of the partition function $\widehat{Z}^{(N_{\rm f})}$, in particular we have
\begin{equation}
\mathcal{O}^{(N_{\rm f})}(\widehat{m}_1)=\lim_{N\to\infty}\partial_{\widehat{m}_1}{\rm ln}\, Z_N^{(0,N_{\rm f})}=\partial_{\widehat{m}_1}{\rm ln}\,\widehat{Z}^{(N_{\rm f})}\,.
\end{equation}
Inserting the results~\eqref{one-flav-f-2} and~\eqref{two-flav-f-2} into this relation we obtain for one and two flavours
\begin{equation}\label{O-one-flav}
\begin{split}
\mathcal{O}^{(N_{\rm f}=1)}(\widehat{m})
=2\;\frac{\displaystyle \int_{-1}^1dx \;\sqrt{1-x^2}I_1\big(2\widehat{m}\sqrt{1-x^2}\big)\exp(-2\widehat{\mu}^2x^2)}
{\displaystyle \int_{-1}^1dx\; I_0\big(2\widehat{m}\sqrt{1-x^2}\big)\exp(-2\widehat{\mu}^2x^2)}
\end{split}
\end{equation}
and
\begin{equation}\label{O-two-flav}
\begin{split}
\mathcal{O}^{(N_{\rm f}=2)}(\widehat{m}_1)=\frac{\displaystyle \int_{-1}^1dx_1\int_{-1}^1dx_2\frac{x_1-x_2}{x_1+x_2}\exp[-2\widehat{\mu}^2(x_1^2+x_2^2)]\partial_{\widehat{m}_1}F(x_1,x_2;\widehat{m}_1,\widehat{m}_2)}
{\displaystyle \int_{-1}^1dx_1\int_{-1}^1dx_2\frac{x_1-x_2}{x_1+x_2}\exp[-2\widehat{\mu}^2(x_1^2+x_2^2)]F(x_1,x_2;\widehat{m}_1,\widehat{m}_2)},
\end{split}
\end{equation}
respectively, where the dependence of $\mathcal{O}^{(N_f=2)}$ on $\widehat{m}_2$ is made implicit for brevity. The function for the two-flavour case is
\begin{equation}
F(x_1,x_2;\widehat{m}_1,\widehat{m}_2)=\frac{I_0\big(2\widehat{m}_{1}\sqrt{1-x_1^2}\big)I_0\big(2\widehat{m}_{2}\sqrt{1-x_2^2}\big)-I_0\big(2\widehat{m}_{1}\sqrt{1-x_2^2}\big)I_0\big(2\widehat{m}_{2}\sqrt{1-x_1^2}\big)}{\widehat{m}_2^2-\widehat{m}_1^2}.
\end{equation}
Applying the derivative $\partial_{\widehat{m}_1}$ in the two-flavour case explicitly yields a lengthy expression. This statement is even true for its limits $\widehat{\mu}\to0$ and $\widehat{\mu}\to\infty$; therefore we omit these expressions.

\begin{figure}[t!]
	\centering 
	\includegraphics[width=\textwidth]{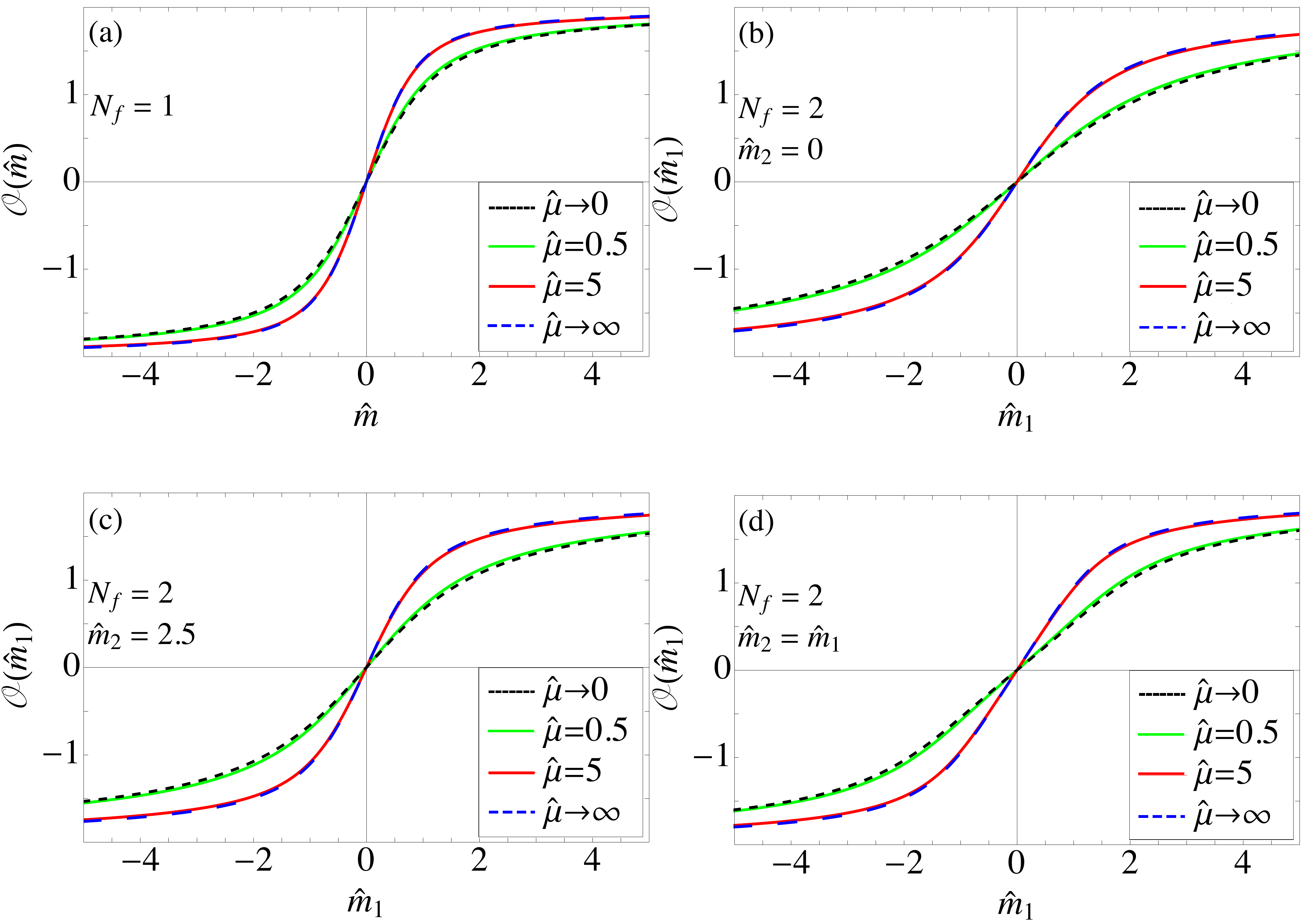}
	\caption{\label{fig.chiral.unquenched} The quantity $\mathcal{O}^{(N_{\rm f}}(\widehat{m}_1)$ for the one-flavour theory (plot (a)) and for the two-flavour theory as a function of the first quark mass, while the second quark mass is fixed at $\widehat{m}_2=0$ (plot (b)),  $\widehat{m}_2=2.5$ (plot (c)), and  $\widehat{m}_2=\widehat{m}_1$ (plot (d)). The solid red and green curves are given by the analytical results~\eqref{O-one-flav} ($N_{\rm f}=1$) and~\eqref{O-two-flav} ($N_{\rm f}=2)$ whilst the black and blue dashed curves are given by the limits $\widehat{\mu}\to0$ and $\widehat{\mu}\to\infty$, respectively. We have employed the values $\widehat{\mu}=0.5$ and $\widehat{\mu}=5$ to emphasize when the limiting behaviours emerge. In the one-flavour case we used the limiting functions~\eqref{O.onef.0} and~\eqref{O.onef.inf}. For the theory with two flavours we  employed Eqs.~\eqref{O.twof0.0} and~\eqref{O.twof0.inf} for the vanishing second quark mass and  the limits~\eqref{O.twofm.0} and~\eqref{O.twofm.inf} when the two quark masses are equal. As in Fig.~\ref{fig.chiral.quenched} we have chosen the top and bottom of the frame to indicate the limits $2{\rm sign}(\widehat{m})$ and  $2{\rm sign}(\widehat{m}_1)$ for large masses $\widehat{m}$ and  $\widehat{m}_1$.}
\end{figure}

The behaviour for the one-flavour case is shown in Fig.~\ref{fig.chiral.unquenched}(a). It is apparent that the GUE limit
\begin{equation}\label{O.onef.0}
\lim_{\widehat{\mu}\to0}\mathcal{O}^{(N_{\rm f}=1)}(\widehat{m})=\frac{2}{\tanh(2\widehat{m})}-\frac{1}{\widehat{m}}
\end{equation}
as well as the chGUE limit
\begin{equation}\label{O.onef.inf}
\lim_{\widehat{\mu}\to\infty}\mathcal{O}^{(N_{\rm f}=1)}(\widehat{m})=2\frac{I_1(2\widehat{m})}{I_0(2\widehat{m})}
\end{equation}
are uniformly approached and, surprisingly, well-achieved for moderate values of $\widehat{\mu}$, namely $\widehat{\mu}=0.5$ and $\widehat{\mu}=5$, respectively. This has to be seen in contrast to the quenched theory in Sec.~\ref{sec:quenched} where no uniform limit to GUE exists. The dynamical quark mass pushes the spectrum away from the origin so that the region on the scale $\widehat{\mu}$ is almost void of eigenvalues.

This behaviour can be also seen for the two-flavour case, which can be observed for instance in Fig.~\ref{fig.chiral.unquenched}(c) where we have chosen $\widehat{m}_2$. The slope at the origin of $\mathcal{O}(\widehat{m}_1)$ flattens out when the second mass decreases, cf. Fig.~\ref{fig.chiral.unquenched}(b) for vanishing mass $\widehat{m}_2$. Considering the duality of flavour and topological charge, the case $\widehat{m}_2=0$ can be identified with the $\nu=1$ case of chGUE. Then the observable $\mathcal{O}^{(N_{\rm f}=2)}(\widehat{m}_1)$  simplifies to
\begin{align}
&\mathcal{O}^{(N_{\rm f}=2)}(\widehat{m}_1)\Big|_{\widehat{m}_2=0}=
\notag \\
& \quad -2\Bigg(
\scalebox{0.8}{$\displaystyle 
\frac{1}{\widehat{m}_1} - \frac{\displaystyle \int_{-1}^1dx_1\int_{-1}^1dx_2\frac{x_1-x_2}{x_1+x_2}e^{-2\widehat{\mu}^2(x_1^2+x_2^2)}\Big[\sqrt{1-x_1^2}I_1\Big(2\widehat{m}_1\sqrt{1-x_1^2}\Big)-\sqrt{1-x_2^2}I_1\Big(2\widehat{m}_1\sqrt{1-x_2^2}\Big)\Big]}{\displaystyle \int_{-1}^1dx_1\int_{-1}^1dx_2\frac{x_1-x_2}{x_1+x_2}e^{-2\widehat{\mu}^2(x_1^2+x_2^2)}\Big[I_0\Big(2\widehat{m}_1\sqrt{1-x_1^2}\Big)-I_0\Big(2\widehat{m}_1\sqrt{1-x_2^2}\Big)\Big]}$}\Bigg).
\end{align}
This expression has the asymptotic behaviour
\begin{equation}\label{O.twof0.inf}
\begin{split}
&\lim_{\widehat{\mu}\to\infty}\mathcal{O}^{(N_{\rm f}=2)}(\widehat{m}_1)\Big|_{\widehat{m}_2=0}=2\left[\frac{I_0(2\widehat{m}_1)}{I_1(2\widehat{m}_1)}-\frac{1}{\widehat{m}_1}\right]
\end{split}
\end{equation}
for the chGUE limit and
\begin{equation}\label{O.twof0.0}
\begin{split}
&\lim_{\widehat{\mu}\to0}\mathcal{O}^{(N_{\rm f}=2)}(\widehat{m}_1)\Big|_{\widehat{m}_2=0}=\frac{1}{\tanh\widehat{m}_1}-\frac{4}{\widehat{m}_1}+\frac{\widehat{m}_1\tanh\widehat{m}_1}{\widehat{m}_1-\tanh\widehat{m}_1}
\end{split}
\end{equation}
for the GUE limit.

Also illustrated in Fig.~\ref{fig.chiral.unquenched}(d) is the case with degenerate masses $\widehat{m}_1=\widehat{m}_2$. Despite the lengthy expression of $\mathcal{O}^{(N_{\rm f}=2)}(\widehat{m}_1)$ for general $\widehat{\mu}$, its limits are relatively simple, namely
\begin{equation}\label{O.twofm.inf}
\begin{split}
&\lim_{\widehat{\mu}\to\infty}\mathcal{O}^{(N_{\rm f}=2)}(\widehat{m}_1)\Big|_{\widehat{m}_2=\widehat{m}_1}=\frac{1}{\widehat{m}_1}\frac{I_1^2(2\widehat{m}_1)}{I_0^2(2\widehat{m}_1)-I_1^2(2\widehat{m}_1)}
\end{split}
\end{equation}
and
\begin{equation}\label{O.twofm.0}
\begin{split}
&\lim_{\widehat{\mu}\to0}\mathcal{O}^{(N_{\rm f}=2)}(\widehat{m}_1)\Big|_{\widehat{m}_2=\widehat{m}_1}=2\frac{1+4\widehat{m}_1^2-\cosh(4\widehat{m}_1)+\widehat{m}_1\sinh(4\widehat{m}_1)}{\widehat{m}_1\cosh(4\widehat{m}_1)-8\widehat{m}_1^3-\widehat{m}_1}.
\end{split}
\end{equation}
Again the results for the values $\widehat{\mu}=5$ and $\widehat{\mu}=0.5$ almost perfectly match the asymptotics.

As for the quenched theory, the Gaussian term in the integrals~\eqref{O-one-flav} and~\eqref{O-two-flav} as well as the dependence on $\widehat{\mu}^2$ help the convergence. Furthermore, the repulsion from the dynamical quark masses shifts the region where we have a good a agreement to larger values of $\widehat{\mu}$. This can be understood from the analytical results~\eqref{O-one-flav} and~\eqref{O-two-flav} because only modified Bessel functions of the first kind are involved which grow exponentially. In the quenched case~\eqref{O.quenched}, those exponential contributions are compensated by the modified Bessel functions of the second kind.

\section{Conclusions}\label{sec:conclusion}

We continued our study~\cite{Kanazawa:2018,Kanazawa:2018kbo} of the spectral statistics of a Gaussian random matrix model that interpolates between the GUE and the chGUE while preserving chirality. In the present work we investigated the statistics on the scale of the local mean level spacing about the origin (hard edge scaling) which agrees with the Dirac spectra in several QCD-like theories in the $\varepsilon$-regime. We gave a detailed account of approximations involved in showing such correspondences.

Additionally, we gave full details of the derivation of the microscopic level density~\eqref{level.density}, that we already presented in the letter~\cite{Kanazawa:2018}. We even carried further the analysis of the level density and identified a scale-free function~\eqref{level-dens.mu0} in the limit of vanishing coupling constant $\widehat{\mu}\to0$. This limit is a persistent deviation from the level density of the GUE which is a constant on the local scale. This deviation results from the protected chirality and essentially reflects the fact that the smallest singular value $\lambda>0$ always feels a residual interaction with its ``mirror charge'' $-\lambda<0$. The scale-free function~\eqref{level-dens.mu0} shows a single peak  that can be interpreted as this singular value and has also been observed for the staggered Dirac operator in three-dimensional QCD. Therefore we are confident that our results may help in fixing the low energy constant corresponding to the term $\Tr(U\tau_3 U^\dagger \tau_3)^2$ in the effective Lagrangian of the Nambu-Goldstone bosons, cf. Eq.~\eqref{eff.Lag.stag}.

Moreover, we anlytically evaluated  an observable~\eqref{O.def} which corresponds to quark bilinear condensates in QCD-like theories. We derived explicit expressions of this quantity for the quenched theory and the unquenched theory with one and two flavours. We observed that in the presence of dynamical quarks the spectrum is pushed away from the origin so that both limits, GUE ($\widehat{\mu}\to0$) and chGUE ($\widehat{\mu}\to\infty$), are approached uniformly in the unquenched theory. In contrast, the quenched theory exhibits a non-uniform asymptotics to the GUE result. As it has been the case for the quenched microscopic level density, one can make out a scale free function~\eqref{O.q.0} on the scale $\widehat{\mu}$ also for this observable.

\acknowledgments 

MK acknowledges support by the German research council (DFG) via the CRC 1283: ``Taming uncertainty and profiting from randomness and low regularity in analysis, stochastics and their applications''.

\appendix
\section{Details of Some Calculations}\label{app:calc}

In this appendix we show detailed computations of several results presented in the main text. The normalization~\eqref{norm-SUSY} of the supersymmetric integral~\eqref{susy:partition.b} is derived in Appendix~\ref{sec:norm-susy}. The one- and two-flavour partition functions for fermions and bosons are evaluated in Appendices~\ref{sec:ferm-part} and~\ref{sec:boson-part}, respectively. In these two sections we also consider the partition functions with an arbitrary number of flavours and show how the Pfaffian forms~\eqref{part-ferm-gen} and~\eqref{part-bos-gen} can be directly obtained from the large $N$ results~\eqref{eq:Zuniv} and~\eqref{partition-b.1}. In Appendix~\ref{sec:susy-part} we explain the main steps of the calculation from Eq.~\eqref{partition-bf.1} to Eq.~\eqref{Susy-limit}.

\subsection{Derivation of the Normalization~(\ref{norm-SUSY})}\label{sec:norm-susy}

To obtain the second line of Eq.~\eqref{norm-SUSY}, we first integrate over the complex Grassmann variables $\eta$ which only appear in the measure and the superdeterminant:
\begin{align}
\!\!\!\! 
\int \hspace*{-0.05cm}d\eta d\eta^*\;\Sdet\hspace*{-0.15cm}\left(\begin{array}{cc} U'_\mathrm{bb} & \eta^\dagger \\ \eta & U'_\mathrm{ff} \end{array}\right)^{N+2(k_\mathrm{f}-k_\mathrm{b})}\hspace*{-0.6cm}& = \det {U'_\mathrm{bb}}^{N+2(k_\mathrm{f}-k_\mathrm{b})}\int d\eta d\eta^* \det(U'_\mathrm{ff}-\eta{U'_\mathrm{bb}}^{-1}\eta^\dagger)^{-N-2(k_\mathrm{f}-k_\mathrm{b})}
\notag \\
& =\frac{\det {U'_\mathrm{bb}}^{N-2k_\mathrm{b}}}{\det {U'_\mathrm{ff}}^{N+2k_\mathrm{f}}}\int d\eta d\eta^*\det(\1_{2k_\mathrm{b}}-\eta^\dagger\eta)^{N+2(k_\mathrm{f}-k_\mathrm{b})}
\notag \\
&=\frac{\det {U'_\mathrm{bb}}^{N-2k_\mathrm{b}}}{\det {U'_\mathrm{ff}}^{N+2k_\mathrm{f}}}\frac{\displaystyle \int_{\mathrm{U}(2k_\mathrm{b})}\!\!\!d\mu(U)\;e^{\Tr U}\det U^{-N+2k_\mathrm{b}}}{\displaystyle \int_{\mathrm{U}(2k_\mathrm{b})}\!\!\!d\mu(U)\; e^{\Tr U}\det U^{-N+2(k_\mathrm{b}-k_\mathrm{f})}}
\notag \\
&=\prod_{j=1}^{2k_\mathrm{b}}\frac{(N-j+2k_\mathrm{f})!}{(N-j)!}\frac{\det {U'_\mathrm{bb}}^{N-2k_\mathrm{b}}}{\det {U'_\mathrm{ff}}^{N+2k_\mathrm{f}}}.
\label{a1.1}
\end{align}
In the first step we rescaled $\eta\to U'_\mathrm{ff}\eta U'_\mathrm{bb}$ while keeping $\eta^\dagger$ fixed, which is possible for Grassmann variables. The integrals over $U$ in the third line of Eq.~\eqref{a1.1} are obtained by expressing the determinant $\det(\1_{2k_\mathrm{b}}-\eta^\dagger\eta)$ as
\begin{equation}
\det(\1_{2k_\mathrm{b}}-\eta^\dagger\eta)^{N+2(k_\mathrm{f}-k_\mathrm{b})}=\frac{\displaystyle \int_{\mathrm{U}(2k_\mathrm{b})}\!\!\!d\mu(U)\; \exp[\Tr (\1_{2k_\mathrm{b}}-\eta^\dagger\eta)U]\det U^{-N+2(k_\mathrm{b}-k_\mathrm{f})}}{\displaystyle \int_{\mathrm{U}(2k_\mathrm{b})}\!\!\!d\mu(U)\; e^{\Tr U}\det U^{-N+2(k_\mathrm{b}-k_\mathrm{f})}}.
\end{equation}
The measure $d\mu(U)$ can be chosen as the normalized Haar measure. Afterwards, we have integrated over $\eta$ because it is simply a Gaussian. We have got the last line of Eq.~\eqref{a1.1} via the Selberg-type integral~\cite{Mehta_book}
\begin{align}
\int_{\mathrm{U}(n)}\!\!\!d\mu(U)\,\exp(\Tr U)\det U^{-l}=&\frac{1}{n!}\oint dz\;\exp(\Tr z)\det z^{-l}\frac{|\Delta_n(z)|^2}{(2\pi i)^n\det z}
\notag \\
=&\frac{1}{n!}\oint dz\;\Delta_n(z) \det\left[\frac{l!}{(l+b-1)!}(-\partial_a)^{b-1}\frac{e^{z_a}}{z_a^{l+1}}\right]_{a,b=1,\ldots,n}
\notag \\
=&\prod_{j=0}^{n-1}\frac{j!}{(l+j)!}
\label{Selb-comp}
\end{align}
with $n,l\in\mathbb{N}$.  We recall that the Vandermonde determinant is given as
\begin{equation}\label{Vandermonde}
\Delta_{n}(x)=\prod_{1\leq a<b\leq n}(x_b-x_a)=\det[x_a^{b-1}]_{a,b=1,\ldots,n}.
\end{equation}
The integrals over $U'_\mathrm{bb}$ and $U'_\mathrm{ff}$ follow from similar formulas as~\eqref{Selb-comp}, i.e.
\begin{align}
\int d U'_\mathrm{bb}\, e^{-\sqrt{N}\Tr U'_\mathrm{bb}L_\mathrm{bb}\tau_1} \det {U'_\mathrm{bb}}^{N-2k_\mathrm{b}}& = \frac{(-1)^{Nk_\mathrm{b}}}{(2k_\mathrm{b})!}\prod_{j=0}^{2k_\mathrm{b}-1}\frac{\pi^{j}}{j!}\int_{\mathbb{R}_+^{2k_\mathrm{b}}} dx\;e^{-\sqrt{N}\Tr x} \det x^{N-2k_\mathrm{b}}\Delta_{2k_\mathrm{b}}^2(x)
\notag \\
&= (-N)^{-Nk_\mathrm{b}}\prod_{j=0}^{2k_ \mathrm{b}-1}\pi^{j}(N-1-j)!\;,
\\
\int_{\mathrm{U}(2k_\mathrm{f})}\!\!\! d U'_\mathrm{ff}\; e^{\sqrt{N}\Tr U'_\mathrm{ff}\tau_1} \det {U'_\mathrm{ff}}^{-N-2k_\mathrm{f}}&=\frac{(-1)^{Nk_\mathrm{f}}}{(2k_\mathrm{f})!}\prod_{j=0}^{2k_\mathrm{f}-1}\frac{\pi^{j}}{j!}\oint dz\; e^{\sqrt{N}\Tr z} \det z^{-N-2k_\mathrm{f}}\Delta_{2k_\mathrm{f}}^2(z)\notag \\
&= (2\pi)^{2k_\mathrm{f}}(-N)^{Nk_\mathrm{f}}\prod_{l=0}^{2k_\mathrm{f}-1}\frac{\pi^{l}}{(N+l)!}\;.
\end{align}
Putting everything together we arrive at the second line of Eq.~\eqref{norm-SUSY}.

\subsection{Derivation of Eqs.~(\ref{one-flav-f-2}) and~(\ref{two-flav-f-2}) }\label{sec:ferm-part}

The next goal of this appendix is to evaluate the group integral involved in the fermionic partition function~\eqref{eq:Zuniv}. Here we consider the general case of an arbitrary number $k_{\rm f}$ of flavours. We achieve this by using the parametrization
\begin{equation}\label{para.ferm}
 U\tau_3U^\dagger=\diag(U_1,U_2)\left(\begin{array}{cc} \cos \vartheta & \sin \vartheta \\ \sin \vartheta & -\cos \vartheta \end{array}\right)\diag(U_1^\dagger,U_2^\dagger)
\end{equation}
with $U_1,U_2\in\mathrm{U}(k_\mathrm{f})$ and $\vartheta=\diag(\vartheta_1,\ldots,\vartheta_{k_\mathrm{f}})\in[0,\pi]^{k_\mathrm{f}}$. The Haar measure can be calculated with the help of the invariant length element
\begin{align}
\Tr [d( U\tau_3U^\dagger)]^2 & = 2\Tr d\vartheta^2+\Tr[U_1^\dagger dU_1,\cos\vartheta]^2+\Tr[U_2^\dagger dU_2,\cos\vartheta]^2
\notag \\
&\quad +2\Tr(U_1^\dagger dU_1\sin\vartheta-\sin\vartheta U_2^\dagger dU_2)(U_2^\dagger dU_2\sin\vartheta-\sin\vartheta U_1^\dagger dU_1)\;,
\label{inv.ferm}
\end{align}
which yields
\begin{equation}\label{Haar.ferm}
d\mu(U)\propto \Delta_{k_\mathrm{f}}^2(\cos\vartheta)\left(\prod_{j=1}^{k_\mathrm{f}}\sin\vartheta_j d\vartheta_j\right)d\mu(U_1)d\mu(U_2)
\end{equation}
up to normalization. The normalization constant $K$ can be calculated as follows,
\begin{equation}
K^{-1}=\int_{[0,\pi]^{k_\mathrm{f}}}\Delta_{k_\mathrm{f}}^2(\cos\vartheta)\left(\prod_{j=1}^{k_\mathrm{f}}\sin\vartheta_j d\vartheta_j\right)=\int_{[-1,1]^{k_\mathrm{f}}}\Delta_{k_\mathrm{f}}^2(x)\prod_{j=1}^{k_\mathrm{f}}dx_j=2^{k_\mathrm{f}^2}k_\mathrm{f}!\prod_{j=0}^{k_\mathrm{f}-1}\frac{(j!)^3}{(k_\mathrm{f}+j)!}\;.
\end{equation}
The unitary matrices $U_1$ and $U_2$ drop out in the quadratic term which is proportional to $\widehat{\mu}^2$, see Eq.~\eqref{eq:Zuniv}. Thus we are left with a Berezin-Karpelevic integral~\cite{BK,Wettig}
\begin{align}
&\int_{\mathrm{U}(k_\mathrm{f})}d\mu(U_1)\int_{\mathrm{U}(k_\mathrm{f})}d\mu(U_2)
\exp(\Tr U\tau_3 U^\dagger \widehat{\kappa}\tau_1)
\notag \\
=\;&\int_{\mathrm{U}(k_\mathrm{f})}d\mu(U_1)\int_{\mathrm{U}(k_\mathrm{f})}d\mu(U_2)
\exp(\Tr U_1\sin\vartheta U_2^\dagger \widehat{\kappa}+\Tr U_2\sin\vartheta U_1^\dagger \widehat{\kappa})
\notag\\
=\;& \left(\prod_{j=0}^{k_\mathrm{f}-1}[j!]^2\right)\frac{\det[I_0(2\widehat{\kappa}_{a}\sin\vartheta_b)]_{a,b=1,\ldots,k_\mathrm{f}}}{\Delta_{k_\mathrm{f}}(\sin^2\vartheta)\Delta_{k_\mathrm{f}}(\widehat{\kappa}^2)}
\label{BK-int}
\end{align}
with $I_\nu$ the modified Bessel function of the first kind.  The fermionic partition function is then
\begin{equation}\label{partition.ferm}
\begin{split}
Z_N^{(0,k_\mathrm{f})}\overset{N\gg 1}{\approx}&N^{(N+k_\mathrm{f})k_\mathrm{f}}e^{-Nk_\mathrm{f}}e^{\widehat{\mu}^2k_\mathrm{f}}\frac{1}{\Delta_{k_\mathrm{f}}(\widehat{\kappa}^2)}\\
&\times\frac{1}{k_\mathrm{f}!}\int_{[-1,1]^{k_\mathrm{f}}} \left(\prod_{j=1}^{k_\mathrm{f}}e^{-2\widehat{\mu}^2x_j^2}dx_j\right)\frac{\Delta_{k_\mathrm{f}}^2(x)}{\Delta_{k_\mathrm{f}}(x^2)}\det\left[I_0\Big(2\widehat{\kappa}_{a}\sqrt{1-x_b^2}\Big)\right]_{a,b=1,\ldots,k_\mathrm{f}}.
\end{split}
\end{equation}
The Vandermonde determinants can be combined via the Schur Pfaffian identity~\cite{Schur}
\begin{equation}\label{Pfaff-ident}
\frac{\Delta_{k_\mathrm{f}}^2(x)}{\Delta_{k_\mathrm{f}}(x^2)}=\left\{\begin{array}{cl} \displaystyle \Pf\left[\frac{x_b-x_a}{x_b+x_a}\right]_{a,b=1,\ldots,k_\mathrm{f}} & \mathrm{for}\ k_\mathrm{f}\ \mathrm{even}, \\ \Pf\left[\begin{array}{c|c} 0 & 1\ \cdots\ 1  \\ \hline \begin{array}{c} -1 \\ \vdots \\ -1 \end{array} & \displaystyle\frac{x_b-x_a}{x_b+x_a} \end{array}\right]_{a,b=1,\ldots,k_\mathrm{f}} &  \mathrm{for}\ k_\mathrm{f}\ \mathrm{odd}, \end{array}\right.
\end{equation}
where the index $a$ labels the rows and $b$ the columns.
Here we normalize the Pfaffian as
\begin{equation}
\Pf\left[\begin{array}{c|c|c} \begin{array}{cc} 0 & 1 \\ -1 & 0 \end{array} &  & {\bf 0}  \\ \hline & \ddots & \\ \hline {\bf 0} & & \begin{array}{cc} 0 & 1 \\ -1 & 0 \end{array} \end{array}\right]=1.
\end{equation}
With the help of de Bruijn's integration theorem~\cite{deBruijn} we already see at this step that the fermionic partition functions with an arbitrary number $k_\mathrm{f}$ of flavours can be reduced to those with $k_\mathrm{f}=1,2$.  In particular we can read off from Eq.~\eqref{partition.ferm} the one-flavour result~\eqref{one-flav-f-2} and the two-flavour one~\eqref{two-flav-f-2}.

\subsection{Derivation of Eqs.~(\ref{one-flavour-b-2}) and (\ref{two-flavour-b-2}) }\label{sec:boson-part}

As for the fermionic partition function, we can calculate the group integral in Eq.~\eqref{partition-b.1} for a general number of bosonic flavours $k_{\rm b}$ and then read off the result for $k_{\rm b}=1,2$. The integration over $U$ can be done after choosing a parametrization very similar to Eq.~\eqref{para.ferm}, namely
\begin{equation}\label{paramet.b}
H=UU^\dagger=U\tau_2U^{-1}\tau_2=\diag(G^\dagger,G^{-1})\left(\begin{array}{cc} \cosh\vartheta & \sinh\vartheta \\ \sinh\vartheta & \cosh\vartheta \end{array}\right)\diag(G,(G^\dagger)^{-1})
\end{equation}
with $\vartheta=\diag(\vartheta_1,\ldots,\vartheta_{k_{\rm b}})\in\mathbb{R}^{k_{\rm b}}$ and $G\in{\rm Gl}_{\mathbb{C}}(k_{\rm b})/[{\rm U}(1)]^{k_{\rm b}}$. The division with respect to $[{\rm U}(1)]^{k_{\rm b}}$ is due to the invariance under $G\to\Phi G$ with $\Phi$ an arbitrary unitary, diagonal matrix. One can readily check that the matrix $H$ is indeed Hermitian, positive definite and $H\tau_2$ is self-inverse. The invariant length element becomes
\begin{align}
\RE\,\Tr\left[U^{-1}dU,\tau_2\right]^2 & =\RE\,\Tr(d(U\tau_2U^{-1}))^2
\notag\\
& = -2\Tr d\vartheta^2-\Tr[\sinh\vartheta,A]^2-\Tr[\sinh\vartheta,A^\dagger]^2
\notag \\
& \quad -2\Tr(\cosh\vartheta A^\dagger+A\cosh\vartheta)(\cosh\vartheta A+A^\dagger\cosh\vartheta)
\end{align}
with $A=dG\,G^{-1}$. Neglecting the normalization for the moment, the partition function is
\begin{align}
Z_N^{(k_{\rm b},0)}& \overset{N\gg 1}{\propto} 
e^{-\widehat{\mu}^2k_{\rm b}}\int_{\mathbb{R}^{k_{\rm b}}}\left(\prod_{j=1}^{k_{\rm b}}\cosh\vartheta_j\exp[-2\widehat{\mu}^2 \sinh^2\vartheta_j]d\vartheta_j\right)\Delta^2_{k_{\rm b}}(\sinh\vartheta)
\notag \\
&\qquad \times\biggl(\int d\hat{\mu}(G)\exp\left[-\Tr\Big(G^\dagger\cosh\vartheta G+G^{-1}\cosh\vartheta(G^\dagger)^{-1}\Big)L_{\rm bb}\widehat{\kappa}\right]\biggl)
\notag\\
& ~~= e^{-\widehat{\mu}^2k_{\rm b}}\int_{\mathbb{R}^{k_{\rm b}}}\left(\prod_{j=1}^{k_{\rm b}}\exp[-2\widehat{\mu}^2 x_j^2]dx_j\right)\Delta^2_{k_{\rm b}}(x)
\notag \\
&\qquad \times\biggl(\int d\hat{\mu}(G)\exp\left[-\Tr\Big(G^\dagger\sqrt{1+x^2} G+G^{-1}\sqrt{1+x^2}(G^\dagger)^{-1}\Big)L_{\rm bb}\widehat{\kappa}\right]\biggl)
\label{partition-b.2}
\end{align}
with $x=\diag(x_1,\ldots,x_{k_{\rm b}})=\sinh\vartheta$.
The normalization of the remaining non-compact coset integral is fixed in the limit $L_{\rm bb}\kappa=\kappa_0\1_{k_{\rm b}}\to\infty$ at $\widehat{\mu}=0$. Before coming to this problem we want to calculate the non-compact coset integral.

The integral over $G$ is the non-compact counterpart of the Berezin-Karpelevic integral~\eqref{BK-int}. Indeed one can show that it satisfies the same differential equation as the compact one, i.e.
\begin{equation}
(\Delta_V-\Tr WW^\dagger)\int d\hat{\mu}(U_1,U_2)\exp(-\Tr U_1 VU_2 W-\Tr U_2^{-1} V^\dagger U_1^{-1} W^\dagger)=0,
\end{equation}
regardless of whether $U_1$ and $U_2$ are unitary matrices or $G=U_1=U_2^\dagger$ is a complex matrix. The two matrices $V$ and $W$ and independent and $k_{\rm b}\times k_{\rm b}$ dimensional  and $\Delta_V=\sum_{a,b}\partial_{V_{ab}}\partial_{V^*_{ba}}$ is the Laplace operator with respect to $V$. After performing a singular value decomposition of $V$ with $\lambda=\diag(\lambda_1,\ldots,\lambda_{k_{\rm b}})$ its singular values, the singular value part of the Laplace operator becomes
\begin{equation}
\Delta_V=\frac{1}{\Delta_{k_{\rm b}}(\lambda^2)}\sum_{j=1}^{k_{\rm b}}(\partial_{\lambda_j}^2+\lambda_j^{-1}\partial_{\lambda_j})\Delta_{k_{\rm b}}(\lambda^2)
\end{equation}
such that the differential equation factorizes. There are two fundamental solutions of the eigenvalue equations of the operator $\partial_{\lambda_j}^2+\lambda_j^{-1}\partial_{\lambda_j}$, either the modified Bessel function of the first kind $I_0$ or the modified Bessel function of the second kind $K_0$. Since Eq.~\eqref{partition-b.2} decays exponentially for large $\kappa$, we can exclude the first  case and we have for the non-compact coset integral
\begin{equation}\label{non.comp.BK-int}
\begin{split}
\int d\hat{\mu}(G)e^{-\Tr(G^\dagger\cosh\vartheta G+G^{-1}\cosh\vartheta(G^\dagger)^{-1})L_{\rm bb}\widehat{\kappa}}
\propto\frac{\det[K_0(2L_a\widehat{\kappa}_{a}\cosh\vartheta_b)]_{a,b=1,\ldots,k_{\rm f}}}{\Delta_{k_{\rm f}}(\cosh^2\vartheta)\Delta_{k_{\rm f}}(\widehat{\kappa}^2)}\,.
\end{split}
\end{equation}
We plug this result into Eq.~\eqref{partition-b.2} and arrive at
\begin{equation}\label{partition-b.3}
\begin{split}
Z_N^{(k_{\rm b},0)}\overset{N\gg 1}{\propto}&e^{-\widehat{\mu}^2k_{\rm b}}\int_{\mathbb{R}^{k_{\rm b}}}\left(\prod_{j=1}^{k_{\rm b}}e^{-2\widehat{\mu}^2x_j^2}dx_j\right)\frac{\Delta^2_{k_{\rm b}}(x)}{\Delta_{k_{\rm f}}(\widehat{\kappa}^2)\Delta_{k_{\rm b}}(x^2)}\det\Big[K_0\Big(2L_a\widehat{\kappa}_{a}\sqrt{1+x_b^2}\Big)\Big]_{a,b=1,\ldots,k_{\rm f}},
\end{split}
\end{equation}
which is very similar to the result~\eqref{partition.ferm} of the fermionic partition function.

Let us calculate the normalization constant in the limit $L_{\rm bb}\kappa=\kappa_0\1_{k_{\rm b}}\to\infty$ at $\widehat{\mu}=0$. For this purpose we consider Eq.~\eqref{partition-b.1} and expand around the unique saddle point $H=UU^\dagger=U\tau_2U^{-1}\tau_2\approx\1_{2k_{\rm b}}$, cf. Eq.~\eqref{paramet.b}, or, equivalently taking the expansion $U\approx\1_{2k_{\rm b}}+\delta H/\sqrt{\kappa_0}+\delta H^2/(2\kappa_0)$ with $\delta H=\delta H^\dagger=-\tau_2\delta H\tau_2$. Therefore we can write $\delta H=\delta H_1\tau_1+\delta H_3\tau_3 $ with two independent Hermitian $k_{\rm b}\times k_{\rm b}$ matrices $\delta H_1$ and $\delta H_3$. Here we used again the abbreviated tensor notation $H_j\tau_j\equiv H_j\otimes\tau_j$. The metric is then
\begin{equation}
\RE\,\Tr[U^{-1}dU,\tau_2]^2=\RE\,\Tr(dH\tau_2)^2\overset{\kappa_0\gg1}{\approx}-\frac{4}{\kappa_0}\Tr d\delta H^2=-\frac{8}{\kappa_0^2}\Tr (d\delta H_1^2+d\delta H_3^2)
\end{equation} 
and, hence, the measure becomes
\begin{equation}
d\hat{\mu}(U)\overset{\kappa_0\gg1}{\approx}2^{k_{\rm b}(k_{\rm b}-1)}\left(\frac{8}{\kappa_0}\right)^{k_{\rm b}^2}d\delta H
\end{equation}
with $d\delta H$ the flat Lebesgue measure. Therefore we have the asymptotic behaviour of the partition function
\begin{align}
Z_{N\gg1}^{(k_{\rm b},0)}(\kappa=\kappa_0L_{\rm bb};\widehat{\mu}=0)&\overset{\kappa_0\gg 1}{\approx} 
N^{-Nk_{\rm b}}\left(\frac{N}{4\pi}\right)^{k_{\rm b}^2} e^{Nk_{\rm b}}\,2^{k_{\rm b}(k_{\rm b}-1)}\left(\frac{8}{\kappa_0}\right)^{k_{\rm b}^2}
\notag \\
&\qquad \times\int d\delta H\exp\left[-2\kappa_0k_{\rm b}-4\Tr(\delta H_1^2+\delta H_3^2)\right]
\notag\\
&~\;=N^{-Nk_{\rm b}}\left(\frac{N}{2\kappa_0}\right)^{k_{\rm b}^2} e^{Nk_{\rm b}}e^{-2\kappa_0k_b}.
\label{partition-b.4}
\end{align}
This result has to be compared with the asymptotics of the intermediate result~\eqref{partition-b.3}. Suppose $C$ is the constant we are looking for. Then
\begin{align}
& Z_{N\gg1}^{(k_{\rm b},0)}(\kappa=\kappa_0L_{\rm bb};\widehat{\mu}=0)
\notag \\
& \approx \frac{C}{\prod_{j=0}^{k_{\rm b}}(2\kappa_0)^jj!}\int_{\mathbb{R}^{k_{\rm b}}}\left(\prod_{j=1}^{k_{\rm b}}dx_j\right)\frac{\Delta^2_{k_{\rm b}}(x)}{\Delta_{k_{\rm b}}(x^2)}\det\Big[\partial_{\kappa_0}^{a-1}K_0\Big(2L_a\widehat{\kappa}_{0}\sqrt{1+x_b^2}\Big)\Big]_{a,b=1,\ldots,k_{\rm f}}
\notag \\
& \!\!\overset{\kappa_0\gg1}{\approx} \frac{C}{\prod_{j=0}^{k_{\rm b}}(-\kappa_0)^jj!}\left(\frac{\pi}{4\kappa_0}\right)^{k_{\rm b}/2}
\int_{\mathbb{R}^{k_{\rm b}}}\left(\prod_{j=1}^{k_{\rm b}}\frac{\exp\Big[-2\kappa_0\sqrt{1+x_j^2}\Big]}{(1+x_j^2)^{1/4}}dx_j\right)\frac{\Delta^2_{k_{\rm b}}(x)\Delta_{k_{\rm b}}(\sqrt{1+x^2})}{\Delta_{k_{\rm b}}(x^2)}
\notag \\
& \!\!\!\! \overset{x\to x/\sqrt{\kappa_0}}{\approx} \frac{C}{\prod_{j=0}^{k_{\rm b}}(-2\kappa_0^2)^jj!}\left(\frac{\pi}{4\kappa_0^2}\right)^{k_{\rm b}/2}\int_{\mathbb{R}^{k_{\rm b}}}\left(\prod_{j=1}^{k_{\rm b}}\exp[-x_j^2]dx_j\right)\Delta^2_{k_{\rm b}}(x)
\notag \\
& = (-1)^{k_{\rm b}(k_{\rm b}-1)/2}\frac{\pi^{k_{\rm b}}k_{\rm b}!}{(2\kappa_0)^{k_{\rm b}^2}}e^{-2\kappa_0k_{\rm b}}C\;.
\label{partition-b.5}
\end{align}
Now we can identify $C$ by setting this result equal to Eq.~\eqref{partition-b.4}.

The final result of this subsection is
\begin{equation}\label{partition-b.6}
\begin{split}
Z_N^{(k_{\rm b},0)}\overset{N\gg 1}{\approx}&(-1)^{k_{\rm b}(k_{\rm b}-1)/2}\frac{N^{(k_{\rm b}-N)k_{\rm b}}}{\pi^{k_{\rm b}}} e^{Nk_{\rm b}}e^{-\widehat{\mu}^2k_{\rm b}}\\
&\times\frac{1}{k_{\rm b}!}\int_{\mathbb{R}^{k_{\rm b}}}\left(\prod_{j=1}^{k_{\rm b}}e^{-2\widehat{\mu}^2x_j^2}dx_j\right)\frac{\Delta^2_{k_{\rm b}}(x)}{\Delta_{k_{\rm f}}(\widehat{\kappa}^2)\Delta_{k_{\rm b}}(x^2)}\det\Big[K_0\Big(2L_a\widehat{\kappa}_{a}\sqrt{1+x_b^2}\Big)\Big]_{a,b=1,\ldots,k_{\rm f}}.
\end{split}
\end{equation}
Using the Schur Pfaffian identity~\eqref{Pfaff-ident}, it can be again shown that also this partition function can be reduced to a Pfaffian comprising partition functions of only one and two flavours. Moreover the results~\eqref{one-flavour-b-2} and~\eqref{two-flavour-b-2} immediately follow when setting $k_{\rm b}=1,2$, respectively.

\subsection{Derivation of Eq.~(\ref{Susy-limit}) }\label{sec:susy-part}

We start our calculation with the second line of Eq.~\eqref{partition-bf.1}. Before we take the limit $N\to\infty$ we integrate over the Grassmann variables. For this purpose we introduce a unitary  matrix $S\in{\rm U}(2)$ as
\begin{equation}
\det\big(\1_{2}-{U'}_{\rm bb}^{-1}\eta^\dagger {U'}_{\rm ff}^{-1}\eta\big)^N=\frac{\displaystyle \int_{{\rm U}(2)} dS \exp\big(N\Tr S-N\Tr S{U'}_{\rm bb}^{-1}\eta^\dagger {U'}_{\rm ff}^{-1}\eta\big)\det S^{-N-2}}{\displaystyle \int_{{\rm U}(2)} dS \exp(N\Tr S)\det S^{-N-2}}\;.\!\!\!
\end{equation}
Afterwards, the integration over the Grassmann variables is a simple Gaussian integral which produces a new determinant,
\begin{align}
Z_N^{(1,1)}=\;&\frac{N!(N+1)!}{16\pi^7 N^{2N}} \int d{U'}_{\rm bb} d{U'}_{\rm ff}dS ~\det\big(N\1_{4}-\widehat{\mu}^2\tau_3\otimes\tau_3+NS{U'}_{\rm bb}^{-1}\otimes{U'}_{\rm ff}^{-1}\big)
\notag \\
&\times\exp[-L({U'}_{\rm bb},\widehat{\kappa}_{\rm b})+L({U'}_{\rm ff},\widehat{\kappa}_{\rm f})+N\Tr S]\det S^{-N-2}.
\label{partition-bf.2}
\end{align}
The first determinant can be expanded as follows.
\begin{align}
&\det\big(N\1_{4}-\widehat{\mu}^2\tau_3\otimes\tau_3+NS{U'}_{\rm bb}^{-1}\otimes{U'}_{\rm ff}^{-1}\big)
\notag 
\\
=\;&(N^2-\widehat{\mu}^4)^2+N(N^2-\widehat{\mu}^4)(N \Tr S{U'}_{\rm bb}^{-1}\Tr{U'}_{\rm ff}^{-1}+\widehat{\mu}^2\Tr S{U'}_{\rm bb}^{-1}\tau_3\Tr{U'}_{\rm ff}^{-1}\tau_3)
\notag \\
&+\frac{N^2}{2}\biggl[(N\Tr S{U'}_{\rm bb}^{-1}\Tr{U'}_{\rm ff}^{-1}+\widehat{\mu}^2\Tr S{U'}_{\rm bb}^{-1}\tau_3\Tr{U'}_{\rm ff}^{-1}\tau_3)^2
\notag \\
&-\Tr(NS{U'}_{\rm bb}^{-1}\otimes{U'}_{\rm ff}^{-1}+\widehat{\mu}^2S{U'}_{\rm bb}^{-1}\tau_3\otimes{U'}_{\rm ff}^{-1}\tau_3)^2\biggl]
\notag \\
&+N^4\det(S{U'}_{\rm bb}^{-1}{U'}_{\rm ff}^{-1})^2\left(\Tr S^{-1}{U'}_{\rm bb}\Tr{U'}_{\rm ff}-\frac{\widehat{\mu}^2}{N}\Tr S^{-1}{U'}_{\rm bb}\tau_3\Tr{U'}_{\rm ff}\tau_3+1\right)\,.
\label{determinant-expansion}
\end{align}
The integrand of Eq.~\eqref{partition-bf.2} is invariant under ${U'}_{\rm ff}\to\tau_1{U'}_{\rm ff}\tau_1$ apart from the determinant where there are some terms which change the sign. Those terms drop out and the remaining terms are only a polynomial in $\widehat{\mu}^4$. Moreover the integral over $S$ would be invariant under $S\to VSV^\dagger$ for all $V\in{\rm U}(2)$ without the determinant~\eqref{determinant-expansion}. Thus, we can also symmetrize those symmetry breaking terms as follows,
\begin{equation}
\begin{split}
\Tr S{U'}_{\rm bb}^{-1}\rightarrow\ & \frac{1}{2}\Tr S\Tr{U'}_{\rm bb}^{-1},\quad \Tr S^{-1}{U'}_{\rm bb}\rightarrow \frac{1}{2}\Tr S^{-1}\Tr{U'}_{\rm bb},\\
\Tr^2 S{U'}_{\rm bb}^{-1}\rightarrow\ &\frac{1}{3}\left[(\Tr^2 S-\det S)(\Tr^2{U'}_{\rm bb}^{-1}-\det{U'}_{\rm bb}^{-1})+3\det S{U'}_{\rm bb}^{-1}\right],\\
 \Tr^2 S{U'}_{\rm bb}^{-1}\tau_3 \rightarrow\ &\frac{1}{3}\left[(\Tr^2 S-\det S)(\Tr^2{U'}_{\rm bb}^{-1}\tau_3+\det{U'}_{\rm bb}^{-1})-3\det S{U'}_{\rm bb}^{-1}\right].
\end{split}
\end{equation}
Since we have also traces of squares of $2\times 2$ matrices in Eq.~\eqref{determinant-expansion} we additionally need the identity $\Tr B^2=\Tr^2 B-2\det B$, which holds for any $2\times 2$ matrix. Then the average over $S$ can be performed with the help of the integral
\begin{equation}
\int_{{\rm U}(2)} dS ~\Tr^j S\, {\det}^{-l} S \exp(N\Tr S)=\frac{4\pi^3(2l-4)!}{(l-2)!(l-1)!(2l-j-4)!}N^{2l-j-4}
\end{equation}
for any two integers $j\geq0$ and $l\geq j/2+2$. Collecting everything, the partition function is
\begin{align}
Z_N^{(1,1)}=
\;&\frac{1}{4\pi^4 } \int d{U'}_{\rm bb} d{U'}_{\rm ff}~\exp[-L({U'}_{\rm bb},\widehat{\kappa}_{\rm b})+L({U'}_{\rm ff},\widehat{\kappa}_{\rm f})]
\notag \\
&\times\biggl\{(N^2-\widehat{\mu}^4)^2+N^2 \Big((N^2-\widehat{\mu}^4) \Tr {U'}_{\rm bb}^{-1}\Tr{U'}_{\rm ff}^{-1}-2(N^2+\widehat{\mu}^4) \det{U'}_{\rm bb}^{-1}{U'}_{\rm ff}^{-1}\Big)
\notag \\
&+N\biggl[N^2\left((N-1)\Tr^2{U'}_{\rm bb}^{-1}\det{U'}_{\rm ff}^{-1}+(N+1)\Tr^2{U'}_{\rm ff}^{-1}\det{U'}_{\rm bb}^{-1}\right)
\notag \\
&-\widehat{\mu}^4\left((N-1)\Tr^2{U'}_{\rm bb}^{-1}\tau_3\det{U'}_{\rm ff}^{-1}+(N+1)\Tr^2{U'}_{\rm ff}^{-1}\tau_3\det{U'}_{\rm bb}^{-1}\right)\biggl]
\notag \\
&+N^2(N^2-1)\det({U'}_{\rm bb}^{-1}{U'}_{\rm ff}^{-1})^2\left(\Tr {U'}_{\rm bb}\Tr{U'}_{\rm ff}+1\right)\biggl\}\;.
\label{partition-bf.3}
\end{align}

In the next step we choose the coordinates
\begin{align}
{U'}_{\rm bb}=\;&Le^{-\vartheta_1\tau_3/2}e^{-\vartheta_2\tau_2/2}\tau_1e^{z_1\1_2+z_2 \tau_1}e^{\vartheta_2\tau_2/2}e^{\vartheta_1\tau_3/2}
\notag \\
=\; &Le^{z_1}\left[\sinh z_2\1_2+\cosh z_2\left(\begin{array}{cc} i \sinh\vartheta_2 &  e^{-\vartheta_1} \cosh\vartheta_2\\ e^{\vartheta_1} \cosh\vartheta_2  &  -i \sinh\vartheta_2 \end{array}\right)\right],
\\
{U'}_{\rm ff}=\;&e^{-i\varphi_1\tau_3/2}e^{-i\varphi_2\tau_2/2}\tau_1e^{i(z_3\1_2+z_4 \tau_1)}e^{i\varphi_2\tau_2/2}e^{i\varphi_1\tau_3/2}
\notag \\
=\;&e^{iz_3}\left[i\sin z_4\1_2+\cos z_4\left(\begin{array}{cc} -\sin\varphi_2 &  e^{-i\varphi_1} \cos\varphi_2 \\ e^{i\varphi_1} \cos\varphi_2  &   \sin\varphi_2 \end{array}\right)\right],
\end{align}
with $z_1,z_2,\vartheta_1,\vartheta_2\in\mathbb{R}$, $\varphi_1\in[-\pi,\pi]$, $\varphi_2,z_3,z_4\in[-\pi/2,\pi/2]$, where the $L$ in $U'_{\rm bb}$ is necessary to ensure its positivity condition. In these coordinates the Lagrangians become
\begin{align}
-L({U'}_{\rm bb},L\widehat{\kappa}_{\rm b})=\;&-2L\widehat{\kappa}_{\rm b}e^{z_1}\cosh z_2\cosh\vartheta_1\cosh\vartheta_2
\notag \\
&-\widehat{\mu}^2e^{2z_1}\left(1+2\cosh^2z_2\sinh^2\vartheta_2\right)-Ne^{2z_1}\cosh\left(2z_2\right)+2Nz_1\;, 
\\
L({U'}_{\rm ff},\widehat{\kappa}_{\rm f})=\; &2\widehat{\kappa}_{\rm f}e^{iz_3}\cos z_4\cos\varphi_1\cos\varphi_2
\notag \\
&+\widehat{\mu}^2e^{2iz_3}\left(1-2\cos^2z_4\sin^2\varphi_2\right)+Ne^{2iz_3}\cos\left(2z_4\right)-2iN z_3\;.
\end{align}
The corresponding measures are
\begin{equation}
\begin{split}
dU'_{\rm bb}=\;&2e^{4z_1}\cosh^2z_2\cosh\vartheta_2 ~d\vartheta_1 d\vartheta_2 dz_1 dz_2,\\
dU'_{\rm ff}=\;&2e^{4i z_3}\cos^2z_4\cos\varphi_2 ~d\varphi_1 d\varphi_2 dz_3 dz_4.
\end{split}
\end{equation}
Thus the partition function is
\begin{align}
Z_N^{(1,1)}=\;&\frac{1}{\pi^4 } \int  d\vartheta_1 d\vartheta_2 dz_1 dz_2 d\varphi_1 d\varphi_2 dz_3 dz_4 dz_4~\exp[-L({U'}_{\rm bb},\widehat{\kappa}_{\rm b})+L({U'}_{\rm ff},\widehat{\kappa}_{\rm f})]
\notag \\
&\times e^{4z_1+4i z_3}\cos^2z_4\cos\varphi_2\cosh^2z_2\cosh\vartheta_2
\notag \\
&\times\biggl\{(N^2-\widehat{\mu}^4)^2+N^2 \Big(4iL(N^2-\widehat{\mu}^4)e^{-z_1-i z_3}\sinh z_2\sin z_4-2(N^2+\widehat{\mu}^4)e^{-2z_1-2i z_3}\Big)
\notag \\
&-4Ne^{-2z_1-2i z_3}\biggl[N^2\Big((N-1)\sinh^2 z_2-(N+1)\sin^2 z_4\Big)
\notag \\
&+\widehat{\mu}^4\Big((N-1)\cosh^2 z_2\sinh^2\vartheta_2-(N+1)\cos^2 z_4\sin^2\varphi_2\Big)\biggl]
\notag \\
&+N^2(N^2-1)e^{-4z_1-4i z_3}\left(4iLe^{z_1+i z_3}\sinh z_2\sin z_4+1\right)\biggl\}\;.
\label{partition-bf.4}
\end{align}
The terms proportional to $\sin z_4$ vanish because they are odd around the origin while we integrate symmetrically,
\begin{align}
Z_N^{(1,1)}=\;&\frac{1}{\pi^4 } \int  d\vartheta_1 d\vartheta_2 dz_1 dz_2 d\varphi_1 d\varphi_2 dz_3 dz_4~ \exp[-L({U'}_{\rm bb},\widehat{\kappa}_{\rm b})+L({U'}_{\rm ff},\widehat{\kappa}_{\rm f})]
\notag \\
&\times e^{4z_1+4i z_3}\cos^2z_4\cos\varphi_2\cosh^2z_2\cosh\vartheta_2
\notag \\
&\times\biggl\{(N^2-\widehat{\mu}^4)^2-2N^2(N^2+\widehat{\mu}^4)e^{-2z_1-2i z_3}+N^2(N^2-1)e^{-4z_1-4i z_3}
\notag \\
&-4Ne^{-2z_1-2i z_3}\biggl[N^2\Big((N-1)\sinh^2 z_2-(N+1)\sin^2 z_4\Big)
\notag \\
&+\widehat{\mu}^4\Big((N-1)\cosh^2 z_2\sinh^2\vartheta_2-(N+1)\cos^2 z_4\sin^2\varphi_2\Big)\biggl]\biggl\}
\notag\\
=\;&\frac{4}{\pi^3 } \int  d\vartheta_2 dz_1 dz_2 d\varphi_2 dz_3 dz_4~ e^{-Ne^{2z_1}\cosh\left(2z_2\right)+Ne^{2iz_3}\cos\left(2z_4\right)}
\notag \\
&\times \exp\big[-\widehat{\mu}^2e^{2z_1}\left(1+2\cosh^2z_2\sinh^2\vartheta_2\right)+\widehat{\mu}^2e^{2iz_3}\left(1-2\cos^2z_4\sin^2\varphi_2\right)\big]
\notag \\
&\times e^{2(2+N)z_1+2i(2-N) z_3}\cos^2z_4\cos\varphi_2\cosh^2z_2\cosh\vartheta_2
\notag \\
&\times K_0(2L\widehat{\kappa}_{\rm b}e^{z_1}\cosh z_2\cosh\vartheta_2)I_0(2\widehat{\kappa}_{\rm f}e^{iz_3}\cos z_4\cos\varphi_2)
\notag \\
&\times\biggl\{N^4(1-e^{-2z_1-2i z_3})^2-2N^2\widehat{\mu}^4(1+e^{-2z_1-2i z_3})-N^2e^{-4z_1-4i z_3}+\widehat{\mu}^8
\notag \\
&-4Ne^{-2z_1-2i z_3}\biggl[N^2\Big((N-1)\sinh^2 z_2-(N+1)\sin^2 z_4\Big)
\notag \\
&+\widehat{\mu}^4\Big((N-1)\cosh^2 z_2\sinh^2\vartheta_2-(N+1)\cos^2 z_4\sin^2\varphi_2\Big)\biggl]\biggl\}\;.
\label{partition-bf.5}
\end{align}
In the second step, we expressed the integrals over $\vartheta_1$ and $\varphi_1$ as modified Bessel functions.

For large $N$ we have to perform a saddle point analysis. The angles $z_1$ and $z_2$ take their maximum uniquely at the origin, so that we have to expand in $(z_1,z_2)=(\delta z_1/\sqrt{N},\delta z_2/\sqrt{N})$. For the fermionic part we have two saddle points: namely, the expansions $(z_3,z_4)=(\delta z_3/\sqrt{N},$ $\delta z_4/\sqrt{N})$ and  $(z_3,z_4)=(\pi/2+\delta z_3/\sqrt{N},\pi/2+\delta z_4/\sqrt{N})$. The contribution of the latter is suppressed by a factor $1/N$. Nonetheless, for $(z_3,z_4)=(\delta z_3/\sqrt{N},\delta z_4/\sqrt{N})$ we have to expand the integrand in Eq.~\eqref{partition-bf.5} up to order $\mathcal{O}(1)$ since the leading order term of order $\mathcal{O}(N)$ vanishes after integration over the massive modes $\delta z_j$. The same holds for the order $\mathcal{O}(\sqrt{N})$. This expansion is quite lengthy and we give already the result of the partition function after integration over $\delta z_j$,
\begin{align}
\lim_{N\to\infty}Z_N^{(1,1)}=\;&\frac{1}{8\pi^2}\int_{-\pi}^\pi d\varphi_1\int_{-\pi/2}^{\pi/2} d\varphi_2\int_{\mathbb{R}^2}d\vartheta_1d\vartheta_2\cos\varphi_2\cosh\vartheta_2
\notag \\
&\times \exp[-2 L\widehat{\kappa}_\mathrm{b}\cosh\vartheta_1\cosh\vartheta_2+2 \widehat{\kappa}_\mathrm{f}\cos\varphi_1\cos\varphi_2-2\widehat{\mu}^2(\sin^2\varphi_2+\sinh^2\vartheta_2)]
\notag \\
&\times\biggl[8\widehat{\mu}^4\left(\sinh^2\vartheta_2-\sin^2\varphi_2+1\right)\left(\sinh^2\vartheta_2-\sin^2\varphi_2\right)
\notag \\
&+4\widehat{\mu}^2\biggl(2\left(\sinh^2\vartheta_2-\sin^2\varphi_2+1\right)\left(\widehat{\kappa}_\mathrm{f}\cos\varphi_1\cos\varphi_2+L\widehat{\kappa}_\mathrm{b}\cosh\vartheta_1\cosh\vartheta_2\right)
\notag \\
&-\frac{\sin^2\varphi_2+\sinh^2\vartheta_2}{2}\biggl)+2\left(\widehat{\kappa}_\mathrm{f}\cos\varphi_1\cos\varphi_2+L\widehat{\kappa}_\mathrm{b}\cosh\vartheta_1\cosh\vartheta_2\right)^2-1\biggl]\;.
\end{align}
After substituting $x=\sin\varphi_2$ and $y=\sinh\vartheta_2$ and integration over $\varphi_1$ and $\vartheta_1$, we arrive at Eq.~\eqref{Susy-limit}.


\begin{thebibliography}{11}

\bibitem{handbook:2010}
G.~Akemann, J.~Baik, and P.~Di~Francesco, eds., {\em {The Oxford Handbook of
  Random Matrix Theory}}.
\newblock Oxford University Press, Oxford, 2011.

\bibitem{Shuryak:1992pi}
E.~V. Shuryak and J.~J.~M. Verbaarschot, {\it {Random matrix theory and
  spectral sum rules for the Dirac operator in QCD}},  {\em Nucl. Phys. A} {\bf
  560} (1993) 306--320, [\href{http://arxiv.org/abs/hep-th/9212088}{{\tt
  arXiv:hep-th/9212088}}].

\bibitem{Verbaarschot:1993pm}
J.~J.~M. Verbaarschot and I.~Zahed, {\it {Spectral density of the QCD Dirac
  operator near zero virtuality}},  {\em Phys. Rev. Lett.} {\bf 70} (1993)
  3852--3855, [\href{http://arxiv.org/abs/hep-th/9303012}{{\tt
  arXiv:hep-th/9303012}}].

\bibitem{Verbaarschot:1994ip}
J.~J.~M. Verbaarschot and I.~Zahed, {\it {Random matrix theory and QCD in
  three-dimensions}},  {\em Phys. Rev. Lett.} {\bf 73} (1994) 2288--2291,
  [\href{http://arxiv.org/abs/hep-th/9405005}{{\tt arXiv:hep-th/9405005}}].

\bibitem{Gasser:1987ah}
J.~Gasser and H.~Leutwyler, {\it {Thermodynamics of Chiral Symmetry}},  {\em
  Phys. Lett. B} {\bf 188} (1987) 477--481.

\bibitem{Leutwyler:1992yt}
H.~Leutwyler and A.~V. Smilga, {\it {Spectrum of Dirac operator and role of
  winding number in QCD}},  {\em Phys. Rev. D} {\bf 46} (1992) 5607--5632.
  
\bibitem{Gasser:1983yg}
J.~Gasser and H.~Leutwyler, {\it {Chiral Perturbation Theory to One Loop}},
  {\em Annals Phys.} {\bf 158} (1984) 142--210.

\bibitem{Leutwyler:1993iq}
H.~Leutwyler, {\it {On the foundations of chiral perturbation theory}},  {\em
  Annals Phys.} {\bf 235} (1994) 165--203,
  [\href{http://arxiv.org/abs/hep-ph/9311274}{{\tt arXiv:hep-ph/9311274}}].  

\bibitem{Luz:2006vu}
M.~Luz, {\it {Determining F(pi) from spectral sum rules}},  {\em Phys. Lett. B}
  {\bf 643} (2006) 235--239, [\href{http://arxiv.org/abs/hep-lat/0607022}{{\tt
  arXiv:hep-lat/0607022}}].

\bibitem{Kanazawa:2009ks}
T.~Kanazawa, T.~Wettig, and N.~Yamamoto, {\it {Chiral Lagrangian and spectral
  sum rules for dense two-color QCD}},  {\em JHEP} {\bf 08} (2009) 003,
  [\href{http://arxiv.org/abs/0906.3579}{{\tt arXiv:0906.3579}}].

\bibitem{Kanazawa:2011tt}
T.~Kanazawa, T.~Wettig and N.~Yamamoto, 
{\it {Singular values of the Dirac operator in dense QCD-like theories}}, {\em JHEP} {\bf 12} (2011) 007, [\href{http://arxiv.org/abs/arXiv:1110.5858}{{\tt arXiv:1110.5858}}].

\bibitem{Bialas:2010hb}
P.~Bialas, Z.~Burda, and B.~Petersson, {\it {Random matrix model for QCD$_3$
  staggered fermions}},  {\em Phys. Rev. D} {\bf 83} (2011) 014507,
  [\href{http://arxiv.org/abs/1006.0360}{{\tt arXiv:1006.0360}}].

\bibitem{Fyodorov:1997}
Y.~V.~Fyodorov, B.~A.~Khoruzhenko, and H.-J.~Sommers, {\it {Almost-Hermitian Random Matrices: Eigenvalue Density in the Complex Plane}},  {\em Phys.Lett. A} {\bf 226} (1997) 46--52,
  [\href{https://arxiv.org/abs/cond-mat/9606173}{{\tt arXiv:cond-mat/9606173}}];
  {\it Almost-Hermitian Random Matrices: Crossover from Wigner-Dyson to Ginibre eigenvalue statistics}, {\em Phys. Rev. Lett.} {\bf 79} (1997) 557--560,
  [\href{https://arxiv.org/abs/cond-mat/9703152}{{\tt arXiv:cond-mat/9703152}}].

\bibitem{Fyodorov:1997b}
Y.~V.~Fyodorov, {\it {Almost-Hermitian Random Matrices: Applications to the Theory of Quantum Chaotic Scattering and Beyond}},  {\em NATO ASI Conference: ``Supersymmetry and Trace formula: Chaos and Disorder'', Cambridge 1997},
  [\href{https://arxiv.org/abs/chao-dyn/9712006}{{\tt arXiv:chao-dyn/9712006}}].

\bibitem{Sommers}
H.~J.~Sommers, A.~Crisanti, H.~Sompolinsky, and Y.~Stein,
{\it Spectrum of large random asymmetric matrices},
{\em Phys. Rev. Lett.} {\bf 60} (1988) 1895--1898.

\bibitem{Akemann:2001bf}
G.~Akemann, {\it {Microscopic correlations of non-Hermitian Dirac operators in
  three-dimensional QCD}},  {\em Phys. Rev. D} {\bf 64} (2001) 114021,
  [\href{http://arxiv.org/abs/hep-th/0106053}{{\tt arXiv:hep-th/0106053}}].

\bibitem{Akemann:2007rf}
G.~Akemann, {\it {Matrix Models and QCD with Chemical Potential}},  {\em Int.
  J. Mod. Phys. A} {\bf 22} (2007) 1077--1122,
  [\href{http://arxiv.org/abs/hep-th/0701175}{{\tt arXiv:hep-th/0701175}}].
  
\bibitem{Kanazawa:2013}
T.~Kanazawa, {\it Dirac spectra in dense QCD}, {\em Springer Theses} {\bf 124}, Springer Japan, Tokyo (2013).
  
\bibitem{Kanazawa:2018}
T.~Kanazawa and M.~Kieburg, {\it {Symmetry Transition Preserving Chirality in QCD: A Versatile Random Matrix Model}},   	{\em Phys. Rev. Lett.} {\bf 120} (2018) 242001, [\href{https://arxiv.org/abs/1803.04122}{{\tt arXiv:1803.04122}}].

\bibitem{Kanazawa:2018kbo}
T.~Kanazawa and M.~Kieburg, {\it {GUE-chGUE transition preserving chirality at finite matrix size}}, 
{\em J. Phys. A} {\bf 51} (2018) 345202, 
[\href{https://arxiv.org/abs/1804.03985}{{\tt arXiv:1804.03985}}].

\bibitem{Zirnbauer}
M.~Zirnbauer, {\it The supersymmetry method of random matrix theory}, {\em Encyclopedia of Mathematical Physics} {\bf 5} (2006) 151, ed. J.-P.~Franoise, G.~L.~Naber, and S.~T.~Tsou, (Oxford: Elsevier),
  [\href{https://arxiv.org/abs/math-ph/0404057}{{\tt arXiv:math-ph/0404057}}].

\bibitem{Guhr}
T.~Guhr, {\it {Supersymmetry}},  {\em Chapter 7 in}~\cite{handbook:2010} (2011),
  [\href{https://arxiv.org/abs/1005.0979}{{\tt arXiv:1005.0979}}].
  
\bibitem{Bernard:1993}
C.~W.~Bernard and M.~F.~L.~Golterman, {\it Partially quenched gauge theories and an application to staggered fermions}, {\em Phys.Rev. D} {\bf 49} (1994) 486--494,
  [\href{https://arxiv.org/abs/hep-lat/9306005}{{\tt arXiv:hep-lat/9306005}}].

\bibitem{Golterman:2009}
 M.~F.~L.~Golterman, {\it Applications of chiral perturbation theory to lattice QCD}, {\em conference proceedings at the International
                        School, 93rd Session, Les Houches, France, August 3-28, (2009)}, {\it Modern perspectives in lattice QCD: Quantum field theory and high performance computing} (2009) 423--515,
  [\href{https://arxiv.org/abs/0912.4042}{{\tt arXiv:0912.4042}}].

\bibitem{Mehta_book}
M.~L.~Mehta, {\em {Random Matrices}}, Academic Press, Amsterdam, 3rd~ed. (2004).

\bibitem{Akemann:2011}
G.~Akemann and T.~Nagao, {\it {Random Matrix Theory for the Hermitian Wilson Dirac Operator and the chGUE-GUE Transition}},  {\em JHEP} {\bf 2011} (2011) 60,
  [\href{https://arxiv.org/abs/1108.3035}{{\tt arXiv:1108.3035}}].
  
\bibitem{Damgaard:2010cz}
P.~H.~Damgaard, K.~Splittorff and J.~J.~M.~Verbaarschot, 
{\it Microscopic Spectrum of the Wilson Dirac Operator}, 
{\em Phys. Rev. Lett.} {\bf 105} (2010) 162002, [\href{https://arxiv.org/abs/1001.2937}{\tt arXiv:1001.2937}].

\bibitem{Susskind:1976jm}
L.~Susskind, {\it Lattice fermions}, {\em Phys. Rev. D} {\bf 16} (1977) 3031--3039.

\bibitem{Damgaard:1998}
P.~H.~Damgaard, U.~M.~Heller, A.~Krasnitz, and T.~Madsen, {\it A Quark-Antiquark Condensate in Three-Dimensional QCD}, {\em Phys. Lett. B} {\bf 440} (1998) 129--135,
[\href{https://arxiv.org/abs/hep-lat/9803012}{{\tt arXiv:hep-lat/9803012}}].

\bibitem{Damgaard:2002} 
 P.~H.~Damgaard, U.~M.~Heller, R.~Niclasen, and B.~Svetitsky, {\it Patterns of spontaneous chiral symmetry breaking in vector like gauge theories}, {\em Nucl. Phys. B} {\bf 633} (2002) 97--113,
  [\href{https://arxiv.org/abs/hep-lat/0110028}{{\tt arXiv:hep-lat/0110028}}].

\bibitem{Kieburg:2014}
M. Kieburg, J. J. M. Verbaarschot, and S. Zafeiropoulos, {\it A classification of 2-dim Lattice Theory}, {\em PoS LATTICE 2013} (2013) 337,
  [\href{https://arxiv.org/abs/1310.6948}{{\tt arXiv:1310.6948}}];
{\it Dirac Spectra of 2-dimensional QCD-like theories}, {\em Phys. Rev. D} {\bf 90} (2014) 085013,
  [\href{https://arxiv.org/abs/1405.0433}{{\tt arXiv:1405.0433}}].

\bibitem{Kieburg:2017rrk}
M.~Kieburg and T.~R. W\"{u}rfel, {\it {Shift of symmetries of naive and
  staggered fermions in QCD-like lattice theories}},  {\em Phys. Rev. D} {\bf
  96} (2017) 034502,
  [\href{http://arxiv.org/abs/1703.08083}{{\tt arXiv:1703.08083}}];
   {\it {Global Symmetries of Naive and Staggered
  Fermions in Arbitrary Dimensions}}, EPJ Web of Conferences {\bf 175} (2018) 04006, 
  \href{http://arxiv.org/abs/1710.03049}{{\tt arXiv:1710.03049}}.

\bibitem{Bott}
R.~Bott, {\it The Stable Homotopy of the Classical Groups}, {\em Annals of Mathematics} {\bf 70} (1959) 313--337; {\it The 
Periodicity Theorem for the Classical  Groups and some of  its Applications},  {\em Advances in Mathematics} {\bf 4} (1970) 353--411.

\bibitem{Osborn:2011}
J.~C.~Osborn, {\it Taste breaking in staggered fermions from random matrix theory}, {\em Nucl. Phys. Proc. Suppl.} {\bf 129} (2004) 886--888,
  [\href{https://arxiv.org/abs/hep-lat/0309123}{{\tt hep-lat/0309123}}];
   {\it Staggered chiral random matrix theory}, {\em Phys.Rev. D} {\bf 83} (2011) 034505,
  [\href{https://arxiv.org/abs/1012.4837}{{\tt arXiv:1012.4837}}];
  {\it Chiral random matrix theory for staggered fermions}, {\em PoS} {\bf LATTICE2011} (2011) 110,
  [\href{https://arxiv.org/abs/1204.5497}{{\tt arXiv:1204.5497}}].

\bibitem{Berezin}
F.~A.~Berezin, {\it {Introduction to Superanalysis}}, D. Reidel Publishing Company, Dordrecht, 1st
ed. (1987).

\bibitem{Sommers:2007}
H.-J.~Sommers, {\it {Superbosonization}},  {\em Acta Phys. Pol. B} {\bf 38} (2007) 4105--4110,
  [\href{https://arxiv.org/abs/0710.5375}{{\tt arXiv:0710.5375}}].

\bibitem{Zirnbauer:2008}
P.~Littelmann, H.-J.~Sommers, and M.~R.~Zirnbauer, {\it {Superbosonization of invariant random matrix ensembles}},  {\em Commun. Math. Phys.} 283 (2008) 343,
  [\href{https://arxiv.org/abs/0707.2929}{{\tt arXiv:0707.2929}}].

\bibitem{Kieburg:2008}
M.~Kieburg, H.-J.~Sommers, and T.~Guhr, {\it {Comparison of the superbosonization formula and the generalized Hubbard-Stratonovich transformation}},  {\em J. Phys. A} {\bf 42} (2009) 275206,
  [\href{https://arxiv.org/abs/0905.3256}{{\tt arXiv:0905.3256}}].

\bibitem{Ginsparg:1980ef}
P.~H. Ginsparg, {\it {First Order and Second Order Phase Transitions in Gauge
  Theories at Finite Temperature}},  {\em Nucl. Phys. B} {\bf 170} (1980)
  388--408.

\bibitem{Appelquist:1981vg}
T.~Appelquist and R.~D. Pisarski, {\it {High-Temperature Yang-Mills Theories
  and Three-Dimensional Quantum Chromodynamics}},  {\em Phys. Rev. D} {\bf 23}
  (1981) 2305.

\bibitem{Nadkarni:1982kb}
S.~Nadkarni, {\it {Dimensional reduction in finite-temperature quantum
  chromodynamics}},  {\em Phys. Rev. D} {\bf 27} (1983) 917.

\bibitem{Nadkarni:1988fh}
S.~Nadkarni, {\it {Dimensional Reduction in Finite Temperature Quantum
  Chromodynamics. 2.}},  {\em Phys. Rev. D} {\bf 38} (1988) 3287.

\bibitem{Landsman:1989be}
N.~P.~Landsman, {\it {Limitations to Dimensional Reduction at High
  Temperature}},  {\em Nucl. Phys. B} {\bf 322} (1989) 498--530.

\bibitem{Kajantie:1995dw}
K.~Kajantie, M.~Laine, K.~Rummukainen, and M.~E.~Shaposhnikov, {\it {Generic
  rules for high temperature dimensional reduction and their application to the
  standard model}},  {\em Nucl. Phys. B} {\bf 458} (1996) 90--136,
  [\href{http://arxiv.org/abs/hep-ph/9508379}{{\tt arXiv:hep-ph/9508379}}].

\bibitem{Braaten:1995jr}
E.~Braaten and A.~Nieto, {\it {Free energy of QCD at high temperature}},  {\em
  Phys. Rev. D} {\bf 53} (1996) 3421--3437,
  [\href{http://arxiv.org/abs/hep-ph/9510408}{{\tt arXiv:hep-ph/9510408}}].

\bibitem{Chandrasekharan:1998yx}
S.~Chandrasekharan, D.~Chen, N.~H. Christ, W.-J. Lee, R.~Mawhinney, and P.~M.
  Vranas, {\it {Anomalous chiral symmetry breaking above the QCD phase
  transition}},  {\em Phys. Rev. Lett.} {\bf 82} (1999) 2463--2466,
  [\href{http://arxiv.org/abs/hep-lat/9807018}{{\tt arXiv:hep-lat/9807018}}].

\bibitem{Stephanov:1996he}
M.~A. Stephanov, {\it {Chiral symmetry at finite T, the phase of the Polyakov
  loop and the spectrum of the Dirac operator}},  {\em Phys. Lett. B} {\bf 375}
  (1996) 249--254, [\href{http://arxiv.org/abs/hep-lat/9601001}{{\tt
  arXiv:hep-lat/9601001}}].

\bibitem{Chandrasekharan:1995gt}
S.~Chandrasekharan and N.~H. Christ, {\it {Dirac spectrum, axial anomaly and
  the QCD chiral phase transition}},  {\em Nucl. Phys. Proc. Suppl.} {\bf 47}
  (1996) 527--534, [\href{http://arxiv.org/abs/hep-lat/9509095}{{\tt
  arXiv:hep-lat/9509095}}].

\bibitem{Bornyakov:2008bg}
V.~G. Bornyakov, E.~V. Luschevskaya, S.~M. Morozov, M.~I. Polikarpov, E.~M.
  Ilgenfritz, and M.~Muller-Preussker, {\it {The Topological structure of SU(2)
  gluodynamics at $T>0$: An Analysis using the Symanzik action and
  Neuberger overlap fermions}},  {\em Phys. Rev. D} {\bf 79} (2009) 054505,
  [\href{http://arxiv.org/abs/0807.1980}{{\tt arXiv:0807.1980}}].

\bibitem{Bornyakov:2008im}
V.~G. Bornyakov, E.~M. Ilgenfritz, B.~V. Martemyanov, and M.~Muller-Preussker,
  {\it {The Dyonic picture of topological objects in the deconfined phase}},
  {\em Phys. Rev. D} {\bf 79} (2009) 034506,
  [\href{http://arxiv.org/abs/0809.2142}{{\tt arXiv:0809.2142}}].

\bibitem{Shuryak:2017kct}
E.~Shuryak, {\it {Instanton-dyon ensembles reproduce deconfinement and chiral
  restoration phase transitions}},  {\em EPJ Web Conf.} {\bf 175} (2018) 12001,
  [\href{http://arxiv.org/abs/1710.03611}{{\tt arXiv:1710.03611}}].

\bibitem{Cherman:2016hcd}
A.~Cherman, T.~Sch\"afer, and M.~\"Unsal, {\it {Chiral Lagrangian from Duality and
  Monopole Operators in Compactified QCD}},  {\em Phys. Rev. Lett.} {\bf 117}
  (2016), 081601, [\href{http://arxiv.org/abs/1604.06108}{{\tt
  arXiv:1604.06108}}].

\bibitem{Kanazawa:2017mgw}
T.~Kanazawa, M.~\"Unsal, and N.~Yamamoto, {\it {Phases of circle-compactified QCD
  with adjoint fermions at finite density}},  {\em Phys. Rev. D} {\bf 96}
  (2017), 034022, [\href{http://arxiv.org/abs/1703.06411}{{\tt
  arXiv:1703.06411}}].

\bibitem{Unsal:2007vu}
M.~\"Unsal, {\it {Abelian duality, confinement, and chiral symmetry breaking in
  QCD(adj)}},  {\em Phys. Rev. Lett.} {\bf 100} (2008) 032005,
  [\href{http://arxiv.org/abs/0708.1772}{{\tt arXiv:0708.1772}}].

\bibitem{Unsal:2007jx}
M.~\"Unsal, {\it {Magnetic bion condensation: A New mechanism of confinement and
  mass gap in four dimensions}},  {\em Phys. Rev. D} {\bf 80} (2009) 065001,
  [\href{http://arxiv.org/abs/0709.3269}{{\tt arXiv:0709.3269}}].

\bibitem{Unsal:2008eg}
M.~\"Unsal, {\it {Quantum phase transitions and new scales in QCD-like
  theories}},  {\em Phys. Rev. Lett.} {\bf 102} (2009) 182002,
  [\href{http://arxiv.org/abs/0807.0466}{{\tt arXiv:0807.0466}}].

\bibitem{Son:2000xc}
D.~T.~Son and M.~A.~Stephanov, {\it QCD at finite isospin density}, {\em Phys. Rev. Lett.} {\bf 86} (2001) 592--595, 
  [\href{http://arxiv.org/abs/hep-ph/0005225}{{\tt arXiv:hep-ph/0005225}}].

\bibitem{Splittorff:2000mm}
K.~Splittorff, D.~T.~Son, and M.~A.~Stephanov, {\it QCD-like theories at finite baryon and isospin density}, {\em Phys. Rev. D} {\bf 64} (2001) 016003, 
  [\href{http://arxiv.org/abs/hep-ph/0012274}{{\tt arXiv:hep-ph/0012274}}].

\bibitem{deForcrand:2007uz}
P.~de Forcrand, M.~A.~Stephanov, and U.~Wenger, {\it On the phase diagram of QCD at finite isospin density}, {\em PoS} {\bf LAT2007} (2007) 237, 
  [\href{http://arxiv.org/abs/0711.0023}{{\tt arXiv:0711.0023}}].
  
\bibitem{BES}
B.~B.~Brandt, G.~Endr\"odi, and S.~Schmalzbauer, {\it QCD phase diagram for nonzero isospin-asymmetry}, {\em Phys. Rev. D} {\bf 97} (2018) 054514, 
  [\href{http://arxiv.org/abs/1712.08190}{{\tt arXiv:1712.08190}}].


\bibitem{Kogut:2000ek}
J.~B.~Kogut, M.~A.~Stephanov, D.~Toublan, J.~J.~M.~Verbaarschot, and A.~Zhitnitsky, {\it QCD-like theories at finite baryon density}, {\em Nucl. Phys. B} {\bf 582} (2000) 477--513, 
  [\href{http://arxiv.org/abs/hep-ph/0001171}{{\tt arXiv:hep-ph/0001171}}].

\bibitem{Kogut:2001na}
J.~B.~Kogut, D.~K.~Sinclair, S.~J.~Hands, and S.~E.~Morrison, {\it Two-color QCD at nonzero quark-number density}, {\em Phys. Rev. D} {\bf 64} (2001) 094505, 
  [\href{http://arxiv.org/abs/hep-lat/0105026}{{\tt arXiv:hep-lat/0105026}}].

\bibitem{Dunne:2002vb}
G.~V.~Dunne and S.~M.~Nishigaki, {\it Two-color QCD in 3D at finite baryon density}, {\em Nucl. Phys. B} {\bf 654} (2003) 445--465, 
  [\href{http://arxiv.org/abs/hep-ph/0210219}{{\tt arXiv:hep-ph/0210219}}].

\bibitem{Bilgici:2008qy}
E.~Bilgici, F.~Bruckmann, C.~Gattringer, and C.~Hagen, {\it {Dual quark
  condensate and dressed Polyakov loops}},  {\em Phys. Rev. D} {\bf 77} (2008)
  094007, [\href{http://arxiv.org/abs/0801.4051}{{\tt arXiv:0801.4051}}].

\bibitem{DeGrand:2006qb}
T.~DeGrand and R.~Hoffmann, {\it {QCD with one compact spatial dimension}},
  {\em JHEP} {\bf 02} (2007) 022,
  [\href{http://arxiv.org/abs/hep-lat/0612012}{{\tt arXiv:hep-lat/0612012}}].

\bibitem{Gattringer:2002tg}
C.~Gattringer and S.~Schaefer, {\it {New findings for topological excitations
  in SU(3) lattice gauge theory}},  {\em Nucl. Phys. B} {\bf 654} (2003) 30--60,
  [\href{http://arxiv.org/abs/hep-lat/0212029}{{\tt arXiv:hep-lat/0212029}}].
  
\bibitem{Misumi:2014raa}
T.~Misumi and T.~Kanazawa, {\it {Adjoint QCD on $\mathbb{R}^3\times S^1$ with
  twisted fermionic boundary conditions}},  {\em JHEP} {\bf 06} (2014) 181,
  [\href{http://arxiv.org/abs/1405.3113}{{\tt arXiv:1405.3113}}].

\bibitem{Fukushima:2008su}
K.~Fukushima, {\it Characteristics of the eigenvalue distribution of the Dirac operator in dense two-color QCD}, {\em JHEP} {\bf 07} (2008) 083,
  [\href{http://arxiv.org/abs/0806.1104}{{\tt arXiv:0806.1104}}].

\bibitem{Hub}
J.~Hubbard, {\it Calculation of Partition Functions}, {\em Phys. Rev. Lett.} {\bf 3} (1959) 77--78.
  
\bibitem{Strat}
R.~L.~Stratonovich, {\it On a Method of Calculating Quantum Distribution Functions}, {\em Soviet Physics Doklady} {\bf 2} (1957) 416.

\bibitem{Osborn:2004}
J.~C.~Osborn, {\it {Universal results from an alternate random matrix model for QCD with a baryon chemical potential}},  {\em Phys. Rev. Lett.} {\bf 93} (2004) 222001,
  [\href{https://arxiv.org/abs/hep-th/0403131}{{\tt arXiv:hep-th/0403131}}].

\bibitem{Stephanov:1996ki}
M.~A.~Stephanov, {\it {Random matrix model of QCD at finite density and the
  nature of the quenched limit}},  {\em Phys. Rev. Lett.} {\bf 76} (1996)
  4472--4475, [\href{http://arxiv.org/abs/hep-lat/9604003}{{\tt
  arXiv:hep-lat/9604003}}].

\bibitem{Damgaard:1997pw}
P.~H.~Damgaard and S.~M.~Nishigaki, {\it Universal massive spectral correlators and QCD in three-dimensions}, {\em Phys. Rev. D} {\bf 57} (1998) 5299--5302, [\href{http://arxiv.org/abs/hep-th/9711096}{{\tt arXiv:hep-th/9711096}}].

\bibitem{Szabo:2005gi}
R.~J.~Szabo, {\it Finite volume gauge theory partition functions in three dimensions}, {\em Nucl. Phys. B} {\bf 723} (2005) 163--197, [\href{http://arxiv.org/abs/hep-th/0504202}{{\tt arXiv:hep-th/0504202}}].

\bibitem{Verbaarschot}
J.~J.~M.~Verbaarschot, {\it {Quantum Chromodynamics}},  {\em Chapter 32 in}~\cite{handbook:2010} (2011),
  [\href{https://arxiv.org/abs/0910.4134}{{\tt arXiv:0910.4134}}].

\bibitem{Pfaffian}
A.~Borodin, {\it Determinantal Point Processes},   {\em Chapter 11 in}~\cite{handbook:2010} (2011),
  [\href{https://arxiv.org/abs/0911.1153}{{\tt arXiv:0911.1153}}].

\bibitem{Harish}
Harish-Chandra, {\it Invariant Differential operators on a semisimple Lie algebra}, {\em Proc. Natl. Acad. Sci. USA} {\bf 42} (1956) 252--253.

\bibitem{IZ}
C. Itzykson and J. B. Zuber, {\it The planar approximation II}, {\em J. Math. Phys.} {\bf 21} (1980) 411--421.

\bibitem{noncomp1}
Y.~V.~Fyodorov and E.~Strahov, {\it Characteristic polynomials of random Hermitian matrices and Duistermaat-Heckman localisation on non-compact Kaehler manifolds}, {\em Nucl.Phys. B} {\bf 630} (2002) 453--491,
  [\href{https://arxiv.org/abs/math-ph/0201045}{{\tt arXiv:math-ph/0201045}}].

\bibitem{noncomp2}
M.~Kieburg, J.~J.~M.~Verbaarschot, and S.~Zafeiropoulos, {\it Spectral Properties of the Wilson Dirac Operator and random matrix theory}, {\em Phys. Rev. D} {\bf 88} (2013) 094502,
  [\href{https://arxiv.org/abs/1307.7251}{{\tt arXiv:1307.7251}}].

\bibitem{NIST:DLMF}
``{\it NIST Digital Library of Mathematical Functions}.'' Release 1.0.15 of
  2017-06-01. F. W. J. Olver, A. B. Olde Daalhuis, D. W. Lozier, B. I.
  Schneider, R. F. Boisvert, C. W. Clark, B. R. Miller, and B. V. Saunders,
  eds.
\newblock \url{http://dlmf.nist.gov/}.

\bibitem{BK}
F.~A.~Berezin and F.~I.~Karpelevich, {\it Zonal spherical functions and Laplace operators on some symmetric spaces}, {\em Doklady Akad. Nauk. SSSR} {\bf 118}  (1958) 9--12.

\bibitem{Wettig}
T.~Guhr and T.~Wettig, {\it An Itzykson-Zuber-like Integral and Diffusion for Complex Ordinary and Supermatrices}, {\em J. Math. Phys.} {\bf 37} (1996) 6395--6413
[\href{https://arxiv.org/abs/hep-th/9605110}{{\tt arXiv:hep-th/9605110}}].

\bibitem{Schur}
 I.~Schur, {\it \"Uber die Darstellung der symmetrischen und der alternirenden Gruppe durch gebrochene lineare Substitutionen}, {\em J. Reine Angew. Math.} {\bf 139} (1911) 155--250.

\bibitem{deBruijn}
N.~G.~de Bruijn, {\it On some multiple integrals involving determinants}, {\em J. Indian Math. Soc.} {\bf 19} (1955) 133--151.
  
  
  

\end{thebibliography}
\end{document}